\documentclass[manuscript]{aastex61}

\hypersetup{linkcolor=red,citecolor=blue,filecolor=cyan,urlcolor=magenta}

\shorttitle{SIGGMA. II}
\shortauthors{Liu et al.}
\setwatermarkfontsize{100 pt}

\usepackage[flushleft]{threeparttable}
\usepackage{threeparttablex}
\usepackage{graphicx}
\usepackage[caption = false]{subfig}
\usepackage{float}
\usepackage{longtable}
\newcommand{\hii}{{\rm H\,}{{\sc ii}}}
\newcommand{\hi}{{\rm H\,}{{\sc i}}}
\newcommand{\kms}{\,km\,s$^{-1}$}
\newcommand{\kpc}{\,kpc}
\newcommand{\ghz}{\,GHz}
\newcommand{\um}{\mu m}
\begin{document}

\title{Survey of Ionized Gas of the Galaxy, Made with the Arecibo telescope (SIGGMA): Inner Galaxy Data Release}

\correspondingauthor{Bin Liu}
\email{bliu@nao.cas.cn}
\author[0000-0002-1311-8839]{Bin Liu}
\affil{National Astronomical Observatories, Chinese Academy of Sciences, Beijing 100012, China}
\affil{CAS Key Laboratory of FAST, NAOC, Chinese Academy of Sciences}
\affil{West Virginia University, Morgantown, WV 26506, USA}
\author{L. D. Anderson}
\affil{West Virginia University, Morgantown, WV 26506, USA}
\affil{Adjunct Astronomer at the Green Bank Observatory, P.O. Box 2, Green Bank WV 24944, USA}
\affil{Center for Gravitational Waves and Cosmology, West Virginia University, Chestnut Ridge Research Building, Morgantown, WV 26505,
USA}
\author{Travis McIntyre}
\affil{New Mexico Legislative Finance Committee, Santa Fe, NM 87501, USA}
\author{D. Anish Roshi}
\affil{National Radio Astronomy Observatory\footnote{The National Radio Astronomy Observatory is a facility of the National Science Foundation operated under a cooperative agreement by Associated Universities, Inc.}, 520 Edgemont Rd, Charlottesville, VA 22903, USA}
\author{Ed Churchwell}
\affil{University of Wisconsin-Madison, Madison, WI 53706, USA}
\author{Robert Minchin}
\affil{Arecibo Observatory, HC03 Box 53995, Arecibo 00612, Puerto Rico}
\author{Yervant Terzian}
\affil{Cornell University, Ithaca, NY 14853, USA}

\begin{abstract}
The Survey of Ionized Gas of the Galaxy, Made with the Arecibo telescope (SIGGMA) provides a fully-sampled view of the radio recombination line (RRL) emission from the portion of the Galactic plane visible by Arecibo.
Observations use the Arecibo L-band Feed Array (ALFA), which has a FWHM beam size of 3$\arcmin$.4. 
Twelve hydrogen RRLs from H163$\alpha$ to H174$\alpha$ are located within the instantaneous bandpass from 1225\,MHz to 1525\,MHz.
We provide here cubes of average (``stacked'') RRL emission for the inner Galaxy region $32\degr \le \ell \le 70\degr$, $|b|\le1.5\degr$, with an angular resolution of 6$\arcmin$.
The stacked RRL rms at 5.1\kms\ velocity resolution is $\sim0.65$\,mJy\,beam$^{-1}$, making this the most sensitive large-scale fully-sampled RRL survey extant.
We use SIGGMA data to catalogue 319 RRL detections in the direction of 244 known \hii\ regions, and 108 new detections in the direction of 79 \hii\ region candidates. 
We identify 11 Carbon RRL emission regions, all of which are spatially coincident with known \hii\ regions. 
We detect RRL emission in the direction of 14 of the 32 supernova remnants (SNRs) found in the survey area.
This RRL emission frequently has the same morphology as the SNRs. The RRL velocities give kinematic distances in agreement with those found in the literature, indicating that RRLs may provide an additional tool to constrain distances to SNRs.
Finally, we analyze the two bright star-forming complexes: W49 and W51. 
We discuss the possible origins of the RRL emission in directions of SNRs W49B and W51C.
\end{abstract}


\section{Introduction} \label{sec_intr}
Radio recombination lines (RRLs) are ideal tracers of the ionized interstellar medium (ISM).
In contrast to observations at optical and near infrared (IR) wavelengths, RRLs are essentially unaffected by interstellar extinction.  They therefore provide a probe of ionized gas in the Galactic plane, sampling the emission from \hii\ regions, photo-dissociation regions (PDRs), planetary nebulae (PNe), supernova remnants (SNRs), and diffuse ionized gas.  From the observed RRL parameters, one can derive the physical properties of the emitting regions, such as the electron temperature, electron density, emission measure (EM) from non-LTE models, the gas kinematics, and information about thermal and turbulent motions.

There have been many pointed RRL surveys of individual Galactic \hii\ regions, beginning in the 1970s \citep[e.g.][]{Reifenstein1970,Wilson1970,Churchwell1978}.
\citet{Lockman1989} cataloged 462 \hii\ regions in the northern sky by observing RRLs at 3\,cm wavelength toward Galactic continuum sources.
\citet{Caswell1987} carried out a comprehensive \hii\ survey of 316 \hii\ regions in the southern sky using H109$\alpha$ \& H110$\alpha$.
The Green Bank Telescope (GBT) \hii\ Region Discovery Survey \citep[GBT HRDS;][]{Bania2010,Anderson2011,Anderson2015} is the most recent RRL survey of targeted sources.
Its sample was selected by examining the IR and radio emission of candidate \hii\ regions. 
With the sensitivity and powerful back-end of the GBT, the HDRS has doubled the number of known \hii\ regions north of $\delta=-45\degr$, which is $340\degr \leq \ell \leq 240\degr$ at $b = 0\degr$.

There are few unbiased surveys of RRL emission toward the Galactic plane \citep[e.g.,][]{Lockman1976,Anantharamaiah1986,Roshi2000}.
\citet{Alves2010,Alves2012,Alves2015} used RRLs within the bandpass of the \hi\ Parkes All-Sky Survey \citep[HIPASS;][]{Staveley1996}, which covers $196\degr \leq \ell \leq 360\degr$ in the third and fourth quadrants, and $0\degr \leq \ell \leq 52\degr$ in the first quadrant, both within $|b| \leq 5\degr$.
They observed H166$\alpha$, H167$\alpha$, and H168$\alpha$ at L-band with a spatial resolution of $14.4\arcmin$ and a velocity resolution of 20\kms. 
Their typical rms noise in the stacked spectra is 4.5\,mK (6.4\,mJy\,beam$^{-1}$).
The HI/OH/RRL survey with the Very Large Array \citep[THOR;][]{Beuther2016} also produced RRL maps, with a smoothed angular resolution of $40\arcsec$ and a velocity resolution of 10\kms. Their rms noise is $\sim2.3$\,mJy\,beam$^{-1}$ by stacking 10 RRLs.
Being an interferometric survey, the THOR RRL data are not sensitive to emission from diffuse ionized gas.

We introduced the Survey of Ionized Gas in the Galaxy, made with the Arecibo Telescope (SIGGMA) in \citet[hereafter ``Paper I'']{Liu2013}.
SIGGMA is an RRL survey at 1.4\ghz\ that fully sampled the entire Galactic plane observable with the 305-m William E. Gordon Telescope at the Arecibo Observatory.
With an rms noise of $\sim1$\,mJy\,beam$^{-1}$, SIGGMA remains the most sensitive fully-sampled RRL survey.

In this paper, we present SIGGMA data from the inner Galaxy ($32 \degr \leq \ell \leq 70 \degr$).
These inner-Galaxy data were collected commensally with the Arecibo pulsar survey ``P-ALFA'' \citep{Cordes2006, Lazarus2015}.
In the outer Galaxy ($175 \degr \leq \ell \leq 207 \degr $), SIGGMA was observed commensally with the ALFA zone of avoidance (ZOA) survey \citep[][]{McIntyre2015}, which is still ongoing.
The survey data from the outer Galaxy will be discussed in a future paper. 
We describe the SIGGMA observations and data reduction in Section~\ref{sec_obs_data} and Section~\ref{sec_cata} gives the \hii\ region catalogues as the initial results of the survey. 
The distribution of RRL emission and derived free-free emission along the Galactic disk are presented in Section~\ref{sec_gal}.
Section~\ref{sec_crrl} and~\ref{sec_snr} discuss C-RRL regions and RRL detections toward SNRs.
We present details of RRL investigations toward two major star-forming complexes in the survey area, W49 and W51, in Section~\ref{sec_indi}.
The summary follows in Section~\ref{sec_summ}.

\section{Observations and data reduction} \label{sec_obs_data}

We describe the observations and data reduction in fully Paper~I, and only discuss the most important details here.
We used the seven-beam Arecibo L-band Feed Array (ALFA)\footnote{See \url{http://www.naic.edu/alfa} for more information.} receiver for the SIGGMA observations.
ALFA's beam pattern consists of six beams surrounding a central beam.
The average full width at half maximum (FWHM) of the seven beams is $\sim3\arcmin.4$.
Three pointings of the ALFA beam pattern together form a beam-sampled grid pattern (see Figure~1 of Paper~I).
The data are calibrated in the normal fashion by position switching. 
We use off-positions that have the same declination, but are separated by five minutes in right ascension from the on-positions.
After matching the on- and off-pairs, we correct the spectra using  $\frac{T_{ON}-T_{OFF}}{T_{OFF}}\times T_{sys}$, where $T_{ON}$ is the on-source spectrum and $T_{OFF}$ is the off-source spectrum.
The on-off calibration corrects the bandpass gain curve, but because the off locations are only offset by one beam pattern, the survey data are insensitive to structures larger than $1\degr.25$ in the R.A. direction.
Therefore, unless there is a strong spatial intensity gradient, RRL emission from the warm ionized medium (WIM) is difficult to detect with SIGGMA data.
We then apply a gain calibration of 11\,K\,Jy$^{-1}$ for the central beam (Beam 0) and 8.5\,K\,Jy$^{-1}$ for Beams 1-6\footnote{See \url{http://www.naic.edu/alfa/performance/}}.
The survey properties are given in Table~\ref{tab_survey}.

Section~\ref{sec_obs} gives the observational status and shows the current survey sky coverage of the survey.
In Section~\ref{sec_stack}, we describe the details of RRL stacking and show the quality of the spectra.
Section~\ref{sec_cube} explains the method we use to produce data cubes.
We discuss our radio frequency interference (RFI) removal technique in Section~\ref{sec_rfi}.

\begin{table}[tbhp]
\centering
\caption{Properties of the survey products \label{tab_survey}}
\begin{threeparttable}
\begin{tabular}{cc}
\hline
\hline
 Parameter & Value\\
\hline
$\ell$  & $32\degr$ to $70\degr$ \\
$b$  & $-1\degr.5$ to $+1\degr.5$ \\
FWHM & $6\arcmin$ \\
Spectral resolution & 5.1\kms \\
Velocity range & $-300$ to $+300$\kms\\
Integration time & 270\,s \\
rms noise & 0.65\,mJy beam$^{-1}$ ($\sim6.5$\,mK) \\
\hline
\end{tabular}
\end{threeparttable}
\end{table}

\subsection{Observational progress} \label{sec_obs}
\begin{figure}[hbtp]
\epsscale{1.}
\plotone{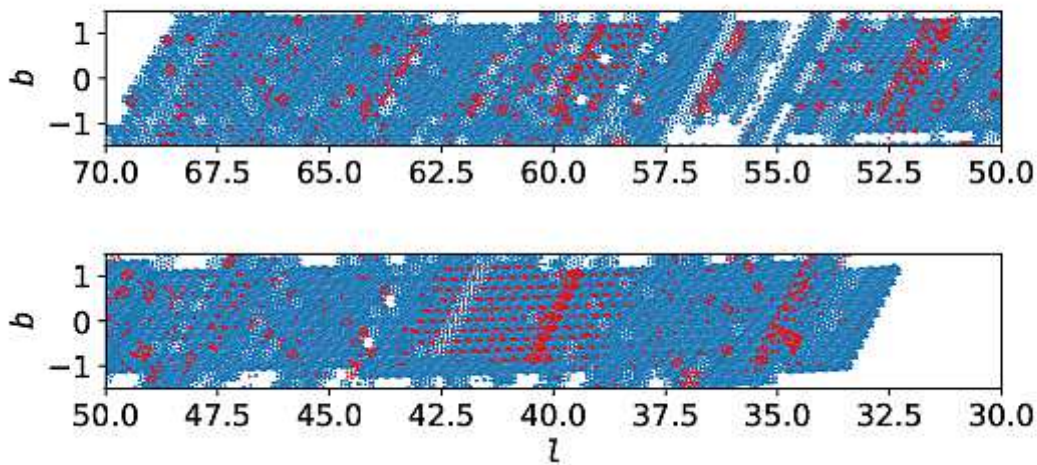}
\caption{SIGGMA sky coverage within the inner Galactic plane ($|b| \leq 1\degr.5$).
Each blue dot stands for a pointing with good spectral quality.
The red dots correspond to the pointings where the data are not usable and hence are not included in the final data products.
Blank areas are yet to be observed.
}\label{fig_skycover}
\end{figure}

The inner Galaxy observations shown in Figure~\ref{fig_skycover} are over 96\% complete.
The regions not yet observed are mostly located at Galactic latitudes $|b|>1\degr$.  The red dots show the pointings with unusable data.  The grid pattern of unusable data in the Galactic longitude range between 38$\degr$ and 43$\degr$ comes from ALFA Beam-6, which was malfunctioning in April 2011.
The slant red streaks indicate unusable data due to strong RFI or high system temperature.

As mentioned in Section~\ref{sec_intr}, P-ALFA is the primary survey for the inner Galactic plane observations, and SIGGMA is secondary.
This means that SIGGMA has no direct control over which locations are observed.  Although the observations are still on-going, the areas that are not yet covered, as well as those in the latitude range of $|1.5|\degr$ to $|2|\degr$, will not be completed soon.  Nor will positions with unusable data be re-observed.
Since the areas of missing data are of less interest (having fewer expected RRL detections), we are releasing the survey data at the current stage.

\subsection{RRL stacking} \label{sec_stack}

\begin{figure}
\begin{minipage}{.48\textwidth}
\centering
\plotone{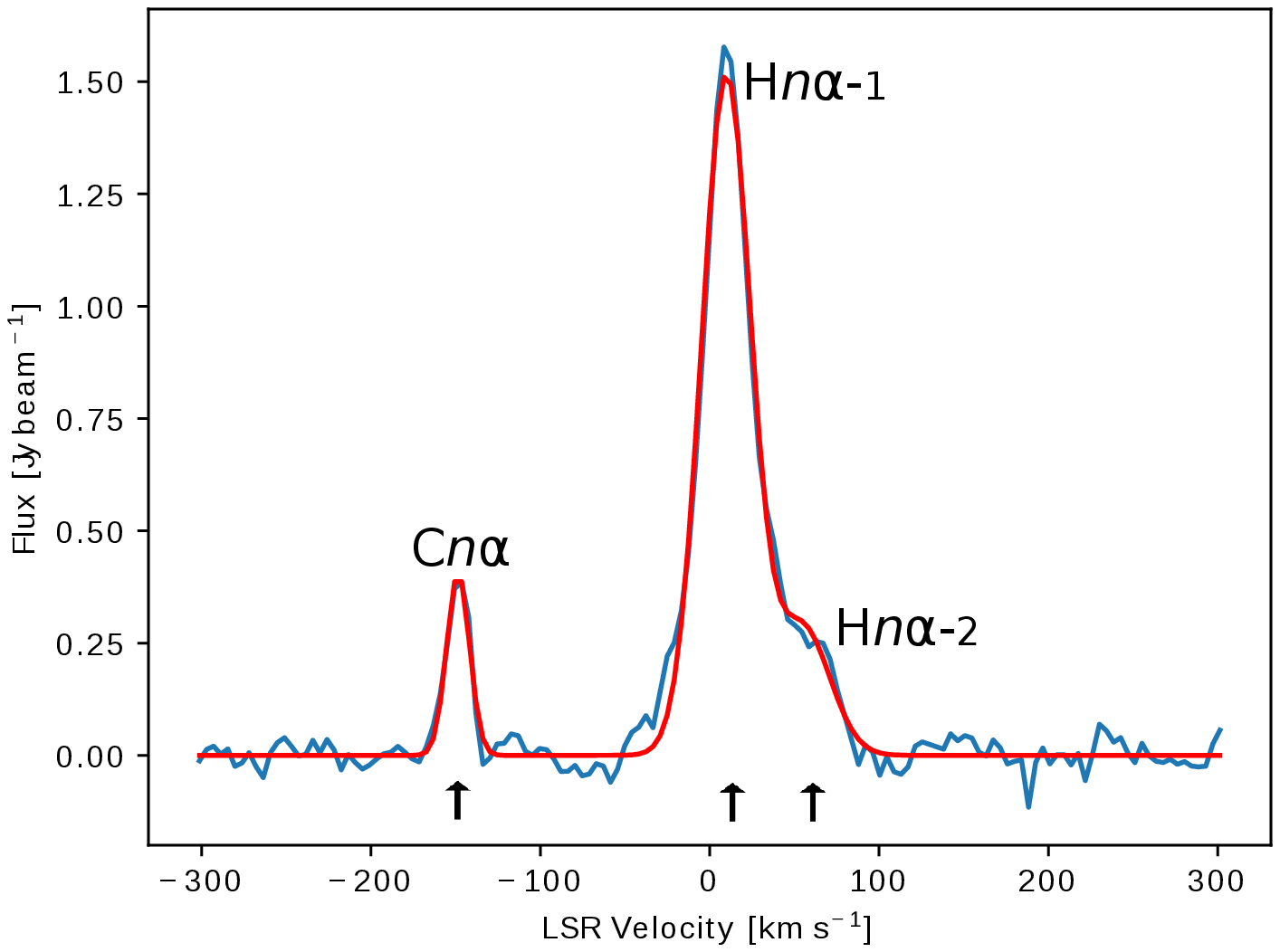}
\caption{The stacked H$n\alpha$ spectrum from the peak W49A region, with Gaussian fits.
We label the two velocity components of H$n\alpha$ by H$n\alpha$-1 and H$n\alpha$-2.
}\label{fig_spec}
\end{minipage}
\hfill
\begin{minipage}{.48\textwidth}
\centering
\plotone{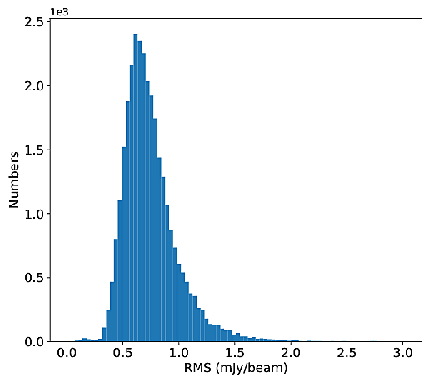}
\caption{The rms distribution of the final spectra, calculated from the RFI- and line-free channels of spectra ($-300$ to $-200$\kms\ and $+200$ to $+300$\kms)
}\label{fig_rms}
\end{minipage}
\end{figure}

We divide the corrected bandpass from each pointing into twelve H$n\alpha$ sub-bands (H$163\alpha - $H$174\alpha$).
The frequency resolution is 21\,kHz, giving a velocity resolution of 4.2 to 5.1\kms.  
The velocity range of each sub-band is $-300$ to $+300$\kms, which covers the observed velocity distribution of Galactic gas, provides an adequate number of data points for baseline fitting, and also encompasses the helium (offset by $-122$\kms\ from that of hydrogen) and carbon (offset by $-150$\kms from that of hydrogen) RRLs.
After performing a 3rd order polynomial fitting to each spectrum individually, we resample the twelve segments to a common velocity grid of 5.1\kms\ resolution and then average all segments to achieve a higher signal-to-noise ratio ($S/N$).
The stacked spectra frequently have irregular baseline ripples, especially when there is strong continuum emission within the beam.
We fit a 5th order polynomial baseline to all stacked spectra after masking data from velocities between $-50$ and $+150$\kms.
As an example, we show the final H$n\alpha$ spectrum toward the center of W49A in Figure~\ref{fig_spec}.

There are $\sim33,000$ raw stacked spectra in total.
Figure~\ref{fig_rms} shows the rms noise distribution of the stacked spectra, which peaks at $\sim0.65$ mJy\, beam$^{-1}$ ($\sim6.5$\,mK, obtained by applying an average gain of 10 K\,Jy$^{-1}$). 
This agrees well with the expected rms of $\sim0.5$ mJy\,beam$^{-1}$ given by Paper~I. 
The extension toward higher values in the rms distribution is caused by positions with bright continuum emission.

\begin{figure}[htbp]
\epsscale{0.5}
\plotone{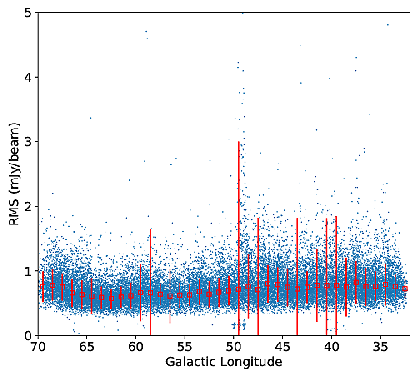}
\plotone{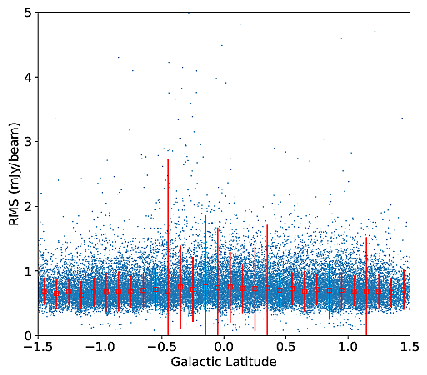}
\caption{The rms distribution versus Galactic longitude (left) and latitude (right).
The open red box is the median of the 0$\degr$.1 binned rms values and the red bars are their standard deviations.
}\label{fig_rms_glat}
\end{figure}

Figure~\ref{fig_rms_glat} gives the rms distribution as a function of Galactic longitude and latitude.
Spectra within $|b| \leq 0\degr.5$ are more likely to have high rms due to the higher density of continuum sources.
We see this pattern in the plot of rms versus Galactic longitude, as the higher rms values correspond to locations of large massive star formation complexes.

\subsection{Data cubes} \label{sec_cube}
After stacking, we use the software package Gridzilla \citep{Barnes2001} to grid the stacked spectra into data cubes.
Each cube is $5\degr \times 3\degr$ with a pixel size of $1\arcmin \times 1 \arcmin$, and has 5.1\kms\ channel spacing covering $-300$ to $+300$\kms.
The beam weighting is set to be proportional with the beam FWHM of $3\arcmin.4$.
We apply a Gaussian smoothing with the kernel FWHM of $6\arcmin$, $\sim2$ beam widths to fill in gaps where we lack usable data.
The spectrum of each pixel is then generated from the weighted median of the stacked spectra within a cut-off radius of $4\arcmin$.
These cubes of stacked RRL emission are the primary data product of SIGGMA, and are the basis of all analysis that follows in this paper.

\subsection{Bad data removal} \label{sec_rfi}
RFI is prevalent in the SIGGMA data, and RFI removal is the first step of our data reduction process.
We first excise strong and broad RFI from the whole bandpass by removing spectral values more than 5 times the spectral rms about the median value, which is obtained using a median filter of width 300 channels.
The spectral rms is calculated from RFI free channels in the velocity range of $-300$ to $-200$\kms\ and $+200$ to $+300$\kms.
During line stacking (Section~\ref{sec_stack}), we apply a second median filter to the twelve sub-bands. 
This median filtering is done to remove any weak and narrow RFI that remains.
We use filter criteria of 15 channels in width and 10 times of the spectral rms. 
The RRL signal remains unaffected with this filtering.

\textbf{
After stacking, some spectra show strong baseline ripples or abnormal noise not identified by the median filter method.
We set an rms threshold of 3\,mJy to remove such stacked spectra from further analyses.
Spectra with strong continuum sources also have high rms noise. 
So as to not exclude such spectra, which also frequently have true line emission, we calculate the ratio of the rms from the central portion of the spectra (from $-200$ to $+200$\kms) to that of the edges.
We keep the stacked spectra that have an rms greater than 3\,mJy when their center to edge rms ratio is greater than 1.5.
About 7\% of the spectra are so affected.
}

\section{The catalog of \hii\ regions} \label{sec_cata}

SIGGMA provides an unbiased look at the RRL properties of discrete \hii\ regions.
Using {\it Wide-Field Infrared Survey Explorer} ({\it WISE}) data, \citet{Anderson2014} created a catalog of over 8000 Galactic \hii\ regions and \hii\ region candidates by searching for their characteristic mid-infrared (MIR) morphology.  This catalog is called the ``{\it WISE} catalog of Galactic \hii\ regions'' (hereafter simply the ``{\it WISE} catalog'').
The {\it WISE} catalog contains four types of objects, labeled as the ``known'', ``group'', ``candidate'', and ``radio quiet.''
``Known'' regions have previously-detected RRL or H$\alpha$ emission, ``group'' regions are associated with known regions but lack spectroscopic observations, ``candidate'' regions have radio continuum and MIR emission but no ionized gas spectroscopic detections, and ``radio quiet'' regions only have MIR emission with the characteristic \hii\ region morphology.
The {\it WISE} catalog provides the known properties of its sources, such as their radii, distances, \textbf{local standard of rest (LSR)} velocities, etc.

We extract the spectrum toward {\it WISE} catalog \hii\ regions from SIGGMA data by averaging SIGGMA spectra over each source's areal extent as defined in the catalog.
Then we perform an automatic Gaussian fitting on the averaged spectrum of each source. 
We perform these fits over the LSR velocity range from $-10$\kms\ to $+120$\kms, and allow RRL line FWHM values of 12\kms\ to 100\kms. We only count as detections lines with $S/N \geq 5$.  Because the weaker Helium RRLs do not meet this signal-to-noise criterion, even for the brightest regions in the survey, they are naturally excluded.
Carbon RRLs are excluded from the fits by the velocity and line width range we used.
\textbf{
We verify the results of the automatic Gaussian fitting by eye and perform fits to spectra lacking automatic fits that are clearly real detections but whose line parameters fall outside the ranges used above.
}

We use a chi-square test and an F-test to evaluate the fitting of multiple velocity component.
For each spectrum, we perform one-, two-, and three-component Gaussian fits separately, and calculate the reduced chi-square values.
If the one-component Gaussian fit has the smallest reduced chi-square value that is larger than unity, we confirm it as the best fitting.
If the multi-component (two or three) Gaussian fits have the smallest larger-than-unity reduced chi-square value, we use an F-test to assess if it is significantly better than the fits with fewer components.
If the calculated F-test probability is less than 0.05, we confirm it as the best fitting.
Otherwise, we confirm the fit with one fewer components as the best fitting.

\begin{figure}[htbp]
\epsscale{0.6}
\plotone{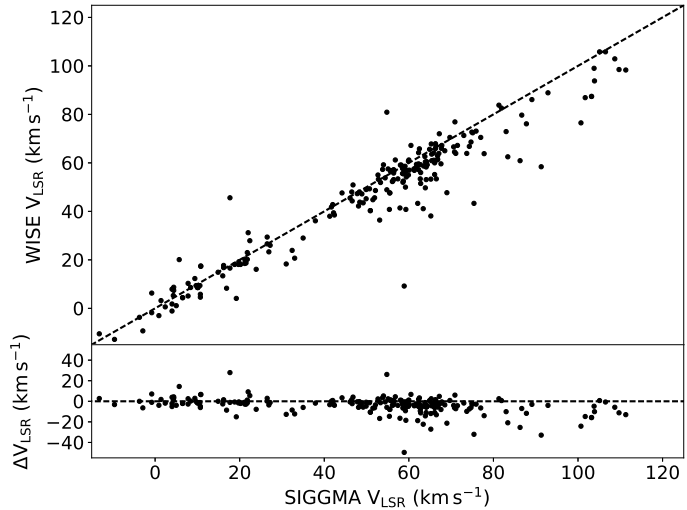}
\caption{Known \hii\ region RRL $V_{LSR}$ values compared with those derived in SIGGMA data.  The dashed line shows a 1:1 relationship.}\label{vlsr_map}
\end{figure}

Within the survey zone of SIGGMA, there are 329 ``known'' \hii\ regions.
We detected H-RRLs in the direction of 244 known \hii\ regions; 61 have two velocity components and 7 have three velocity components.
Figure~\ref{vlsr_map} shows good agreement between SIGGMA line velocities with respect to LSR and those compiled in the WISE catalog.
Table~\ref{tab_hii_k_sample} lists the first 10 rows of RRLs detected toward ``known'' WISE \hii\ regions.
Columns 1, 2, and 3 give the region names with their Galactic coordinates.
Columns 4, 5, and 6 show the fitted line profile parameters. 
Columns 7 and 8 are the rms and $S/N$ of the spectrum.
We provide the full catalogue in Table~\ref{tab_HII_known} in Appendix~\ref{appe:cata}.

We also provide line parameters in the direction of ``candidate'' \hii\ regions, for which no ionized gas spectroscopic detections exist in the literature.
There are 165 ``candidate'' sources observed by SIGGMA, of which we detect RRL emission in the direction of 79.
We give the first 10 samples of SIGGMA detected ``candidate'' \hii\ region RRL properties in Table~\ref{tab_hii_c_sample} (see Table~\ref{tab_HII_candidate} in Appendix~\ref{appe:cata} for the full ``candidate'' catalogue).

There are two main reasons for non-detections towards known \hii\ regions. Three fourths of the 85 non-detected known \hii\ regions have angular sizes less than 2$\arcmin$, while the ALFA beam is 3$\arcmin.4$. The emission from these targets may be beam-diluted. Also, one can see from the actual data coverage of the survey (see Figure~\ref{fig_skycover}), that there are considerable blank areas, and many non-detections may fall within these areas, although they are within the nominal SIGGMA observational zone.
It is possible that RRL detections toward a source come from nearby regions within the telescope beam, or from diffuse ionized gas along the line of sight.
We therefore note in the Tables which SIGGMA detections arise from positions with multiple ``known'' or ``candidate'' \hii\ regions within the ALFA beam.
We also note cases where integrating over the entire \hii\ region also includes positions from overlapping ``known'' or ``candidate'' \hii\ regions.
Sources in the ``group'' sample are found toward known regions, which confuses the interpretation of their spectra.
We therefore do not provide line parameters for group \hii\ regions.
We also do not provide line parameters for the radio quite sample as the SIGGMA line emission is too faint for detections.

Due to baseline structure in SIGGMA data, the line parameters for faint \hii\ regions can be difficult to determine (see Section~\ref{sec_rfi}).
In Table~\ref{tab_HII_known} and~\ref{tab_HII_candidate} there are some items with line width $>50$\kms, which are not typical values toward \hii\ regions. 
Although the detection criterion of $S/N >5$ should result in reliable detections, bad baselines may introduce uncertainty in the derived line parameters.
We checked the broad line profiles by eye and removed the ones that had characteristics similar spectra with bad baselines.
We did keep some broad lines, which may be blended line components not separable with our rather low velocity resolution.

There are also sources with FWHM $\leq$15 \kms\ in Table~\ref{tab_HII_known} and~\ref{tab_HII_candidate}.
\citet{Anderson2018} reported narrow-RRL detections with FWHM $<$10 \kms\ from large angular size \hii\ regions.
They suggested that the narrow lines could come from interactions with molecular clouds or from partially ionized zones within the \hii\ region PDRs.
They also indicated that the narrow-RRL components may be caused by WIM along the line of sight, because the narrow-RRL sources in their study always have multiple velocity components.
Similarly, most of the narrow SIGGMA detections have multiple velocity components, which favors the WIM-related interpretation.
However, it is also possible that the narrow lines are emitted by \hii\ regions, within which the emitting gas has low turbulence.

\begin{table}[htbp]
\centering
\caption{RRL emission from ``known'' \hii\ regions}\label{tab_hii_k_sample}
\begin{threeparttable}
\begin{tabular*}{\textwidth}{l@{\extracolsep{\fill}}*{8}{c}}
\hline
\hline
Name & $l$ & $b$ & Peak& V$_{LSR}$& FWHM& RMS& $S/N$\\
&$^\circ$&$^\circ$&mJy\,beam$^{-1}$&km\,s$^{-1}$&km\,s$^{-1}$&mJy\,beam$^{-1}$& \\
\hline
G032.800$+$00.190     &32.800&	$+$0.190	&9.16	$\pm$ 1.01	&14.9	$\pm$ 1.2	&22.1	$\pm$ 2.8	&1.65	&8.9 \\ 
G032.823$+$00.072$^a$ &32.824&	$+$0.072	&30.40	$\pm$ 2.89	&106.7	$\pm$ 1.2	&25.3	$\pm$ 2.8	&2.76	&19.0\\
G032.835$+$00.017     &32.835&	$+$0.017	&191.52	$\pm$ 22.13	&109.8	$\pm$ 1.8	&33.3	$\pm$ 4.8	&28.51	&13.2\\
G032.982$-$00.338     &32.983&	$-$0.338	&17.08	$\pm$ 2.16	&50.8	$\pm$ 1.2	&20.6	$\pm$ 3.3	&2.56	&10.3\\
G033.051$-$00.078     &33.051&	$-$0.078	&26.21	$\pm$ 1.76	&103.9	$\pm$ 0.9	&28.0	$\pm$ 2.3	&0.77	&61.2\\
G033.080$+$00.073     &33.080&	$+$0.073	&11.65	$\pm$ 0.97	&102.2	$\pm$ 1.1	&27.8	$\pm$ 2.7	&1.22	&17.3\\
G033.142$-$00.088$^a$ &33.142&	$-$0.087	&20.93	$\pm$ 1.52	&103.2	$\pm$ 0.9	&26.4	$\pm$ 2.2	&1.19	&30.9\\
G033.176$-$00.016     &33.176&	$-$0.015	&27.82	$\pm$ 2.09	&105.1	$\pm$ 1.0	&26.2	$\pm$ 2.3	&1.63	&29.9\\
G033.205$-$00.013     &33.205&	$-$0.012	&14.64	$\pm$ 2.52	&106.5	$\pm$ 1.1	&13.1	$\pm$ 2.6	&2.46	&7.4\\
G033.263$+$00.066     &33.263&	$+$0.067	&11.51	$\pm$ 1.06	&111.3	$\pm$ 1.9	&42.2	$\pm$ 4.5	&1.40	&18.2\\
\hline
\end{tabular*}
\begin{tablenotes}
      \small
      \item a. Detections arise from positions with multiple ``known'' or ``candidate'' \hii\ regions within the ALFA beam.
\end{tablenotes}
\end{threeparttable}
\end{table}

\begin{table}[htbp]
\centering
\caption{RRL emission from ``candidate'' \hii\ regions }\label{tab_hii_c_sample}
\begin{threeparttable}
\begin{tabular*}{\textwidth}{l@{\extracolsep{\fill}}*{8}{c}}
\hline
\hline
Name & $l$ & $b$ & Peak& V$_{LSR}$& FWHM& RMS& $S/N$\\
&$^\circ$&$^\circ$&mJy\,beam$^{-1}$&km\,s$^{-1}$&km\,s$^{-1}$&mJy\,beam$^{-1}$& \\
\hline
G033.024$-$00.366     &33.025&$-$0.366&26.15	$\pm$2.02&50.8	$\pm$1.2	&28.2	$\pm$3.0	&2.61	&18.2\\
                      &33.025&$-$0.366&11.14	$\pm$1.76&104.6	$\pm$3.5	&43.1	$\pm$11.0	&2.61	&9.6\\
G033.306$-$00.150$^a$ &33.307&$-$0.150&8.23	$\pm$0.70&106.8	$\pm$1.2	&28.4	$\pm$2.8	&0.55	&27.5\\
G033.734$-$00.316$^a$ &33.735&$-$0.315&5.34	$\pm$0.68&52.7	$\pm$1.2	&20.3	$\pm$3.3	&0.71	&11.6\\
G034.130$-$00.174$^b$ &34.131&$-$0.173&12.77	$\pm$0.78&50.1	$\pm$0.6	&28.1	$\pm$2.0	&0.71	&32.4\\
                      &34.131&$-$0.173&3.12	$\pm$0.89&90.7	$\pm$1.8	&15.7	$\pm$5.3	&0.71	&5.9\\
G034.174$-$00.086$^b$ &34.175&$-$0.085&10.50	$\pm$0.99&48.2	$\pm$0.9	&29.1	$\pm$3.2	&0.80	&24.3\\
                      &34.175&$-$0.085&4.59	$\pm$1.06&92.2	$\pm$1.3	&13.1	$\pm$3.8	&0.80	&7.1\\
G034.190$-$00.063$^b$ &34.191&$-$0.063&10.26	$\pm$0.90&48.9	$\pm$0.8	&30.8	$\pm$2.8	&0.74	&26.4\\
                      &34.191&$-$0.063&4.32	$\pm$0.92&90.5	$\pm$1.2	&14.3	$\pm$3.6	&0.74	&7.6\\
\hline
\end{tabular*}
\begin{tablenotes}
      \small
      \item a. Detections arise from positions with multiple ``known'' or ``candidate'' \hii\ regions within the ALFA beam.
      \item b. Detections integrated over the entire \hii\ region also includes positions from overlapping ``known'' or ``candidate'' \hii\ regions.
\end{tablenotes}
\end{threeparttable}
\end{table}

\section{Distributions along the Galactic Plane} \label{sec_gal}
\subsection{RRL intensity} \label{sec_gal_rrl}
\begin{figure}[htbp]
\epsscale{1.2}
\plotone{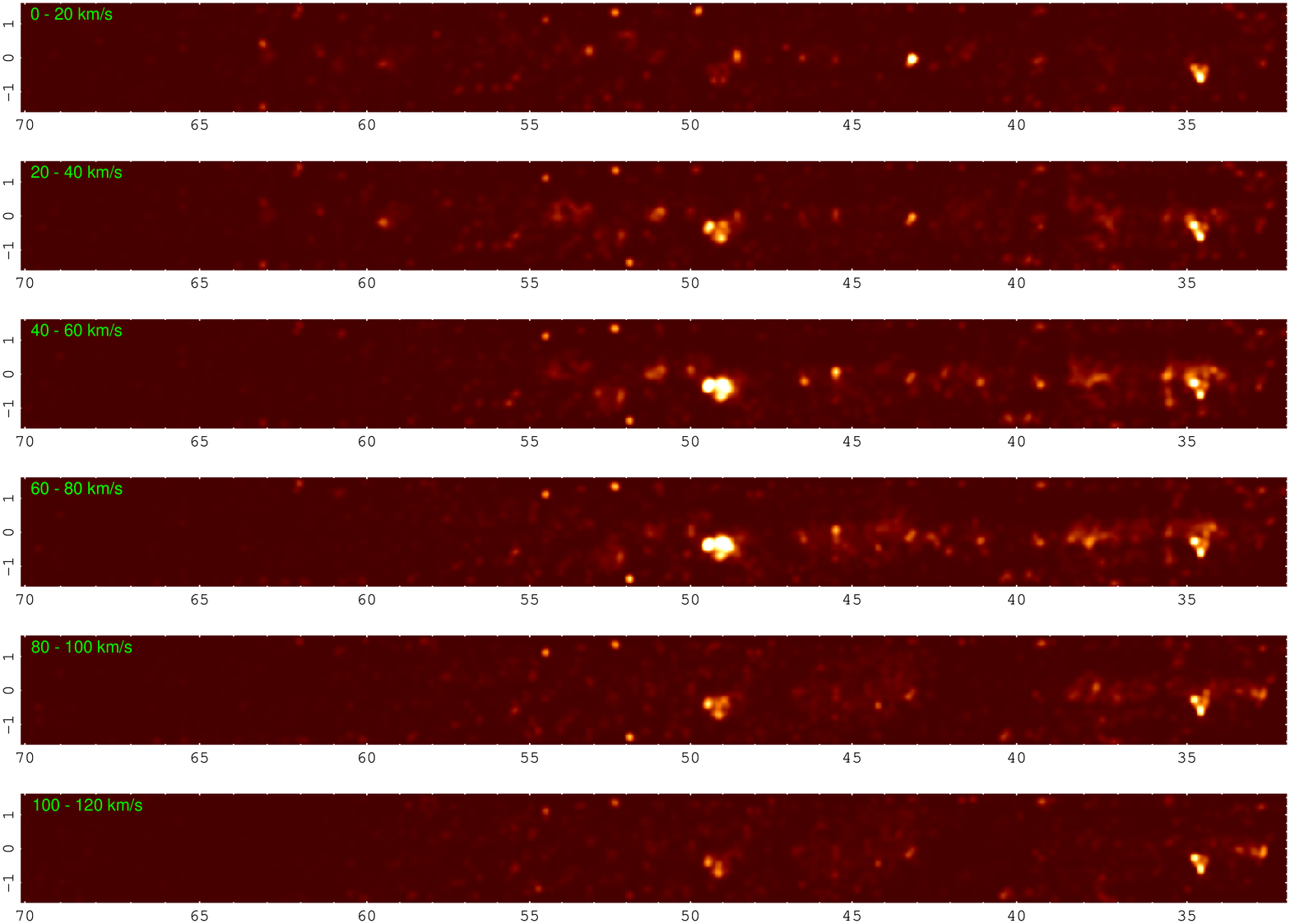}
\caption{Integrated RRL intensity maps over various velocity ranges.
}\label{fig_rrl_map}
\end{figure}

Figure~\ref{fig_rrl_map} shows the RRL integrated intensity maps in 20\kms\ intervals, smoothed to $5\arcmin \times 5\arcmin$ pixels.
We can see both individual bright sources and extended structure in these maps.
RRL detections in the range $65\degr \leq \ell \leq  70\degr$ are rare. 
We only detect one \hii\ region candidate within this range, which is listed Table~\ref{tab_HII_candidate}.
Some bright sources that appear at all velocity ranges are caused by baseline distortions from their strong continuum emission.
The two bright spots centered at G34.8$-$0.3 and G34.6$-$0.6 are introduced by the broad line profile observed toward the SNR W44 (see Section~\ref{sec_snr}).

\begin{figure}[htbp]
\centering
\epsscale{1.2}
\plotone{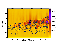}
\caption{Longitude-velocity map of  SIGGMA data (color) with CO contours from the survey of \textbf{\cite{Dame2001}}. The contour levels are 0.01, 1.0, 7.5, 15.0, 22.5, 30.0, 37.5, 45.0, 52.5, and 60.0 K.}\label{lv_map}
\end{figure}

Figure~\ref{lv_map} shows the longitude-velocity diagram of RRL intensity.
We smooth the SIGGMA data cube to $1\degr$ resolution in Galactic longitude, and integrate over $|b| \leq 0\degr.5$.
The vertical stripes are due to bright massive star formation regions.  Since RRLs are $\sim25$\kms\ wide, they have a large vertical extent in this figure.
For comparison with the molecular gas, we overlap the CO $1-0$ emission \citep{Dame2001} integrated over the same Galactic latitude range.
The RRL and CO emission are both bright around massive star formation region complexes, but on the whole there is a poor correlation between the CO and RRL gas.
The strong feature at Galactic longitude of 34$\degr$ is caused by the SNR W44 (see Section~\ref{sec_snr}) and some features in the negative velocity range are from carbon and helium RRLs.

\subsection{Free-free emission} \label{sec_gal_free}
\begin{figure}[htbp]
\centering
\includegraphics[width = 7in]{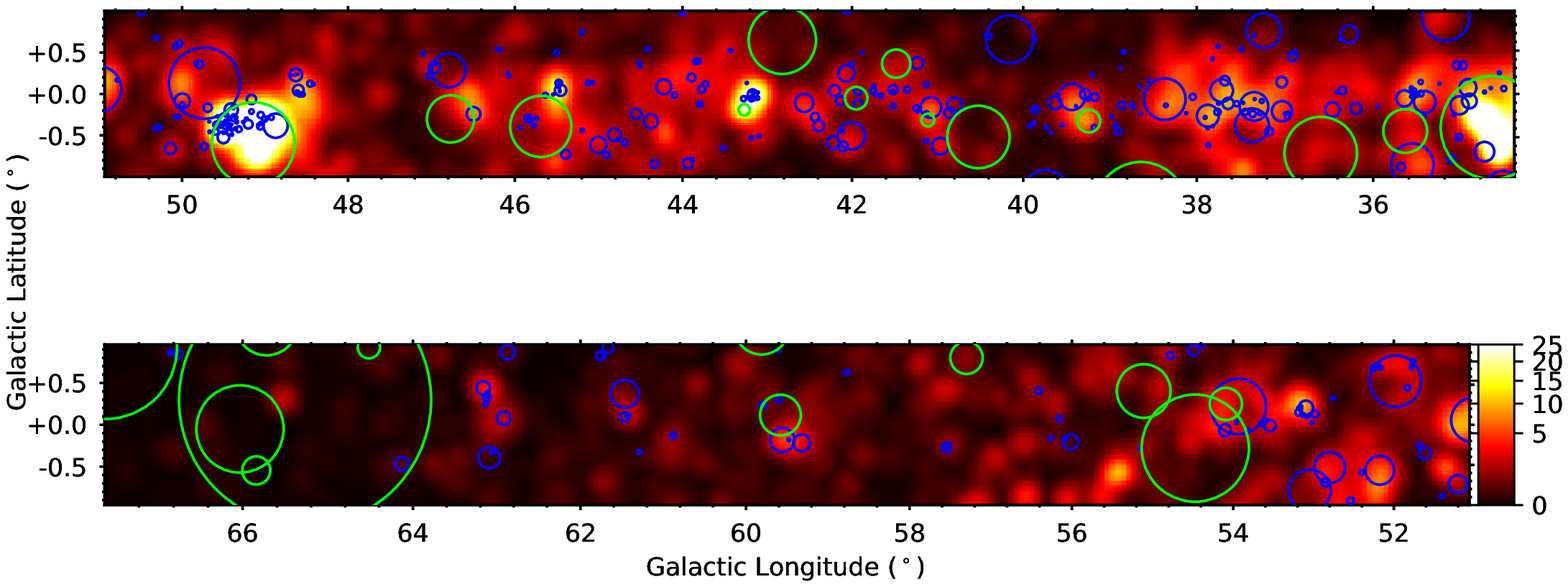}
\caption{Free-free continuum intensity.
$T_{C}$ is calculated following Equation~\ref{equ_temp}, using the SIGGMA H-RRL 0$^{th}$ moment map (integrated over $-20$ to $+120$\kms) and assuming $T_e = 8000 K$.
The unit for the map is in Kelvin.
The blue circles are ``known'' HII regions from the WISE catalog and the green circles are SNRs from the Green catalog (see Section~\ref{sec_snr}).
The sizes of the circles are proportional to the angular sizes of the sources, which are given by the catalogs.
The bright spots between $55\degr \leq \ell \leq 57\degr$ are likely artificial effects due to missing data (see Figure~\ref{fig_skycover}).}
\label{fig_gal_free}
\end{figure}

Although radio continuum data for the Galactic plane at 1.4\,\ghz\ have contributions from both thermal and non-thermal sources, RRL emission reveals only the thermal component.
\citet{Alves2012} described the derivation of the Galactic free-free emission using their RRL observations.
Similarly, one can also calculate the continuum temperature of thermal emission, by assuming an electron temperature, which is expressed as \citep{Mezger1967a, Lockman1978, Quireza2006a}:
\begin{equation}
  T_C=
  \frac{(1+0.08)}{6985}\cdot a(T_e) \cdot T_e^{1.15} \cdot \nu _{GHz}^{-1.1} \cdot \int T_L dV,
  \label{equ_temp}
\end{equation}
where $V$ is in \kms, $a(T_e) \simeq 0.97$ at 1.4\ghz, and $\nu_{GHz}$ is 1.4 in our case.  This equation assumes a helium to hydrogen ionic abundance of 0.08.
Using Equation~\ref{equ_temp} we compute the radio continuum temperature from thermal free-free emission alone using the SIGGMA integrated intensity (moment 0) maps, and assuming an constant electron temperature of 8000\,K.
Figure~\ref{fig_gal_free} shows the result of the thermal emission along the Galactic plane.
The figure is smoothed to $0\degr.5 \times 0\degr.5$ resolution.
The free-free emission intensity follows the distribution of known sources, which are mostly \hii\ regions.
We also compare the SIGGMA $T_C$ map (Figure \ref{fig_gal_free}) with that derived from RRL HIPASS \cite[see Figure 7 in][]{Alves2015}. The overlapping Galactic longitude range for the two surveys is from $52\degr$ to $32\degr$. We find that all of the free-free emission revealed in the RRL HIPASS map has corresponding features in the SIGGMA map, although more detailed and weaker features are identified by SIGGMA due to its higher angular resolution and sensitivity.

\begin{figure}[bhpt]
\centering
\includegraphics[width = 6in]{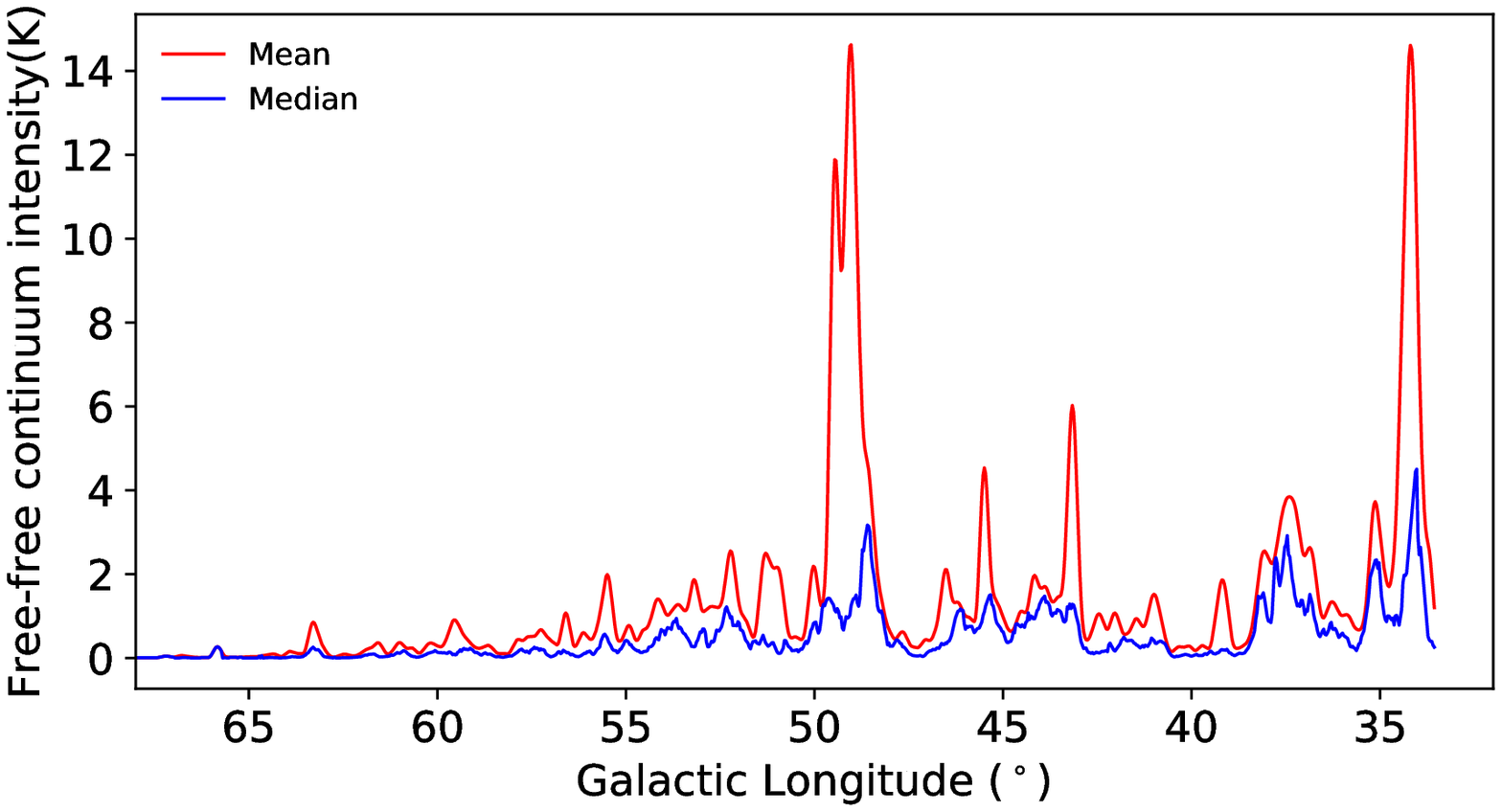}
\caption{Free-free continuum intensity as a function of Galactic longitude. The red line is the mean temperature over $|b|\leq 1\degr$. The blue line is the median value over the same range.}
\label{fig_gal_plot}
\end{figure}

To examine the general trend of the thermal emission as a function of Galactic longitude, we plot the flux averaged over $|b| \leq 1\degr$ (see Figure~\ref{fig_gal_plot}). 
The median flux is less affected by bright star-forming regions, and therefore better traces the overall emission from the Galactic plane.


\section{Carbon RRL emission regions} \label{sec_crrl}

C-RRLs are offset by $\sim-150$\kms\ from H-RRLs for the same principal quantum numbers.
SIGGMA detects H-RRL emission mostly from the velocity range of $\sim-10$ to $+150$\kms (cf.~Figure~\ref{lv_map}).
To identify C-RRL emission regions, we modify the SIGGMA data cubes by shifting the velocity to center at the C-RRL rest frequency.
We create C-RRL moment maps from the shifted cubes and detect C-RRL emission from 11 spatial locations.
We obtained the averaged C-RRL and H-RRL spectra for each region defined by the central emitting area where the C-RRL $S/N \geq 5$.
Table~\ref{tab_crrl} contains information about the C-RRL regions.
In this table, we list the region name, the Galactic coordinate centroid, the angular size, the WISE catalog of \hii\ regions that are spatially coincident, the velocity integrated C-RRL intensity, the C- to H-RRL integrated intensity ratio, and a qualitative assessment of the C-RRL data quality (A=highest quality, C=low quality).
The H-RRL integrated intensity is the product of the fitted line peak intensity and the FWHM.
Since the line widths for C-RRLs are generally $\lesssim10$\kms, the SIGGMA velocity resolution likely broadens the C-RRLs and decreases their intensities.  We thus provide integrated line intensities in Table~\ref{tab_crrl} instead of peak line intensities.

\begin{figure}[htbp]
\centering
\subfloat[SIGGMA spectral grid]{\includegraphics[width = 3.3in]{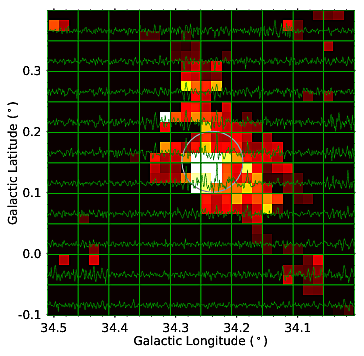}}
\subfloat[Infrared color map  ]{\includegraphics[width = 3.3in]{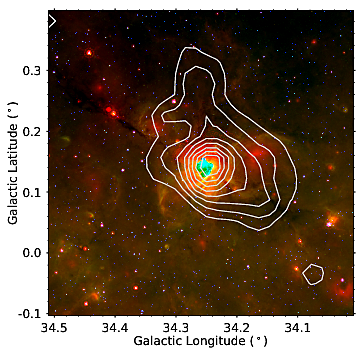}}\\
\subfloat[SIGGMA spectral grid]{\includegraphics[width = 3.3in]{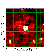}}
\subfloat[Infrared color map  ]{\includegraphics[width = 3.3in]{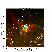}}\\ 
\caption{The C-RRL region G34.2+0.2.
      (a) The C-RRL grid with velocity from $-50$ to $+150$\kms.
	  The background is the integrated C-RRL intensity over 30 to 70\kms.
	  (b) The IR RGB map (r: 24\,$\um$, g: 8\,$\um$, b: 4.5\,$\um$) overlaid with integrated C-RRL contours.
	  The contour levels are 2, 4, 6, 8, 10, 12, 14, 16 mJy\,beam$^{-1}$\kms.
	  (c) The H-RRL grid with velocity from $-50$ to $+150$\kms.
	  The background is the integrated H-RRL intensity over 0 to 100\kms.
	  (d) The IR RGB map (r: 24\,$\um$, g: 8\,$\um$, b: 4.5\,$\um$) overlaid by integrated H-RRL contours.
	  The levels are 0.2, 0.4, 0.6, 0.8, 1, 1.2, 1.4, 1.6 Jy\,beam$^{-1}$\kms.
      \textbf{
      The circles in panels (a) and (c) illustrate the area over which the C-RRL and H-RRL spectra were averaged.
  }
	  }
\label{fig_crrl-g342}
\end{figure}

Carbon RRL emission is thought to arise from PDRs, and can be used to investigate PDR properties \citep[][etc.]{Pankonin1977,Natta1994,Roshi2007}.
In some cases the carbon RRL emission is found to be associated with cold \hi\ regions \citep{Roshi2011}.
Polycyclic aromatic hydrocarbon (PAH) molecules in PDRs emit at 8.0$\um$ \citep{Watson2008,Watson2009}, and we therefore expect a good agreement between C-RRL and 8.0$\um$ emission.
We compare the C-RRL emission distribution with the {\it Spitzer} 8.0\,\micron\ IR data from the Galactic Legacy Infrared Midplane Survey Extraordinaire (GLIMPSE) Legacy Program \citep{Benjamin2003,Churchwell2009}.
Figure~\ref{fig_crrl-g342} gives an example of those C-RRL regions.
We also include in this figure 4.5$\um$ and 24$\um$ {\it Spitzer} data from the GLIMPSE and MIPSGAL surveys \citep{Carey2009}.
This figure shows the spectral grid overlying the integrated C-RRL intensity map, and contours of the integrated C-RRL emission over the {\it Spitzer} three-color map (red: 24\,$\um$, green: 8\,$\um$, blue: 4.5\,$\um$).
For the source G34.2$+$0.2, the integrated C-RRL distribution matches with the corresponding \hii\ region as defined by the IR emission.
Comparing Figure~\ref{fig_crrl-g342} (b) and Figure~\ref{fig_crrl-g342} (d), we can see an offset of the C-RRL emission peak from that of the H-RRL emission.
This may imply the PDR origin of the C-RRL emission.
We note that the offset is less than half of the ALFA beam size.
Thus the uncertainty of the offset may be considerable.
We find an extended C-RRL structure to the north, but the 8.0$\um$ emission is weak there, making the PDR origin of this emission doubtful.
However, the weak distribution above 0.3$\degr$ in Galactic latitude is possibly an artifact of baseline noise. Further observations are needed to confirm that this feature is real.
\textbf{See Figure~set~1 in the online journal for more figures of detected C-RRL regions.}
For G34.7$-$0.6 and G35.1$-$0.7 \textbf{(in Figure~set~1)}, we see the C-RRL emission regions are offset relative to the \hii\ regions.
\cite{Alves2015} has detected stronger C-RRL emission outside the continuum peaks of \hii\ regions towards W41 (G23.4+0.0) and W42 (G25.3-0.1).
The background of spectral grid maps of some C-RRL regions show that many of the C-RRL emission regions have an annular structure, with stronger emission regions surrounding weaker emission regions \textbf{(see figures of G38.9$-$0.4, G53.6$+$0.0, G60.9$-$0.1, G61.4$+$0.1, and G63.1$+0.4$ in Figure~set~1).}
This may imply the emission feature from PDRs, although the morphologies do not agree with that seen at 8.0\,\micron, the wavelength  commonly used to identify PAH emission from PDRs \citep{Watson2009}.

\begin{table}[htbp]
\caption{C-RRL emission regions}\label{tab_crrl}
\begin{threeparttable}
\begin{tabular*}{\textwidth}{cccclclc}
\hline
\hline
Name& $l$& $b$&Radius$^a$&\hii\ Region&$\int I_{CRRL}dV$& C/H$^b$& Quality$^c$\\
&deg.&deg.&deg.&&Jy\kms&&\\
\hline
G34.2$+$0.2& 34.24& $+$0.15& 0.05&G034.256$+$00.136 (NRAO584) &0.07& 0.03 & B\\
G34.7$-$0.6& 34.73& $-$0.58& 0.05&G034.757$-$00.669           &0.15& 0.03 & C\\
G35.1$-$0.7& 35.12& $-$0.71& 0.03&G035.126$-$00.755           &0.12& 0.38 & B\\
G38.9$-$0.4& 38.93& $-$0.42& 0.05&G038.926$-$00.389           &0.03& \nodata$^d$& B\\
G43.2$-$0.0& 43.16& $-$0.00& 0.05&G043.170$-$00.004 (W49A)    &0.58& 0.08 & A\\
G45.5$+$0.0& 45.46& $+$0.07& 0.05&G045.453$+$00.044 (K47)     &0.10& 0.08 & B\\
G49.5$-$0.4& 49.49& $-$0.40& 0.05&G049.484$-$00.391 (W51A)    &0.83& 0.07 & A\\
G53.6$+$0.0& 53.57& $+$0.02& 0.03&G053.541$-$00.011           &0.05& 0.50 & B\\
G60.9$-$0.1& 60.88& $-$0.11& 0.05&G060.881$-$00.135 (S87)     &0.03& 0.30 & A\\
G61.4$+$0.1& 61.44& $+$0.13& 0.05&G061.473$+$00.094 (S88)     &0.24& 0.24 & B\\
G63.1$+$0.4& 63.14& $+$0.38& 0.05&G063.164$+$00.449 (S90)     &0.10& 0.06 & C\\
\hline
\end{tabular*}
\begin{tablenotes}
      \small
      \item a. The radius of the region, which is used to produce the averaged spectrum.
      \item b. Integrated intensity ratio of H-RRL and C-RRLs.
      \item c. Qualitative metric of data quality (A=high quality, C=low quality)
      \item d. There is no valid H-RRL detection since a H-RRL signal is in the off position.
\end{tablenotes}
\end{threeparttable}
\end{table}

\section{RRLs toward Galactic supernova remnants} \label{sec_snr}

SNRs are non-thermal radio sources, and are not expected to have strong RRL emission.
\citet{Downes1974} conducted a survey of H$134\alpha$ toward SNRs using the Effelsberg 100-m telescope and reported few detections save for toward W49B.
Several other SNRs have RRLs detected in their directions, namely: H$166\alpha$ toward W44 \citep{Bignell1973}, H-RRLs toward 3C\,391 and W49B \citep{Cesarsky1973a,Pankonin1975}.
\citet{Cesarsky1976} also reported C-RRL detections (C$157\alpha$) toward five Galactic SNRs.

We examine RRL emission in the direction of Galactic SNRs with SIGGMA data.
\citet[][hereafter the ``Green catalog'']{Green2014} lists 32 SNRs in the SIGGMA survey zone.
Some SNRs with large radii are faint and some are cospatial with one or more \hii\ regions; these are not ideal candidates for SIGGMA RRL detections.
By examining the continuum emission at 1.4\ghz\ from the VLA Galactic Plane Survey \citep[VGPS,][]{Stil2006}, we determine the angular extent and then extract the mean spectrum from SIGGMA \textbf{data for each of the remaining 14 isolated SNRs (see Table~\ref{tab_snr}).}
To avoid RRL signal from diffuse ISM when averaging, we averaged over rather compact regions instead of the entire angular extent given in the Green catalog.
We then fit the averaged spectrum for each source using the same Gaussian fitting technique as we applied to the \hii\ regions.
In Table~\ref{tab_snr}, we list the names of the SNRs, the Galactic coordinate of the central region and its radius, the LSR velocity of the peak H-RRL emission components, the FWHM of the RRL velocity components, and the calculated near and far kinematic distances using the \citet{Reid2014} rotation curve.
For all regions, one of the two kinematic distances agrees well with the distance from the literature.

The line widths of the detected H-RRLs toward SNRs are broad. 
This implies that the temperature and/or turbulent motions of the plasma are higher than that of \hii\ regions.
The broad line width, however, makes the signal  more likely to be confused with baseline ripples.
The high continuum intensity of the SNRs may contribute to baseline issues. 
Uncertainties in derived RRL parameters toward SNRs are therefore larger than those of \hii\ regions.
The formal fitting errors of the line profiles are relatively small compared to the larger uncertainties due to baseline structure; we do not attempt to quantify the parameter uncertainties in Table~\ref{tab_snr}.
Most detected SNRs show a broad line profile, with line widths $>50$\kms.
The agreement between the spectra of the different SNRs implies that despite the baseline uncertainties, these detections are real.

Early studies have posited three possible origins for RRL emission in the direction of SNRs.
\citet{Bignell1973} studied the RRL emission towards W44 and suggested that the observed H$166\alpha$ emission may be due to gas associated with and ionized by the supernova explosion of W44.
By comparing the RRL intensity and the low-frequency absorption toward 3C\,391, \citet{Cesarsky1973b} suggested that the RRL emission could possibly originate in cold ISM ($T_e \leq 400$\,K).
However, \citet{Pankonin1976} investigated RRLs toward W49B and concluded that the spectral lines were likely from low-brightness \hii\ regions along the line of sight.
These former studies on 3C391 and W49B indicate that stimulated emission may be a possible origin for RRLs in the direction of SNRs at frequencies $\leq 2$\,GHz.

\begin{figure}[htbp]
\centering
\subfloat[SIGGMA spectral grid]{\includegraphics[width = 3.5in]{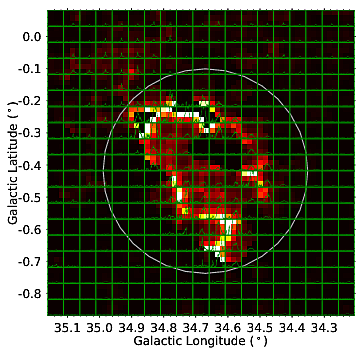}}
\subfloat[VGPS 1.4\ghz\ continuum map]{\includegraphics[width = 3.5in]{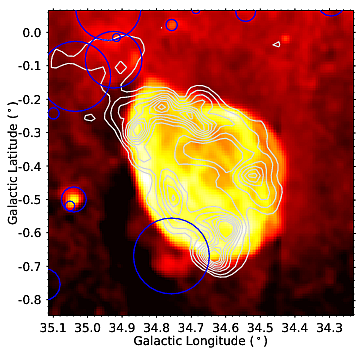}}\\
\caption{The H-RRL emission toward SNR: G34.7-0.4.
          (a) The SIGGMA spectral grid with the velocity range from $-300$ to $+300$\kms.
	  The background is the 0$^{th}$ moment H-RRL map over $0$ to $+150$\kms.
	  The white circle is the central region of the SNR.
	  (b) The VGPS 1.4\ghz\ continuum map overlaid by integrated RRL contours.
	  The contour levels are 0.5, 1, 2, 3, 4, 5, 6, 7 Jy\,beam$^{-1}$\kms.
	  The blue circles indicate the location of known \hii\ regions from the {\it WISE} catalog.
	  }
\label{fig_snr-g347}
\end{figure}

Figure~\ref{fig_snr-g347} shows the RRL detection toward SNR G34.7$-$0.4 (W44) with the SIGGMA spectral grid overlaid on the integrated H-RRL intensity map (left) and VGPS continuum with integrated RRL contours overlaid (right).
The spectral profiles vary over the source.
Within the central region, the lines are broad with strongly varying peak intensities.
At the northeast and southeast edge there are \hii\ regions (blue circles in Figure~\ref{fig_snr-g347} (b)), the spectra are combinations of broad and narrow Gaussian features.
The H-RRL morphology has a ring-like shape around the periphery of the SNR, indicating that a shell of gas around the expanding edge of W44 may be ionized by photons produced by the supernova explosion.
W44 is located adjacent to giant molecular clouds \citep{Denoyer1983} that have shell-like morphologies \citep{Jones1993}.
The interaction between W44 and the dense molecular clouds is revealed by the presence of shocked molecular gas \citep{Seta2004, Reach2005}.
\textbf{Figures for the other SNRs listed in Table~\ref{tab_snr} are given in Figure~set~2 in the online journal.}

Since SIGGMA data are of low frequency (1.4\ghz) RRL emission and thermal emission from SNRs is rare \citep{Cruciani2016}, the SIGGMA-detected RRLs toward SNRs are likely due to stimulated emission from the background synchrotron radiation of the SNRs.
The RRL-emitting gas may be physically associated with the SNRs, or could exist somewhere else along the line of sight.
The morphology of the RRL emitting regions toward the SNR sample, as well as their velocities and line widths, may allow for more detailed studies on the origin of RRLs toward SNRs.

\begin{table}[htbp]
\renewcommand{\arraystretch}{0.8}
\caption{ H-RRL detections toward SNRs}\label{tab_snr}
\begin{threeparttable}
\begin{tabular*}{\textwidth}{l@{\extracolsep{\fill}}*{12}{c}}
\hline
\hline
GName$^a$&Other Name& $l$& $b$& Radius&V$_{LSR}$&FWHM&D$_{N}$&D$_{F}$&D& Quality$^b$\\
&& deg.&deg.& deg.& \kms&\kms&kpc&kpc&kpc&\\
\hline
G33.6$+$0.1&Kes 79 & 33.67& $+$0.03& 0.11 &106.3&26.1 &6.4 &7.5  &7.1$^{[1]}$     &C\\ 
G34.7$-$0.4&W44    & 34.67& $-$0.42& 0.32 &47.7 &114.3& 2.9   &10.8 &2.9$^{[2]}$     &B\\
           &       &      &        &      &49.4 &11.6 &3.0 &10.7 &&\\
           &       &      &        &      &138.0&80.0 & 6.8   & 6.8    &&\\
G35.6$-$0.4&       & 35.60& $-$0.43& 0.13 &53.7 &23.5 &3.2 &10.3 &3.6$^{[3]}$     &A\\ 
G36.6$-$0.7&       & 36.59& $-$0.69& 0.28 &40.6 &64.4 &2.5    &10.8     &\nodata         &B\\
           &       &      &        &      &57.5 &20.2 & 3.4   & 9.9    &&\\
           &       &      &        &      &146.0&71.5 &6.7 &6.7 &&\\
G39.2$-$0.3&3C 396 & 39.23& $-$0.32& 0.08 &12.8 &34.0 &0.8    &9.2     &$\leq11.3^{[4]}$&A\\ 
           &       &      &        &      &61.2 &24.5 &3.7 &9.2  &&\\
G40.5$-$0.5&       & 40.51& $-$0.51& 0.23 &45.3 &36.1 &2.8 &9.8 &\nodata         &C\\
G41.1$-$0.3&3C 397 & 41.12& $-$0.31& 0.05 &62.1 &24.5 &3.8 &8.7 &10.3$^{[5]}$    &A\\ 
G41.5$+$0.4&       & 41.46& $+$0.39& 0.12 &19.2 &18.2 &1.3 &11.2 &10.3$^{[6]}$    &A\\ 
G43.3$-$0.2&W49B   & 43.27& $-$0.19& 0.07 &9.1  &42.5 &0.6 &11.5     &7.5$^{[1]}$,10$^{[7]}$      &A\\ 
           &       &      &        &      &62.7 &24.7 &4.1 &8.0  &&\\
G45.7$-$0.4&       & 45.56& $-$0.35& 0.21 &65.1 &18.6 &4.8 &6.8  &\nodata         &B\\
           &       &      &        &      &86.5 &100.4 &5.8    &5.8     &&\\
G46.8$-$0.3&HC 30  & 46.77& $-$0.27& 0.16 &60.2 &25.3 &4.4 &7.0  &6.4$^{[1]}$     &C\\
G49.2$-$0.7&W51C   & 49.14& $-$0.61& 0.22 &62.3 &41.2 &5.4 &5.4  &6$^{[1]}$       &A\\
           &       &      &        &      &122.8&51.8  &5.4    &5.4     &&\\
G54.4$-$0.3&HC 40  & 54.47& $-$0.29& 0.38 &43.6 &15.5 &4.1 &5.6  &3.3$^{[1]}$,6.6$^{[8]}$     &B\\
G57.2$+$0.8&4C21.53& 57.24& $+$0.82& 0.10 &106.1&39.8 &4.5 &4.5  &\nodata         &B\\ 
\hline
\end{tabular*}
\begin{tablenotes}
      \small
      \item a. Names are defined using the Galactic coordinates given by \cite{Green2014}.
      \item b. Qualitative metric of data quality (A=high quality, C=low quality)
      \item REFERENCES: [1] \citet{Case1998};[2] \citet{Seta1998};[3] \citet{Zhu2013};[4] \citet{Caswell1975}; [5] \citet{Safi-Harb2005}; [6] \citet{Pavlovic2014}; [7] \citet{Zhu2014}; [8] \citet{Ranasinghe2017}.
      \end{tablenotes}
\end{threeparttable}

\end{table}

\section{Individual RRL emission regions}\label{sec_indi}

SIGGMA data can reveal information not apparent in single-pointing RRL observations.
To illustrate this, we present the RRL analysis of two large star-forming regions, W49 and W51 \citep{Westerhout1958}.

\subsection{W49}
The W49 star-forming complex can be separated into sub-regions W49A and W49B \citep{Mezger1967c}.
W49A is among the most luminous star formation complexes in the Galaxy and is located at a distance of 11.11$^{+0.79}_{-0.69}$\,\kpc\ reported by \citet{Zhang2013} from trigonometric parallax measurements of H$_{2}$O masers.
W49B is a SNR whose centroid is $\sim12\arcmin$ from that of W49A \citep{Lacey2001}.
In the W49A region, there are compact radio sources with RRL emission that are embedded in diffuse ionized gas \citep{Depree1997,Depree2004}.
Although W49A has many embedded ultra-compact \hii\ regions, due to beam dilution much of the RRL emission observed by SIGGMA likely arises mostly from diffuse gas.
Also, at L-band, compact \hii\ regions may be optically thick and therefore may not contribute as much to the observed RRL emission as they do at higher frequencies \citep{Kim2001,Wood1989}.

\begin{figure}[htbp]
\centering
\subfloat[Integrated RRL intensity map overlaid with spectral lines.]{\includegraphics[width = 3.0in]{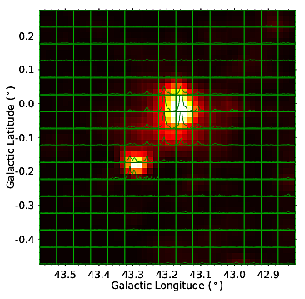}}
\subfloat[H-RRL spectra of W49A and W49B]{\includegraphics[width = 3.0in]{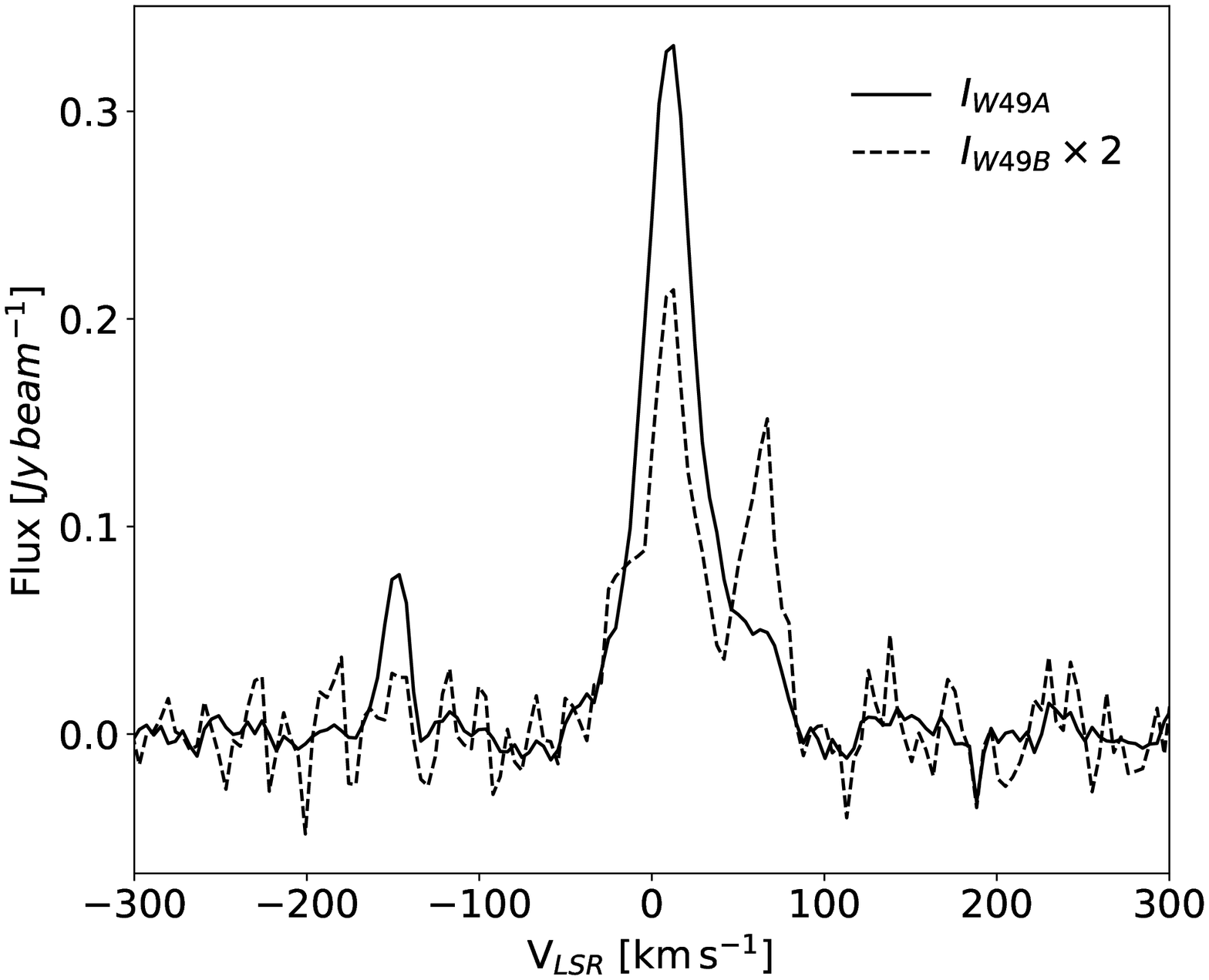}}\\
\subfloat[The 10\kms\ velocity component]{\includegraphics[width = 3.0in]{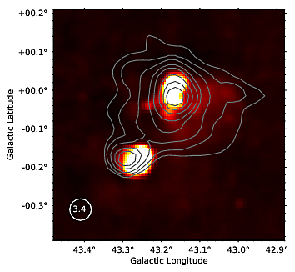}}
\subfloat[The 60\kms\ velocity component]{\includegraphics[width = 3.0in]{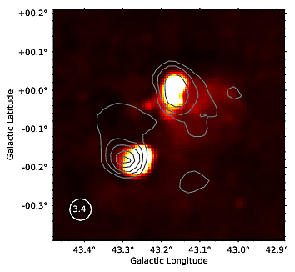}}\\
\caption{SIGGMA observations of the W49 complex. W49A is centered around G43.2+0.0 and W49B is to the southeast centered around G43.3-0.2.
(a) The RRL spectral grid. The background image is of RRL integrated intensity integrated over the velocity range from 0 to 70\kms.
(b) The spectra of W49A and W49B, which are averaged over radius from the {\it WISE} catalog for W49A and the radius given in Table~\ref{tab_snr} for W49B. The line intensity of W49B is enlarged by factor of 2 so that its detail can be seen.
(c) and (d) are the SIGGMA integrated intensity RRL contours overlaid on VGPS 1.4\ghz\ continuum data. The 3.$4\arcmin$ diameter circles show the ALFA beam size. The SIGGMA data were integrated over $0$ to $+20$\kms\ (panel c) and $+50$ to $+70$\kms\ (panel d).
The SIGGMA contour levels are at values of 0.05, 0.1, 0.2, 0.4, 0.6, 1.0, 1.4, and 1.8 Jy\,beam$^{-1}$\kms\ for (c) and 0.05, 0.1, 0.2, 0.4, and 0.6 Jy\,beam$^{-1}$\kms\ for (d).
}\label{w49_spec_map}
\end{figure}

Figure~\ref{w49_spec_map} (a) shows the SIGGMA RRL spectral grid for the W49 complex.
RRL emission is detected over the entire region.
In addition to the known velocity of W49A, $\sim10$\kms, there is a second velocity component around $\sim60$\kms\ found toward both W49A and W49B (Figure~\ref{w49_spec_map} (b)).
This 60\kms\ component has been detected previously.
\citet{Downes1974} reported the RRL detection in the direction of W49B with a LSR velocity of 65\kms.
\citet{Pankonin1975} also detected RRL emission toward W49B at an LSR velocity of 60\kms.
\citet{Anantharamaiah1986} detected a RRL velocity of 63\kms\ toward W49A (G43.2-0.1).
Figure~\ref{w49_spec_map} panels (c) and (d) show the RRL intensity contours of the two velocity components overlaid on top of VGPS continuum data.
The 60\kms\ RRL component is also found toward W49A, although some of its intensity could arise from the line wing of 10\kms\ component.
The peak emission of both velocity components is approximately at the location of maximum VGPS continuum intensity for W49A.
The location of peak H-RRL intensity for both components is offset to the east of the VGPS peak of W49B.

The existence of two H-RRL velocity components toward the W49 complex is puzzling.  The kinematic distance for W49A from the 10\kms\ component, 11.4\kpc\, agrees with the known distance for this region, 11.1\kpc\, so we can assume that the 10\kms\ component comes from gas about 11\kpc\ away.
Previous \hi\ absorption studies suggested W49B lies at a distance of $\sim8.0$\kpc\ \citep{Moffett1994} or $\sim11.4$\kpc\ \citep{Brogan2001}.
Recent work by \citet{Zhu2014} assigned a kinematic distance of $\sim10$\kpc\ for W49B based on the detection of CO emission at a velocity of 40\kms.
As mentioned in Section~\ref{sec_snr}, \citet{Pankonin1976} attributed the 60\kms\ RRL to extended \hii\ regions, which is consistent with the extended emitting morphology seen in SIGGMA data.
It is also possible that the 60\kms\ component is due to the local velocity structure associated with the expanding motion of W49B. In this scenario, one would expect to observe a shell of RRL emission surrounding W49B, as may be seen in Figure~\ref{w49_spec_map} (d).
However, the line width we observe for this velocity component is similar to that of other \hii\ regions; we would expect it to be broadened if the ionized gas is expanding.
Thus we do not believe that this velocity component can be explained by expansion. 

\begin{figure}[htbp]
\centering
\plotone{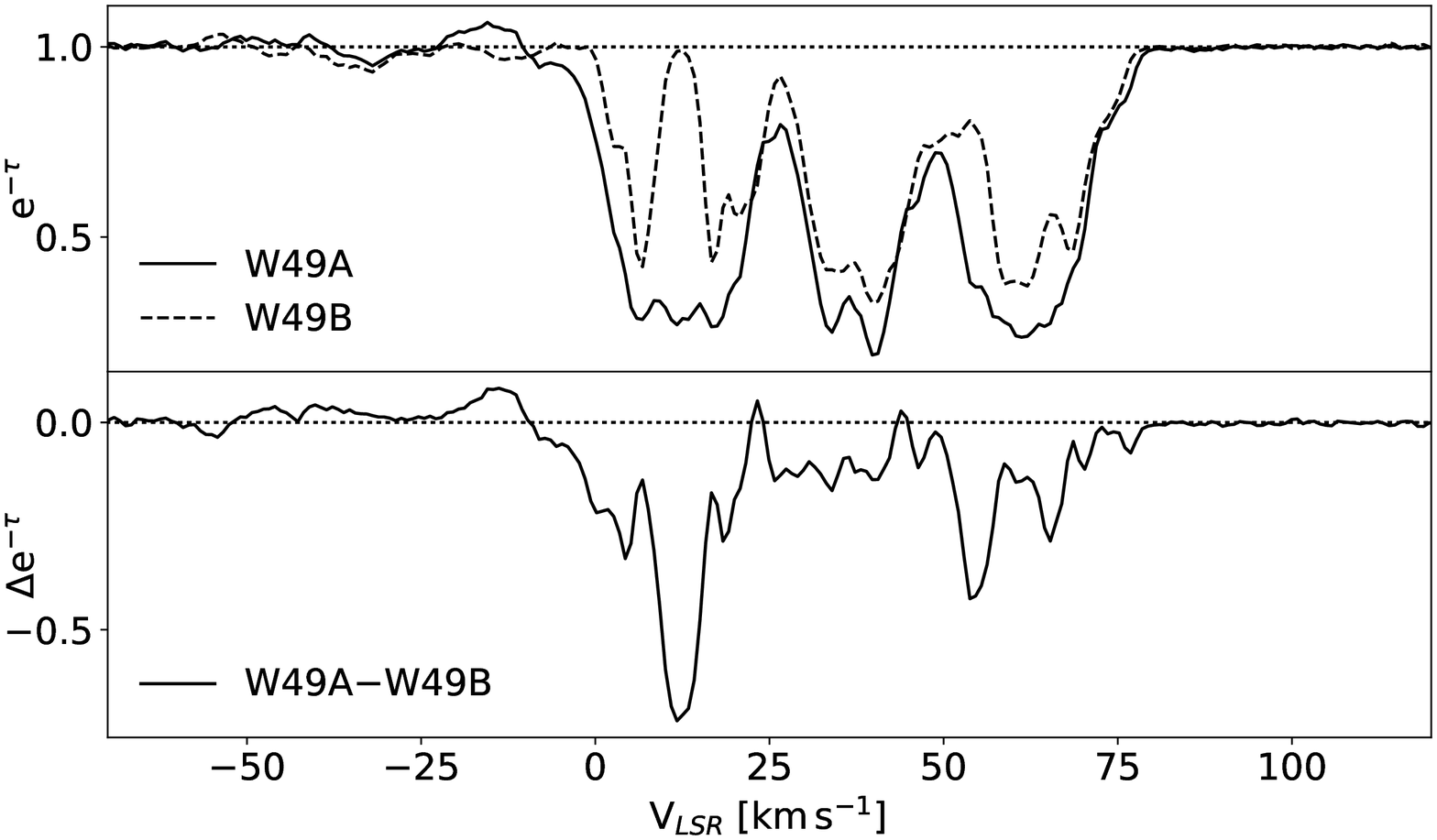}
\caption{\hi\ absorption spectra toward W49A and W49B (top), and the difference in \hi\ absorption between the two regions (bottom). The absorption features in the bottom panel may arise from gas in between W49A and W49B. 
}\label{w49_hi}
\end{figure}

To determine if the 60\kms\ component is associated with W49B, we analyze VGPS \hi\ data in its direction.  We extract the integrated \hi\ VGPS spectrum toward a portion of W49A and W49B.  We did not extract spectra toward the brightest portions of these regions so that the average spectra are not saturated.  We see \hi\ absorption from all gas in foreground to the bright continuum sources.
The upper spectrum in Figure~\ref{w49_hi} shows that there is \hi\ with velocities from 0$-$70\kms\ along the lines of sight toward W49A and W49B.
The lower spectrum is the difference of the \hi\ absorption between W49A and W49B.  The negative dips in the lower panel therefore represent the velocities of gas that is foreground to W49A, but that lies behind W49B.

Essentially all velocities where W49A shows absorption have corresponding absorption for W49B, except near 10\kms\ and possibly near 60\kms.
We conclude from this analysis there is \hi\ at 10\kms\ that lies in between W49A and W49B, and therefore that W49B itself lies foreground to W49A.
The 60\kms\ gas may also lie in between W49A and W49B, although in this case the situation is more unclear due to the smaller difference in the \hi\ spectra.
We see \hi\ absorption at 60\kms\ from both W49A and W49B with different optical depths in the upper panel in Figure~\ref{w49_hi}. A possible scenario is that the 60\kms\ \hi\ gas is located at the tangent point in the foreground of the W49 complex.  At the tangent point, velocity crowding would increase the signal from diffuse gas.
Since we only see the 60\,\kms\ component in the direction of the continuum peaks, it is likely stimulated by W49A and W49B.

\subsection{W51}
The W51 region is another enormous first-quadrant star formation complex.
It is located near the tangent point in the Sagittarius Arm at a distance of 5.41$^{+0.31}_{-0.28}$\kpc\ measured by using maser trigonometric parallax \citep{Sato2010}.  W51 is grouped into three sub-regions: W51A, W51B, and W51C.
W51A and W51B are massive star-forming regions, whereas W51C is a SNR \citep{Subrahmanyan1995}.
Pointed RRL observations toward W51 have been carried out \citep[i.e.][]{Mezger1967b,Terzian1969,Pankonin1979,Roelfsema1992b, Anderson2011}, showing that it can be decomposed into multiple discrete \hii\ regions.

\begin{figure}[htbp]
\centering
\plotone{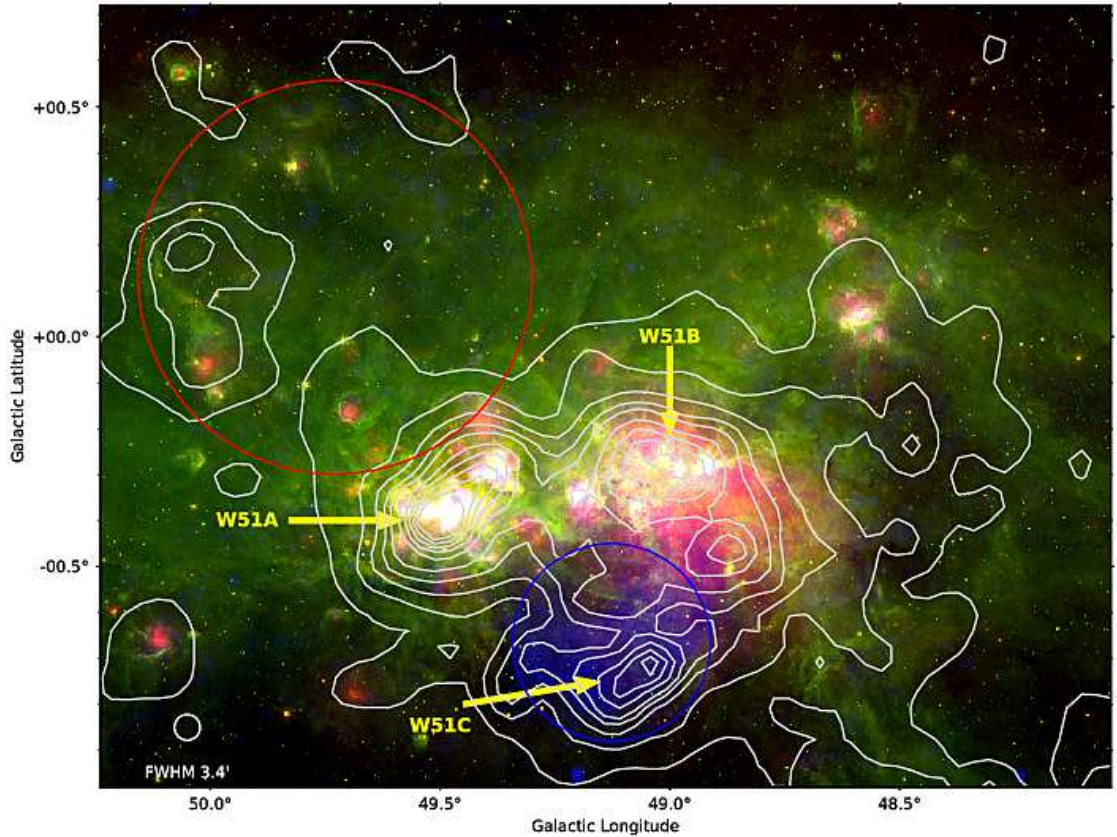}
\caption{Three-color map of W51 
with {\it Spitzer} 24 $\um$ data in red,  {\it Spitzer} 8 $\um$ data in green, and VGPS 1.4\ghz\ continuum data in blue. 
SIGGMA H-RRL integrated over 30\kms\ to 90\kms\ are overlaid at levels of 0.05, 0.15, 0.3, 0.45, 0.6, 0.9, 1.2, 1.5, 1.8, and 2.1 Jy\,beam$^{-1}$\kms.
The blue circle shows the location and extent of the SNR W51C as defined in the Green catalog.}\label{w51_map}
\end{figure}

Similar to W49, much of the SIGGMA RRL emission toward W51 is likely due to diffuse gas rather than from compact \hii\ regions.
Figure~\ref{w51_map} shows the comparison of the integrated RRL contours from SIGGMA, with 1.4\ghz\ continuum from the VGPS (blue), and  IR data from {\it Spitzer} (8$\um$ in green and 24 $\um$ in red).
The IR emission, which arises from dust and PAHs, is strongly associated with the \hii\ regions.
The RRL contours traces the strong IR emission and also shows some RRL emitting regions where the IR emission is not obvious.
To the west of W51B in Figure~\ref{w51_map} (around G48.5$-$0.2), RRLs are detected from diffuse ionized gas, which shows a bubble-like feature in the IR.
\textbf{One can see the spectral grid of C-RRL emission toward W51A and part of W51B in Figure~set~1 (figure of G49.5$-$0.4), and the integrated H-RRL contours of the W51C region in Figure~set~2 (figure of G49.2$-$0.7).}

\begin{figure}[htbp]
\centering
\subfloat[The RRL spectra.]{\includegraphics[width = 3.2in]{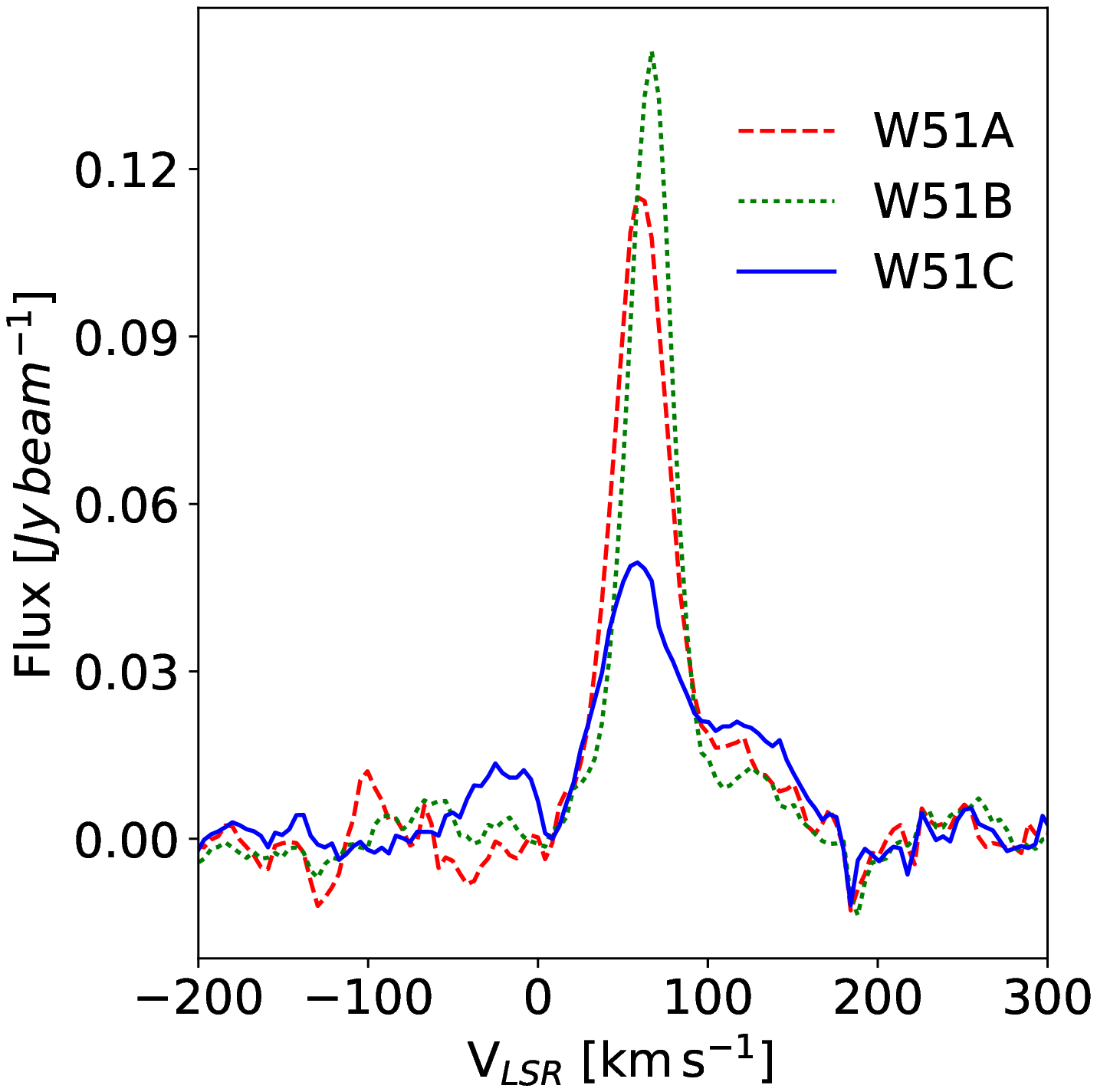}}
\subfloat[The moment maps]{\includegraphics[width = 3.2in]{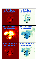}}\\
\caption{The SIGGMA RRL results of W51 region. (a) The averaged spectra of W51A, W51B and W51C. (b) The 0th and 2nd moment maps with velocity ranges of $-30$ to $0$, $+10$ to $+90$, and $+120$ to $+150$\kms. The location and extent of SNR W51C as defined in the Green catalog is marked by the black circles in the images.}\label{w51_rrl}
\end{figure}

As we saw for other SNRs, there is strong RRL emission toward W51C (Figure~\ref{w51_map}), which is bright in 1.4\ghz\ continuum, but weak in the IR.
Figure~\ref{w51_rrl} (a) shows the averaged RRL \textbf{spectra} from W51A, W51B, and W51C.
The tangent velocity in the direction of W51 is $\sim60$\kms.
The LSR velocities of SIGGMA-detected RRLs toward W51A, W51B, and W51C are 62.8\kms, 66.4\kms, and 62.3\kms, indicating that they are all probably part of a large star formation complex in a large molecular cloud.
This agrees well with the comprehensive studies in the literature \citep{Mufson1979, Mehringer1994, Brogan2013}.
Besides the primary $\sim60$\kms\ velocity component, the spectral line toward W51C also shows a broad profile, which is a common feature among the SIGGMA detections toward other SNRs.

\citet{Koo1997a, Koo1997b} studied the interaction between W51C and a nearby molecular cloud.
From the interface between the SNR and the surrounding molecular cloud, they found emission line components of CO and HCO$^{+}$, as well as \hi, at a velocity of around 100\kms.
Coincidentally, the broad RRL component is at 122.8\kms\ toward W51C.
In addition, we can see a bump around $-20$\kms\ in the W51C spectrum (Figure~\ref{w51_rrl} (a)) which shows no indication of being fictitious. 
The $-20$\kms\ component may be C-RRL counterpart of the 122\kms\ H-RRL, although the broad width of this line is not consistent with it being from carbon.

In order to obtain a better understanding of the origin of the RRL in the direction of W51C, we produce 0th and 2nd moment maps by averaging over three separated velocity ranges, which are $-30$ to $0$, $+10$ to $+90$, and $+120$ to $+150$\kms\ (Figure~\ref{w51_rrl} (b)).
From the middle-row maps, we see the $\sim60$\kms\ RRL emitting region is covering the whole region uniformly. 
Towards the SNR W51C, Figure~\ref{w51_map} shows that the morphology of the 60 km/s RRL emitting region has a good correlation with the background continuum.
Thus, the 62.3\kms\ component along the line of sight of W51C might be from the stimulated emission from diffuse \hii\ gas within the W51 complex.
The upper- and lower-row maps show that the $<0$\kms\ and $>+100$\kms\ line emitting components seem to be encompassing the SNR, possibly indicating that RRLs originate from the post-shock ionized gas located at the interface between the SNR and molecular gas.
The $\sim$100\kms\ velocity from the H-RRL is compatible with the velocity of an expanding shell estimated by \cite{Koo1997a} based on their \hi\ analysis.


\section{Summary} \label{sec_summ}
SIGGMA is the most sensitive large-scale RRL survey extent.
The SIGGMA data fully sample the RRL emission from ionized gas along the Galactic plane $32\degr \leq \ell \leq 70\degr$, $|b|\le 1\degr.5$.
The average rms level of the SIGGMA spectra is $\sim0.65$\,mJy\,beam$^{-1}$ ($\sim6.5$\,mK).
Using the survey data observed with a FWHM of 3$\arcmin$.4, we have produced RRL data cubes with a smoothed FWHM of 6$\arcmin$ and a velocity resolution of 5.1\kms.
Comparing with the previous studies within the survey zone, SIGGMA provides RRL maps with the highest angular resolution.
In this paper, we have summarized the current status of the survey, discussed the data quality and processing technique, and presented first survey results.
The survey data are available online in standard fits files format.
\citep[][Dataset: \url{http://doi.org/10.5281/zenodo.1432823}]{liu_bin_2018_1432823}

By matching with the {\it WISE} catalog, we generate RRL spectra in the direction of 244 known and 79 candidate \hii\ regions.
We analyzed the Galactic distribution of ionized emission and derived a thermal continuum map along the Galactic plane.
Due to the observational strategy, SIGGMA data are insensitive to large-scale emission from the WIM.

SIGGMA has catalogued unusually broad (FWHM $\geq$ 50\kms) and narrow (FWHM$\leq$ 15\kms) RRLs toward known and candidate \hii\ regions. Considering the qualities of SIGGMA spectra, it is possible that these line widths are caused by observational artifacts. For example, the broad lines may be due to incomplete baseline removal. The narrow lines, which are almost always found in spectra with multiple components, may be introduced by the calibration. If there is a RRL signal at a similar velocity in both the on- and off-source locations, the on-off spectrum may have a narrow residual. 
This effect can be mediated by calibrating the spectra using a median filtering technique \citep{McIntyre2013}\footnote{see \url{http://www.naic.edu/~astro/aotms/performance/medianfiltering.pdf}}, although this method also introduces undesirable systematic effects into the data, namely a reduction in peak line flux.
Follow-up studies of the narrow and broad lines should be done. If the detections are confirmed, they would imply extreme properties of ionized gas, possibly related to expansion velocity and turbulence.

We detected 11 C-RRL emitting regions, all of which are co-spatial with known \hii\ regions.
The C-RRL distributions of these regions are well-matched with \textbf{their} morphologies at \textbf{8}\,$\um$, which is a good tracer for PDRs.
The comparison supports a PDR origin of C-RRL emission.
The occasional mismatches, where C-RRL was detected with weak IR emission, suggests the possibilities of other C-RRL forming regions far away from \hii\ region molecular cloud interfaces.

Previous studies have reported $\sim5$ RRL detections in the direction of SNRs. 
We detected the RRL emission toward 14 Galactic SNRs within the SIGGMA region, nearly tripling the known sample.
Many of these detections have broad line components $\geq 40$\kms.
Thermal emission is not commonly expected towards SNRs.
The RRL emission may be from ionized gas that is physically associated with the SNRs, or due to foreground diffuse ionized gas along the line of sight.
For SNRs with large FWHM RRL line widths, the former scenario is preferred since the broad line widths may represent the expanding motion of local gas surrounding the SNRs.
The spacial distribution of RRL emission revealed by SIGGMA can be useful to identify its origin. Comparing the morphology of the RRL emitting region with the non-thermal background of SNR, we can discuss the possibility of stimulated RRL emission.

To illustrate the utility of SIGGMA data toward star formation complexes, we discussed results toward the two bright star-forming complexes, W49 and W51.
We detect two RRL velocity components toward the W49 complex.  Our analysis of the SNR W49B data suggest that the 10\kms\ RRL component is unlikely to be associated with the SNR.
Although the origin of the 60\kms\ is not clear, it may be caused by stimulated emission of foreground diffuse ionized gas.
RRLs in the direction of SNR W51C were also detected by SIGGMA, with velocities of $\sim60$\kms\ and $\sim120$\kms.
The $\sim60$\kms\ component toward the SNR W51C is similar to that of W51A and W51B.
Stimulated emission may again explain the strength of the 60\kms\ component toward W51C, and
the high velocity component may indicate the expanding motion of local gas surrounding the SNR.

\acknowledgments
We thank the anonymous referee for thoughtful comments and suggestions which have improved this work.
We are grateful to the Arecibo Observatory staff for their help and hospitality during the SIGGMA project.
The Arecibo Observatory is operated by SRI International under a cooperative agreement with the National Science
Foundation (AST-1100968), and in alliance with Ana G. M\'endez-Universidad Metropolitana, and the Universities Space Research Association.
We also thank the P-ALFA and ALFA-ZOA teams for organizing the commensal observations. 
Bin Liu thanks West Virginia University for hosting him as a postdoctoral researcher during completion of this project, and thanks the China Scholarships Council for financial support from program 201504910306.
Bin Liu is also supported by National Natural Science Foundation of China (Grant No: 11503036) and by the Open Project Program of the Key Laboratory of FAST, NAOC, Chinese Academy of Sciences.

\software{
        IDL (https://www.harrisgeospatial.com/SoftwareTechnology/IDL.aspx),
        Gridzilla \citep{Barnes2001},
        Astropy (http://dx.doi.org/10.1051/0004-6361/201322068),
        Matplotlib (http://dx.doi.org/10.1109/MCSE.2007.55).
}

\figsetstart
\figsetnum{1}
\figsettitle{Figures of detected C-RRL emission regions.}\label{figset:crrl}
\figsetgrpstart
\figsetgrpnum{1.1}
\figsetgrptitle{The C-RRL emission region: G34.7-0.6.}
\figsetplot{G34.7-0.6_ca_spec.eps}
\figsetplot{G34.7-0.6_rgb_contour.eps}
\figsetgrpnote{
    (a) The SIGGMA spectral grid with the velocity range from $-50$ to $+150$\kms.
	The background is the 0$^{th}$ moment C-RRL map over $30$ to $50$\kms.
    The circle illustrates the area over which the C-RRL spectra were averaged.
	The negative spike to the south of the region is likely to be RFI from the off-position that remains in the final spectra.
	(b) The IR RGB color map (r: 24\,$\um$, g: 8\,$\um$, b: 4.5\,$\um$) overlaid by integrated RRL contour.
	The levels are 4, 8, 12, 16, 18, 22 mJy\,beam$^{-1}$\kms.}
	\label{fig_crrl-g347}
\figsetgrpend
\figsetgrpstart
\figsetgrpnum{1.2}
\figsetgrptitle{The C-RRL emission region: G35.1-0.7.}
\figsetplot{G35.1-0.7_ca_spec.eps}
\figsetplot{G35.1-0.7_rgb_contour.eps}
\figsetgrpnote{
	(a) The SIGGMA spectral grid with the velocity range from $-50$ to $+150$\kms.
	The background is the 0$^{th}$ moment C-RRL map over $10$ to $40$\kms.
    The circle illustrates the area over which the C-RRL spectra were averaged.
	(b) The IR RGB color map (r: 24\,$\um$, g: 8\,$\um$, b: 4.5\,$\um$) overlaid by integrated RRL contour.
	The levels are 2, 4, 6, 8, 10, 12, 15, 18 mJy\,beam$^{-1}$\kms.}
	\label{fig_crrl-g351}
\figsetgrpend
\figsetgrpstart
\figsetgrpnum{1.3}
\figsetgrptitle{The C-RRL emission region G38.9-0.4.}
\figsetplot{G38.9-0.4_ca_spec.eps}
\figsetplot{G38.9-0.4_rgb_contour.eps}\\
\figsetgrpnote{
	(a) The SIGGMA spectral grid with the velocity range from $-50$ to $+150$\kms.
	The background is the 0$^{th}$ moment C-RRL map integrated over $20$ to $50$\kms.
    The circle illustrates the area over which the C-RRL spectra were averaged.
	(b) The IR RGB color map (r: 24\,$\um$, g: 8\,$\um$, b: 4.5\,$\um$) overlaid with integrated C-RRL contours.
	The contour levels are 2, 4, 6, 7 mJy\,beam$^{-1}$\kms.}
	\label{fig_crrl-g389}
\figsetgrpend

\figsetgrpstart
\figsetgrpnum{1.4}
\figsetgrptitle{The C-RRL emission region: G43.2-0.0.}
\figsetplot{G43.2-0.0_ca_spec.eps}
\figsetplot{G43.2-0.0_rgb_contour.eps} 
\figsetgrpnote{
	(a) The SIGGMA spectral grid with the velocity range from $-50$ to $+150$\kms.
	The background is the 0$^{th}$ moment C-RRL map over $-20$ to $20$\kms.
    The circle illustrates the area over which the C-RRL spectra were averaged.
	(b) The IR RGB color map (r: 24\,$\um$, g: 8\,$\um$, b: 4.5\,$\um$) overlaid by integrated RRL contour.
	The levels are 5, 10, 25, 50, 100, 200 mJy\,beam$^{-1}$\kms.}
	\label{fig_crrl-g432}
\figsetgrpend
\figsetgrpstart
\figsetgrpnum{1.5}
\figsetgrptitle{The C-RRL emission region: G45.5+0.0.}
\figsetplot{G45.5+0.0_ca_spec.eps}
\figsetplot{G45.5+0.0_rgb_contour.eps} 
\figsetgrpnote{
	(a) The SIGGMA spectral grid with the velocity range from $-50$ to $+150$\kms.
	The background is the 0$^{th}$ moment C-RRL map over $30$ to $70$\kms.
    The circle illustrates the area over which the C-RRL spectra were averaged.
	(b) The IR RGB color map (r: 24\,$\um$, g: 8\,$\um$, b: 4.5\,$\um$) overlaid by integrated RRL contour.
	The levels are 5, 10, 15, 20, 25 mJy\,beam$^{-1}$\kms.}
	\label{fig_crrl-g455}
\figsetgrpend
\figsetgrpstart
\figsetgrpnum{1.6}
\figsetgrptitle{The C-RRL emission region: G49.5-0.4.}
\figsetplot{G49.5-0.4_ca_spec.eps}
\figsetplot{G49.5-0.4_rgb_contour.eps}
\figsetgrpnote{
	(a) The SIGGMA spectral grid with the velocity range from $-50$ to $+150$\kms.
	The background is the 0$^{th}$ moment C-RRL map over $30$ to $70$\kms.
    The circle illustrates the area over which the C-RRL spectra were averaged.
	(b) The IR RGB color map (r: 24\,$\um$, g: 8\,$\um$, b: 4.5\,$\um$) overlaid by integrated RRL contour.
	The levels are 5, 10, 25, 50, 100, 200, 250 mJy\,beam$^{-1}$\kms.}
	\label{fig_crrl-g495}
\figsetgrpend
\figsetgrpstart
\figsetgrpnum{1.7}
\figsetgrptitle{The C-RRL emission region: G53.6+0.0.}
\figsetplot{G53.6+0.0_ca_spec.eps}
\figsetplot{G53.6+0.0_rgb_contour.eps}
\figsetgrpnote{
	(a) The SIGGMA spectral grid with the velocity range from $-50$ to $+150$\kms.
	The background is the 0$^{th}$ moment C-RRL map over $0$ to $+30$\kms.
    The circle illustrates the area over which the C-RRL spectra were averaged.
	(b) The IR RGB color map (r: 24\,$\um$, g: 8\,$\um$, b: 4.5\,$\um$) overlaid by integrated RRL contour.
	The levels are 2, 4, 6, 8 mJy\,beam$^{-1}$\kms.}
	\label{fig_crrl-g536}
\figsetgrpend
\figsetgrpstart
\figsetgrpnum{1.8}
\figsetgrptitle{The C-RRL emission region: G60.9-0.1.}
\figsetplot{G60.9-0.1_ca_spec.eps}
\figsetplot{G60.9-0.1_rgb_contour.eps}
\figsetgrpnote{
	(a) The SIGGMA spectral grid with the velocity range from $-50$ to $+150$\kms.
	The background is the 0$^{th}$ moment C-RRL map over $0$ to $+30$\kms.
    The circle illustrates the area over which the C-RRL spectra were averaged.
	(b) The IR RGB color map (r: 24\,$\um$, g: 8\,$\um$, b: 4.5\,$\um$) overlaid by integrated RRL contour.
	The levels are 2, 4, 6, 8 mJy\,beam$^{-1}$\kms.}
	\label{fig_crrl-g609}
\figsetgrpend
\figsetgrpstart
\figsetgrpnum{1.9}
\figsetgrptitle{The C-RRL emission region: G61.4+0.1.}
\figsetplot{G61.4+0.1_ca_spec.eps}
\figsetplot{G61.4+0.1_rgb_contour.eps}
\figsetgrpnote{
	(a) The SIGGMA spectral grid with the velocity range from $-50$ to $+150$\kms.
	The background is the 0$^{th}$ moment C-RRL map over $0$ to $+30$\kms.
    The circle illustrates the area over which the C-RRL spectra were averaged.
	(b) The IR RGB color map (r: 24\,$\um$, g: 8\,$\um$, b: 4.5\,$\um$) overlaid by integrated RRL contour.
	The levels are 5, 10, 20, 40, 80,160, 320 mJy\,beam$^{-1}$\kms.}
	\label{fig_crrl-g614}
\figsetgrpend
\figsetgrpstart
\figsetgrpnum{1.10}
\figsetgrptitle{The C-RRL emission region: G63.1+0.4.}
\figsetplot{G63.1+0.4_ca_spec.eps}
\figsetplot{G63.1+0.4_rgb_contour.eps}
\figsetgrpnote{
	(a) The SIGGMA spectral grid with the velocity range from $-50$ to $+150$\kms.
	The background is the 0$^{th}$ moment C-RRL map over $-5$ to $+30$\kms.
    The circle illustrates the area over which the C-RRL spectra were averaged.
	(b) The IR RGB color map (r: 24\,$\um$, g: 8\,$\um$, b: 4.5\,$\um$) overlaid by integrated RRL contour.
	The levels are 5, 10, 15, 20, 25, 30 mJy\,beam$^{-1}$\kms.}
	\label{fig_crrl-g631}
\figsetgrpend
\figsetend
\figsetstart
\figsetnum{2}

\figsettitle{Figures of detected H-RRL emission toward Galactic Supernova Remnants.}\label{figset:snr}
\figsetgrpstart
\figsetgrpnum{2.1}
\figsetgrptitle{The H-RRL emission toward SNR: G33.6+0.1.}
\figsetplot{G33.6+0.1_ha_spec.eps}
\figsetplot{G33.6+0.1_vgps_contour.eps}  
\figsetgrpnote{
	(a) The SIGGMA spectral grid with the velocity range from $-300$ to $+300$\kms.
	The background is the 0$^{th}$ moment H-RRL map over $0$ to $+150$\kms.
	The white circle is the central region of the SNR.
	(b) The VGPS 1.4\ghz\ continuum map overlaid by integrated RRL contour.
	The levels are 0.1, 0.15, 0.2, 0.25, 0.3, 0.35 Jy\,beam$^{-1}$\kms.
	The blue circles are location of known \hii\ regions from the WISE catalog.
	}
	\label{fig_snr-g336}
\figsetgrpend
\figsetgrpstart
\figsetgrpnum{2.2}
\figsetgrptitle{The H-RRL emission toward SNR: G35.6-0.4.}
\figsetplot{G35.6-0.4_ha_spec.eps}
\figsetplot{G35.6-0.4_vgps_contour.eps}
\figsetgrpnote{
	(a) The SIGGMA spectral grid with the velocity range from $-300$ to $+300$\kms.
	The background is the 0$^{th}$ moment H-RRL map over $0$ to $+150$\kms.
	The white circle is the central region of the SNR.
	(b) The VGPS 1.4\ghz\ continuum map overlaid by integrated RRL contour.
	The levels are 0.05, 0.07, 0.09, 0.11, 0.13, 0.15 Jy\,beam$^{-1}$\kms.
	The blue circles are location of known \hii\ regions from the WISE catalog.
	}
	\label{fig_snr-g356}
\figsetgrpend
\figsetgrpstart
\figsetgrpnum{2.3}
\figsetgrptitle{The H-RRL emission toward SNR: G36.6-0.7.}
\figsetplot{G36.6-0.7_ha_spec.eps}
\figsetplot{G36.6-0.7_vgps_contour.eps}
\figsetgrpnote{
	(a) The SIGGMA spectral grid with the velocity range from $-300$ to $+300$\kms.
	The background is the 0$^{th}$ moment H-RRL map over $0$ to $+150$\kms.
	The white circle is the central region of the SNR.
	(b) The VGPS 1.4\ghz\ continuum map overlaid by integrated RRL contour.
	The levels are 0.04, 0.07, 0.1, 0.13, 0.16 Jy\,beam$^{-1}$\kms.
	There are no \hii\ regions located in the figure covered region.
	}
	\label{fig_snr-g366}
\figsetgrpend
\figsetgrpstart
\figsetgrpnum{2.4}
\figsetgrptitle{The H-RRL emission toward SNR: G39.2-0.3.}
\figsetplot{G39.2-0.3_ha_spec.eps}
\figsetplot{G39.2-0.3_vgps_contour.eps}
\figsetgrpnote{
	(a) The SIGGMA spectral grid with the velocity range from $-300$ to $+300$\kms.
	The background is the 0$^{th}$ moment H-RRL map over $0$ to $+150$\kms.
	The white circle is the central region of the SNR.
	(b) The VGPS 1.4\ghz\ continuum map overlaid by integrated RRL contour.
	The levels are 0.1, 0.2, 0.4, 0.6, 0.8, 1.0 Jy\,beam$^{-1}$\kms.
	The blue circles are location of known \hii\ regions from the WISE catalog.
	}
	\label{fig_snr-g392}
\figsetgrpend
\figsetgrpstart
\figsetgrpnum{2.5}
\figsetgrptitle{The H-RRL emission toward SNR: G40.5-0.5.}
\figsetplot{G40.5-0.5_ha_spec.eps}
\figsetplot{G40.5-0.5_vgps_contour.eps}
\figsetgrpnote{
	(a) The SIGGMA spectral grid with the velocity range from $-300$ to $+300$\kms.
	The background is the 0$^{th}$ moment H-RRL map over $0$ to $+150$\kms.
	The white circle is the central region of the SNR.
	(b) The VGPS 1.4\ghz\ continuum map overlaid by integrated RRL contour.
	The levels are 0.02, 0.04, 0.06, 0.08, 0.1, 0.15, 0.2 Jy\,beam$^{-1}$\kms.
	There are no \hii\ regions located in the figure covered region.
	}
	\label{fig_snr-g405}
\figsetgrpend
\figsetgrpstart
\figsetgrpnum{2.6}
\figsetgrptitle{The H-RRL emission toward SNR: G41.1-0.3.}
\figsetplot{G41.1-0.3_ha_spec.eps}
\figsetplot{G41.1-0.3_vgps_contour.eps}
\figsetgrpnote{
	(a) The SIGGMA spectral grid with the velocity range from $-300$ to $+300$\kms.
	The background is the 0$^{th}$ moment H-RRL map over $0$ to $+150$\kms.
	The white circle is the central region of the SNR.
	(b) The VGPS 1.4\ghz\ continuum map overlaid by integrated RRL contour.
	The levels are 0.1, 0.2, 0.3, 0.4, 0.5 Jy\,beam$^{-1}$\kms.
	The blue circles are location of known \hii\ regions from the WISE catalog.
	}
	\label{fig_snr-g411}
\figsetgrpend
\figsetgrpstart
\figsetgrpnum{2.7}
\figsetgrptitle{The H-RRL emission toward SNR: G41.5+0.4.}
\figsetplot{G41.5+0.4_ha_spec.eps}
\figsetplot{G41.5+0.4_vgps_contour.eps}
\figsetgrpnote{
	(a) The SIGGMA spectral grid with the velocity range from $-300$ to $+300$\kms.
	The background is the 0$^{th}$ moment H-RRL map over $0$ to $+150$\kms.
	The white circle is the central region of the SNR.
	(b) The VGPS 1.4\ghz\ continuum map overlaid by integrated RRL contour.
	The levels are 0.05, 0.08, 0.11, 0.14, 0.17 Jy\,beam$^{-1}$\kms.
	The blue circles are location of known \hii\ regions from the WISE catalog.
	}
	\label{fig_snr-g415}
\figsetgrpend
\figsetgrpstart
\figsetgrpnum{2.8}
\figsetgrptitle{The H-RRL emission toward SNR: G43.3-0.2.}
\figsetplot{G43.3-0.2_ha_spec.eps}
\figsetplot{G43.3-0.2_vgps_contour.eps}
\figsetgrpnote{
	(a) The SIGGMA spectral grid with the velocity range from $-300$ to $+300$\kms.
	The background is the 0$^{th}$ moment H-RRL map over $0$ to $+150$\kms.
	The white circle is the central region of the SNR.
	(b) The VGPS 1.4\ghz\ continuum map overlaid by integrated RRL contour.
	The levels are 0.25, 0.5, 1, 1.5, 2, 2.5, 3 Jy\,beam$^{-1}$\kms.
	The blue circles are location of known \hii\ regions from the WISE catalog.
	}
	\label{fig_snr-g433}
\figsetgrpend
\figsetgrpstart
\figsetgrpnum{2.9}
\figsetgrptitle{The H-RRL emission toward SNR: G45.7-0.4.}
\figsetplot{G45.7-0.4_ha_spec.eps}
\figsetplot{G45.7-0.4_vgps_contour.eps}
\figsetgrpnote{
	(a) The SIGGMA spectral grid with the velocity range from $-300$ to $+300$\kms.
	The background is the 0$^{th}$ moment H-RRL map over $0$ to $+150$\kms.
	The white circle is the central region of the SNR.
	(b) The VGPS 1.4\ghz\ continuum map overlaid by integrated RRL contour.
	The levels are 0.05, 0.1, 0.15, 0.2, 0.25, 0.3, 0.4 Jy\,beam$^{-1}$\kms.
	The blue circles are location of known \hii\ regions from the WISE catalog.
	}
	\label{fig_snr-g457}
\figsetgrpend
\figsetgrpstart
\figsetgrpnum{2.10}
\figsetgrptitle{The H-RRL emission toward SNR: G46.8-0.3.}
\figsetplot{G46.8-0.3_ha_spec.eps}
\figsetplot{G46.8-0.3_vgps_contour.eps}
\figsetgrpnote{
	(a) The SIGGMA spectral grid with the velocity range from $-300$ to $+300$\kms.
	The background is the 0$^{th}$ moment H-RRL map over $0$ to $+150$\kms.
	The white circle is the central region of the SNR.
	(b) The VGPS 1.4\ghz\ continuum map overlaid by integrated RRL contour.
	The levels are 0.1, 0.15, 0.2, 0.25, 0.3, 0.4 Jy\,beam$^{-1}$\kms.
	The blue circle is the location of a known \hii\ region from the WISE catalog.
	}
	\label{fig_snr-g468}
\figsetgrpend
\figsetgrpstart
\figsetgrpnum{2.11}
\figsetgrptitle{The H-RRL emission toward SNR: G49.2-0.7.}
\figsetplot{G49.2-0.7_ha_spec.eps}
\figsetplot{G49.2-0.7_vgps_contour.eps}
\figsetgrpnote{
	(a) The SIGGMA spectral grid with the velocity range from $-300$ to $+300$\kms.
	The background is the 0$^{th}$ moment H-RRL map over $0$ to $+150$\kms.
	The white circle is the central region of the SNR.
	(b) The VGPS 1.4\ghz\ continuum map overlaid by integrated RRL contour.
	The levels are 0.5, 1, 1.5, 2, 2.5, 3, 3.5, 4 Jy\,beam$^{-1}$\kms.
	The blue circles are location of known \hii\ regions from the WISE catalog.
	}
	\label{fig_snr-g492}
\figsetgrpend
\figsetgrpstart
\figsetgrpnum{2.12}
\figsetgrptitle{The H-RRL emission toward SNR: G54.4-0.3.}
\figsetplot{G54.4-0.3_ha_spec.eps}
\figsetplot{G54.4-0.3_vgps_contour.eps}
\figsetgrpnote{
	(a) The SIGGMA spectral grid with the velocity range from $-300$ to $+300$\kms.
	The background is the 0$^{th}$ moment H-RRL map over $0$ to $+150$\kms.
	The white circle is the central region of the SNR.
	(b) The VGPS 1.4\ghz\ continuum map overlaid by integrated RRL contour.
	The levels are 0.05, 0.15, 0.25, 0.35, 0.45 Jy\,beam$^{-1}$\kms.
	The blue circles are location of known \hii\ regions from the WISE catalog.
	}
	\label{fig_snr-g544}
\figsetgrpend
\figsetgrpstart
\figsetgrpnum{2.13}
\figsetgrptitle{The H-RRL emission toward SNR: G57.2+0.8.}
\figsetplot{G57.2+0.8_ha_spec.eps}
\figsetplot{G57.2+0.8_vgps_contour.eps}
\figsetgrpnote{
	(a) The SIGGMA spectral grid with the velocity range from $-300$ to $+300$\kms.
	The background is the 0$^{th}$ moment H-RRL map over $0$ to $+150$\kms.
	The white circle is the central region of the SNR.
	(b) The VGPS 1.4\ghz\ continuum map overlaid by integrated RRL contour.
	The levels are 0.03, 0.05, 0.07, 0.09 Jy\,beam$^{-1}$\kms.
	There are no \hii\ regions located in the figure covered region.
	}
	\label{fig_snr-g572}
\figsetgrpend
\figsetend

\newpage
\appendix
\section{Catalogues of detected \hii\ regions}\label{appe:cata}
\begin{ThreePartTable}
\begin{longtable}{lccccccc}
\caption{Catalog of 319 RRL detections towards 244 ``known'' HII regions from the WISE catalog\label{tab_HII_known}} \\
\hline
\hline
Name & $l$ & $b$ & Peak& V$_{LSR}$& FWHM& RMS& $S/N$\\
&$^\circ$&$^\circ$&mJy\,beam$^{-1}$&km\,s$^{-1}$&km\,s$^{-1}$&mJy\,beam$^{-1}$& \\
\hline
\endfirsthead
\caption{continued \ldots}\\
\hline
\hline
Name & $l$ & $b$ & Peak& V$_{LSR}$& FWHM& RMS& $S/N$\\
&$^\circ$&$^\circ$&mJy\,beam$^{-1}$&km\,s$^{-1}$&km\,s$^{-1}$&mJy\,beam$^{-1}$& \\
\hline
\endhead
\hline
\endfoot
\hline
\endlastfoot
G032.800$+$00.190     &32.800&	$+$0.190	&9.16	$\pm$ 1.01	&14.9	$\pm$ 1.2	&22.1	$\pm$ 2.8	&1.65	&8.9 \\ 
G032.823$+$00.072$^a$ &32.824&	$+$0.072	&30.40	$\pm$ 2.89	&106.7	$\pm$ 1.2	&25.3	$\pm$ 2.8	&2.76	&19.0\\
G032.835$+$00.017     &32.835&	$+$0.017	&191.52	$\pm$ 22.13	&109.8	$\pm$ 1.8	&33.3	$\pm$ 4.8	&28.51	&13.2\\
G032.982$-$00.338     &32.983&	$-$0.338	&17.08	$\pm$ 2.16	&50.8	$\pm$ 1.2	&20.6	$\pm$ 3.3	&2.56	&10.3\\
G033.051$-$00.078     &33.051&	$-$0.078	&26.21	$\pm$ 1.76	&103.9	$\pm$ 0.9	&28.0	$\pm$ 2.3	&0.77	&61.2\\
G033.080$+$00.073     &33.080&	$+$0.073	&11.65	$\pm$ 0.97	&102.2	$\pm$ 1.1	&27.8	$\pm$ 2.7	&1.22	&17.3\\
G033.142$-$00.088$^a$ &33.142&	$-$0.087	&20.93	$\pm$ 1.52	&103.2	$\pm$ 0.9	&26.4	$\pm$ 2.2	&1.19	&30.9\\
G033.176$-$00.016     &33.176&	$-$0.015	&27.82	$\pm$ 2.09	&105.1	$\pm$ 1.0	&26.2	$\pm$ 2.3	&1.63	&29.9\\
G033.205$-$00.013     &33.205&	$-$0.012	&14.64	$\pm$ 2.52	&106.5	$\pm$ 1.1	&13.1	$\pm$ 2.6	&2.46	&7.4\\
G033.263$+$00.066     &33.263&	$+$0.067	&11.51	$\pm$ 1.06	&111.3	$\pm$ 1.9	&42.2	$\pm$ 4.5	&1.40	&18.2\\
G033.419$-$00.005     &33.419&	$-$0.004	&14.86	$\pm$ 1.71	&100.7	$\pm$ 2.0	&35.6	$\pm$ 4.7	&2.45	&12.3\\
G033.809$-$00.186$^a$ &33.810&	$-$0.185	&25.78	$\pm$ 2.91	&50.9	$\pm$ 1.0	&18.7	$\pm$ 2.4	&3.67	&10.4\\
                      &33.810&	$-$0.185	&14.74	$\pm$ 2.42	&102.8	$\pm$ 2.2	&27.1	$\pm$ 5.1	&3.67	&7.1\\
G033.809$-$00.190$^a$ &33.809&	$-$0.190	&25.78	$\pm$ 2.91	&50.9	$\pm$ 1.0	&18.7	$\pm$ 2.4	&3.67	&10.4\\
                      &33.809&	$-$0.190	&14.74	$\pm$ 2.42	&102.8	$\pm$ 2.2	&27.1	$\pm$ 5.1	&3.67	&7.1\\
G033.882$+$00.057     &33.882&	$+$0.057	&4.98	$\pm$ 0.41	&59.3	$\pm$ 3.2	&77.2	$\pm$ 7.4	&1.23	&12.1\\
G033.914$+$00.107     &33.915&	$+$0.108	&7.72	$\pm$ 0.67	&103.8	$\pm$ 2.5	&57.3	$\pm$ 5.8	&1.09	&18.4\\
G033.941$-$00.039     &33.942&	$-$0.039	&21.87	$\pm$ 1.34	&62.1	$\pm$ 0.6	&19.3	$\pm$ 1.4	&2.04	&16.1\\
G033.987$-$00.012$^a$ &33.988&	$-$0.011	&14.47	$\pm$ 0.50	&56.4	$\pm$ 0.7	&37.9	$\pm$ 1.5	&1.07	&28.4\\
G033.991$-$00.005$^a$ &33.991&	$-$0.004	&14.47	$\pm$ 0.50	&56.4	$\pm$ 0.7	&37.9	$\pm$ 1.5	&1.07	&28.4\\
G034.026$-$00.058     &34.026&	$-$0.058	&22.76	$\pm$ 2.53	&59.1	$\pm$ 1.0	&15.0	$\pm$ 1.5	&0.90	&33.3\\
                      &34.026&	$-$0.058	&11.51	$\pm$ 1.43	&42.5	$\pm$ 2.7	&19.5	$\pm$ 4.5	&0.90	&19.2\\
                      &34.026&	$-$0.058	&5.02	$\pm$ 0.40	&96.5	$\pm$ 5.9	&83.1	$\pm$ 12.2	&0.90	&17.3\\
G034.041$+$00.052$^a$ &34.041&	$+$0.053	&7.45	$\pm$ 0.43	&53.1	$\pm$ 1.9	&36.4	$\pm$ 4.3	&0.86	&17.8\\
G034.045$+$00.053$^a$ &34.046&	$+$0.053	&7.45	$\pm$ 0.43	&53.1	$\pm$ 1.9	&36.4	$\pm$ 4.3	&0.86	&17.8\\
G034.104$-$00.046     &34.105&	$-$0.046	&32.30	$\pm$ 2.54	&49.9	$\pm$ 0.9	&36.5	$\pm$ 3.0	&1.97	&33.9\\
G034.158$+$00.147     &34.158&	$+$0.148	&15.69	$\pm$ 1.87	&61.7	$\pm$ 0.6	&17.2	$\pm$ 1.9	&0.94	&23.6\\
                      &34.158&	$+$0.148	&11.11	$\pm$ 1.90	&49.7	$\pm$ 2.7	&43.6	$\pm$ 5.1	&0.94	&26.6\\
G034.256$+$00.136     &34.256&	$+$0.136	&151.57	$\pm$ 12.16	&54.9	$\pm$ 1.0	&36.9	$\pm$ 2.9	&6.66	&47.2\\
                      &34.256&	$+$0.136	&44.35	$\pm$ 4.21	&114.9	$\pm$ 8.6	&86.4	$\pm$ 17.9	&6.66	&21.2\\
G034.333$+$00.212     &34.333&	$+$0.213	&45.52	$\pm$ 2.47	&66.8	$\pm$ 0.5	&19.0	$\pm$ 1.3	&2.15	&31.5\\
G034.404$+$00.227     &34.404&	$+$0.228	&36.35	$\pm$ 1.30	&64.1	$\pm$ 0.3	&17.4	$\pm$ 0.7	&1.88	&27.6\\
G034.542$+$00.062     &34.542&	$+$0.062	&12.71	$\pm$ 5.74	&54.5	$\pm$ 5.0	&25.9	$\pm$ 6.1	&1.93	&11.5\\
G034.550$-$01.110     &34.550&	$-$1.110	&15.13	$\pm$ 0.89	&49.0	$\pm$ 0.5	&15.4	$\pm$ 1.1	&1.21	&16.7\\
G034.591$+$00.244     &34.591&	$+$0.244	&21.31	$\pm$ 1.81	&61.8	$\pm$ 0.7	&18.9	$\pm$ 2.0	&1.67	&18.9\\
                      &34.591&	$+$0.244	&6.95	$\pm$ 0.81	&114.3	$\pm$ 5.6	&73.9	$\pm$ 14.0	&1.67	&12.2\\
G034.686$+$00.068     &34.686&	$+$0.068	&15.52	$\pm$ 0.85	&58.6	$\pm$ 0.9	&34.6	$\pm$ 2.2	&1.73	&18.1\\
G034.756$+$00.022     &34.757&	$+$0.022	&22.29	$\pm$ 2.08	&53.2	$\pm$ 1.0	&21.8	$\pm$ 2.4	&3.34	&10.6\\
                      &34.757&	$+$0.022	&12.85	$\pm$ 2.52	&89.1	$\pm$ 1.4	&14.7	$\pm$ 3.4	&3.34	&5.0\\
G034.757$-$00.669     &34.757&	$-$0.668	&25.46	$\pm$ 2.24	&48.7	$\pm$ 0.4	&12.4	$\pm$ 1.0	&1.40	&21.9\\
                      &34.757&	$-$0.668	&10.30	$\pm$ 3.26	&27.6	$\pm$ 2.3	&28.5	$\pm$ 9.5	&1.40	&13.4\\
G034.916$-$00.016     &34.916&	$-$0.016	&25.62	$\pm$ 1.28	&48.8	$\pm$ 0.6	&31.3	$\pm$ 1.9	&1.60	&30.7\\
G034.924$-$00.082     &34.925&	$-$0.082	&35.20	$\pm$ 2.89	&49.2	$\pm$ 0.6	&19.0	$\pm$ 1.9	&1.85	&28.4\\
G034.940$+$00.073     &34.940&	$+$0.074	&21.63	$\pm$ 1.26	&52.1	$\pm$ 0.6	&25.2	$\pm$ 1.8	&1.61	&23.0\\
G034.997$+$00.330     &34.997&	$+$0.331	&12.34	$\pm$ 0.55	&59.3	$\pm$ 0.6	&26.5	$\pm$ 1.4	&0.98	&22.1\\
G035.040$-$00.500     &35.040&	$-$0.500	&32.33	$\pm$ 1.43	&48.8	$\pm$ 0.3	&15.7	$\pm$ 0.8	&1.97	&22.2\\
G035.051$-$00.520     &35.051&	$-$0.520	&12.06	$\pm$ 0.98	&46.5	$\pm$ 0.5	&13.7	$\pm$ 1.3	&1.25	&12.2\\
G035.063$+$00.330     &35.064&	$+$0.330	&10.35	$\pm$ 0.63	&56.4	$\pm$ 0.9	&18.2	$\pm$ 1.9	&0.91	&16.5\\
                      &35.064&	$+$0.330	&3.29	$\pm$ 0.61	&80.3	$\pm$ 2.8	&19.4	$\pm$ 6.5	&0.91	&5.4\\
G035.099$-$00.243     &35.099&	$-$0.243	&4.63	$\pm$ 0.39	&57.4	$\pm$ 4.0	&96.3	$\pm$ 9.5	&1.25	&12.4\\
G035.126$-$00.755     &35.126&	$-$0.754	&3.72	$\pm$ 0.54	&49.1	$\pm$ 1.9	&27.3	$\pm$ 4.6	&0.97	&6.8\\
G035.187$+$00.892     &35.187&	$+$0.893	&4.90	$\pm$ 0.24	&88.4	$\pm$ 0.5	&18.4	$\pm$ 1.1	&0.34	&20.9\\
G035.258$+$00.118     &35.259&	$+$0.119	&4.63	$\pm$ 0.33	&65.2	$\pm$ 2.7	&77.3	$\pm$ 6.4	&0.89	&15.6\\
G035.429$-$00.260     &35.430&	$-$0.260	&6.19	$\pm$ 0.60	&52.7	$\pm$ 1.5	&30.5	$\pm$ 3.4	&1.15	&10.1\\
G035.467$+$00.004     &35.467&	$+$0.004	&27.78	$\pm$ 0.74	&58.5	$\pm$ 0.4	&29.0	$\pm$ 0.9	&1.38	&37.0\\
G035.467$+$00.139     &35.468&	$+$0.139	&8.61	$\pm$ 0.56	&54.8	$\pm$ 1.2	&37.7	$\pm$ 2.9	&1.20	&15.1\\
G035.543$+$00.006     &35.543&	$+$0.007	&55.24	$\pm$ 0.98	&55.1	$\pm$ 0.2	&27.2	$\pm$ 0.6	&1.77	&55.6\\
G035.559$-$00.824     &35.560&	$-$0.823	&10.10	$\pm$ 0.58	&60.8	$\pm$ 1.1	&19.3	$\pm$ 1.7	&0.43	&35.5\\
                      &35.560&	$-$0.823	&3.02	$\pm$ 0.65	&42.1	$\pm$ 3.5	&18.5	$\pm$ 5.1	&0.43	&10.4\\
G035.571$+$00.071$^a$ &35.571&	$+$0.072	&29.19	$\pm$ 0.85	&55.5	$\pm$ 0.9	&41.9	$\pm$ 2.0	&1.82	&35.5\\
                      &35.571&	$+$0.072	&8.26	$\pm$ 0.74	&-4.8	$\pm$ 3.7	&53.5	$\pm$ 8.8	&1.82	&11.4\\
G035.576$-$00.032     &35.577&	$-$0.031	&25.40	$\pm$ 1.02	&56.2	$\pm$ 0.5	&25.1	$\pm$ 1.2	&1.77	&24.5\\
G035.579$+$00.065$^a$ &35.579&	$+$0.066	&29.19	$\pm$ 0.85	&55.5	$\pm$ 0.9	&41.9	$\pm$ 2.0	&1.82	&35.5\\
                      &35.579&	$+$0.066	&8.26	$\pm$ 0.74	&-4.8	$\pm$ 3.7	&53.5	$\pm$ 8.8	&1.82	&11.4\\
G035.649$-$00.053     &35.650&	$-$0.052	&44.18	$\pm$ 0.93	&52.2	$\pm$ 0.7	&22.9	$\pm$ 1.1	&0.95	&76.3\\
G036.192$-$00.171     &36.193&	$-$0.171	&4.71	$\pm$ 0.51	&81.7	$\pm$ 3.2	&98.0	$\pm$ 8.8	&0.65	&24.4\\
                      &36.193&	$-$0.171	&4.06	$\pm$ 0.82	&87.8	$\pm$ 1.5	&17.0	$\pm$ 4.3	&0.65	&8.7\\
G036.344$+$00.040     &36.344&	$+$0.040	&7.30	$\pm$ 0.35	&76.3	$\pm$ 2.1	&57.6	$\pm$ 4.7	&0.81	&23.4\\
                      &36.344&	$+$0.040	&2.30	$\pm$ 0.33	&-2.0	$\pm$ 6.9	&64.1	$\pm$ 16.2	&0.81	&7.8\\
G036.405$+$00.021     &36.405&	$+$0.022	&7.39	$\pm$ 0.42	&65.2	$\pm$ 1.3	&46.0	$\pm$ 3.0	&0.99	&17.3\\
G036.459$-$00.183     &36.459&	$-$0.183	&12.96	$\pm$ 0.80	&75.1	$\pm$ 0.8	&25.8	$\pm$ 1.8	&0.64	&35.4\\
                      &36.459&	$-$0.183	&1.91	$\pm$ 0.79	&114.4	$\pm$ 7.0	&40.9	$\pm$ 18.1	&0.64	&6.6\\
G036.870$+$00.462     &36.870&	$+$0.462	&7.51	$\pm$ 0.65	&-26.8	$\pm$ 1.1	&26.9	$\pm$ 2.7	&1.16	&11.4\\
G036.993$-$00.231     &36.993&	$-$0.230	&32.79	$\pm$ 1.00	&86.7	$\pm$ 0.3	&17.9	$\pm$ 0.6	&1.44	&32.9\\
G037.028$-$00.202     &37.029&	$-$0.202	&23.63	$\pm$ 0.39	&85.4	$\pm$ 0.1	&17.4	$\pm$ 0.3	&0.56	&59.8\\
                      &37.029&	$-$0.202	&2.94	$\pm$ 0.34	&39.1	$\pm$ 1.3	&23.3	$\pm$ 3.1	&0.56	&8.6\\
 $^c$                     &37.029&	$-$0.202	&2.48	$\pm$ 0.38	&-40.0	$\pm$ 1.4	&18.5	$\pm$ 3.3	&0.56	&6.5\\
G037.032$+$00.139     &37.032&	$+$0.139	&5.82	$\pm$ 1.04	&43.2	$\pm$ 1.6	&20.0	$\pm$ 4.7	&0.75	&11.9\\
                      &37.032&	$+$0.139	&3.06	$\pm$ 1.18	&84.1	$\pm$ 2.5	&15.3	$\pm$ 7.7	&0.75	&5.5\\
G037.175$-$00.440     &37.176&	$-$0.439	&10.87	$\pm$ 1.31	&38.7	$\pm$ 0.9	&16.6	$\pm$ 2.5	&1.48	&10.2\\
G037.200$-$00.430     &37.200&	$-$0.430	&11.96	$\pm$ 0.77	&41.3	$\pm$ 1.2	&39.5	$\pm$ 2.9	&1.67	&15.4\\
G037.259$-$00.083     &37.259&	$-$0.083	&15.48	$\pm$ 0.48	&55.4	$\pm$ 1.1	&71.0	$\pm$ 2.9	&1.29	&34.6\\
G037.278$-$00.226     &37.278&	$-$0.226	&29.21	$\pm$ 1.17	&41.7	$\pm$ 0.7	&34.1	$\pm$ 1.6	&2.35	&24.8\\
G037.347$-$00.147     &37.347&	$-$0.146	&16.25	$\pm$ 0.38	&53.5	$\pm$ 0.6	&47.8	$\pm$ 2.0	&0.60	&63.9\\
                      &37.347&	$-$0.146	&3.71	$\pm$ 0.64	&85.7	$\pm$ 0.9	&14.3	$\pm$ 2.9	&0.60	&8.0\\
G037.362$-$00.234     &37.363&	$-$0.234	&27.02	$\pm$ 0.61	&41.9	$\pm$ 0.4	&30.1	$\pm$ 0.9	&0.99	&51.3\\
                      &37.363&	$-$0.234	&8.62	$\pm$ 0.71	&85.0	$\pm$ 0.7	&16.4	$\pm$ 1.6	&0.99	&12.1\\
G037.370$-$00.368     &37.370&	$-$0.367	&6.66	$\pm$ 0.15	&36.8	$\pm$ 0.4	&32.2	$\pm$ 1.1	&0.30	&43.6\\
                      &37.370&	$-$0.367	&3.08	$\pm$ 0.15	&81.2	$\pm$ 0.9	&30.5	$\pm$ 2.3	&0.30	&19.6\\
G037.445$-$00.212     &37.446&	$-$0.211	&7.53	$\pm$ 1.23	&73.7	$\pm$ 1.5	&29.2	$\pm$ 4.6	&1.03	&13.5\\
G037.469$-$00.105     &37.469&	$-$0.104	&32.60	$\pm$ 0.91	&63.5	$\pm$ 0.5	&29.9	$\pm$ 1.7	&1.57	&38.7\\
                      &37.469&	$-$0.104	&5.54	$\pm$ 0.86	&21.1	$\pm$ 2.8	&28.6	$\pm$ 6.9	&1.57	&6.4\\
G037.544$-$00.115     &37.544&	$-$0.114	&34.66	$\pm$ 1.50	&54.8	$\pm$ 0.5	&23.8	$\pm$ 1.2	&2.54	&22.8\\
G037.641$-$00.112     &37.642&	$-$0.112	&39.82	$\pm$ 0.79	&55.2	$\pm$ 0.3	&25.6	$\pm$ 0.6	&1.37	&50.4\\
                      &37.642&	$-$0.112	&12.12	$\pm$ 1.08	&89.4	$\pm$ 0.6	&13.4	$\pm$ 1.4	&1.37	&11.1\\
G037.677$+$00.155     &37.678&	$+$0.156	&32.02	$\pm$ 1.13	&92.7	$\pm$ 0.5	&23.5	$\pm$ 1.1	&1.13	&47.1\\
                      &37.678&	$+$0.156	&9.60	$\pm$ 1.05	&53.0	$\pm$ 1.7	&28.3	$\pm$ 4.2	&1.13	&15.5\\
G037.754$+$00.560     &37.754&	$+$0.560	&6.49	$\pm$ 0.56	&96.8	$\pm$ 1.8	&41.5	$\pm$ 4.2	&0.91	&15.6\\
                      &37.754&	$+$0.560	&6.03	$\pm$ 0.73	&31.0	$\pm$ 1.5	&24.3	$\pm$ 3.5	&0.91	&11.1\\
G037.763$-$00.212     &37.763&	$-$0.211	&33.01	$\pm$ 16.47	&62.9	$\pm$ 1.1	&20.2	$\pm$ 3.9	&1.89	&26.8\\
                      &37.763&	$-$0.211	&26.37	$\pm$ 7.67	&79.7	$\pm$ 8.0	&33.9	$\pm$ 8.5	&1.89	&27.8\\
G037.868$-$00.601     &37.868&	$-$0.601	&17.68	$\pm$ 2.86	&69.0	$\pm$ 1.4	&19.0	$\pm$ 3.8	&2.56	&10.3\\
G037.872$-$00.399     &37.872&	$-$0.398	&96.36	$\pm$ 107.25	&69.6	$\pm$ 3.7	&24.8	$\pm$ 4.9	&4.08	&40.2\\
                      &37.872&	$-$0.398	&10.63	$\pm$ 67.32	&54.0	$\pm$ 113.5	&32.5	$\pm$ 91.0	&4.08	&5.1\\
G037.903$-$00.276     &37.904&	$-$0.275	&8.66	$\pm$ 1.19	&65.0	$\pm$ 0.9	&17.1	$\pm$ 2.9	&0.77	&15.9\\
G038.045$-$00.034     &38.045&	$-$0.034	&19.08	$\pm$ 0.87	&63.3	$\pm$ 0.6	&25.1	$\pm$ 1.3	&1.52	&21.5\\
G038.121$-$00.226     &38.121&	$-$0.226	&12.10	$\pm$ 1.05	&57.2	$\pm$ 0.6	&13.6	$\pm$ 1.6	&1.33	&11.5\\
                      &38.121&	$-$0.226	&6.93	$\pm$ 0.94	&81.3	$\pm$ 1.2	&17.2	$\pm$ 3.1	&1.33	&7.4\\
G038.353$-$00.135     &38.353&	$-$0.134	&16.35	$\pm$ 3.93	&69.7	$\pm$ 4.1	&21.5	$\pm$ 20.9	&1.05	&24.8\\
                      &38.353&	$-$0.134	&15.35	$\pm$ 11.81	&87.4	$\pm$ 5.4	&17.8	$\pm$ 4.7	&1.05	&21.1\\
G038.365$-$00.062     &38.365&	$-$0.062	&12.06	$\pm$ 0.64	&61.0	$\pm$ 0.9	&20.9	$\pm$ 1.4	&0.47	&40.1\\
                      &38.365&	$-$0.062	&12.02	$\pm$ 0.41	&84.7	$\pm$ 1.0	&25.1	$\pm$ 1.8	&0.47	&43.8\\
G038.643$-$00.227     &38.643&	$-$0.227	&5.21	$\pm$ 0.93	&77.8	$\pm$ 1.7	&22.4	$\pm$ 5.0	&1.09	&7.7\\
G038.861$-$00.135     &38.861&	$-$0.134	&23.57	$\pm$ 1.96	&68.7	$\pm$ 0.9	&21.1	$\pm$ 2.0	&3.11	&11.9\\
G038.978$-$00.269     &38.978&	$-$0.268	&3.89	$\pm$ 0.63	&62.8	$\pm$ 1.5	&19.2	$\pm$ 3.7	&0.96	&6.1\\
                      &38.978&	$-$0.268	&3.32	$\pm$ 0.52	&15.4	$\pm$ 2.2	&28.4	$\pm$ 5.3	&0.96	&6.3\\
G039.170$-$00.037     &39.171&	$-$0.037	&11.53	$\pm$ 0.55	&21.2	$\pm$ 0.7	&26.2	$\pm$ 1.7	&0.97	&20.7\\
                      &39.171&	$-$0.037	&3.79	$\pm$ 0.49	&67.4	$\pm$ 2.5	&35.4	$\pm$ 6.4	&0.97	&7.9\\
G039.176$-$00.399     &39.176&	$-$0.398	&7.70	$\pm$ 0.66	&58.7	$\pm$ 0.9	&21.4	$\pm$ 2.1	&1.06	&11.5\\
G039.225$-$00.053$^a$ &39.225&	$-$0.053	&32.31	$\pm$ 1.27	&21.8	$\pm$ 0.5	&25.7	$\pm$ 1.2	&2.21	&25.3\\
                      &39.225&	$-$0.053	&8.48	$\pm$ 1.52	&62.5	$\pm$ 1.6	&17.6	$\pm$ 3.8	&2.21	&5.5\\
G039.248$-$00.064$^a$ &39.249&	$-$0.063	&32.31	$\pm$ 1.27	&21.8	$\pm$ 0.5	&25.7	$\pm$ 1.2	&2.21	&25.3\\
                      &39.249&	$-$0.063	&8.48	$\pm$ 1.52	&62.5	$\pm$ 1.6	&17.6	$\pm$ 3.8	&2.21	&5.5\\
G039.388$-$00.143     &39.389&	$-$0.143	&89.22	$\pm$ 24.38	&10.8	$\pm$ 4.0	&38.6	$\pm$ 12.0	&30.32	&6.3\\
G039.544$-$00.366     &39.545&	$-$0.366	&22.99	$\pm$ 2.63	&66.1	$\pm$ 1.2	&21.5	$\pm$ 2.9	&4.21	&8.7\\
                      &39.545&	$-$0.366	&12.59	$\pm$ 2.50	&21.6	$\pm$ 2.3	&23.9	$\pm$ 5.6	&4.21	&5.0\\
G039.607$-$00.021     &39.607&	$-$0.021	&9.17	$\pm$ 0.91	&16.9	$\pm$ 1.3	&25.7	$\pm$ 3.1	&1.59	&10.0\\
                      &39.607&	$-$0.021	&7.11	$\pm$ 0.86	&67.4	$\pm$ 1.7	&29.2	$\pm$ 4.2	&1.59	&8.3\\
G039.630$-$00.107     &39.630&	$-$0.106	&5.78	$\pm$ 0.41	&23.8	$\pm$ 0.9	&24.2	$\pm$ 2.2	&0.68	&14.3\\
                      &39.630&	$-$0.106	&2.52	$\pm$ 0.43	&61.7	$\pm$ 1.9	&21.5	$\pm$ 4.7	&0.68	&5.9\\
G039.864$+$00.645     &39.865&	$+$0.646	&8.42	$\pm$ 1.42	&45.1	$\pm$ 1.6	&19.5	$\pm$ 3.8	&2.17	&5.9\\
G039.873$-$00.177     &39.874&	$-$0.176	&13.56	$\pm$ 1.44	&66.1	$\pm$ 0.9	&16.8	$\pm$ 2.1	&2.04	&9.3\\
G039.883$-$00.346     &39.883&	$-$0.346	&7.77	$\pm$ 1.20	&76.9	$\pm$ 1.9	&25.6	$\pm$ 4.6	&2.10	&6.4\\
G039.900$-$01.321     &39.901&	$-$1.321	&10.41	$\pm$ 0.65	&48.0	$\pm$ 0.7	&23.4	$\pm$ 1.7	&1.09	&15.8\\
G039.924$-$00.378     &39.925&	$-$0.378	&3.03	$\pm$ 0.56	&74.7	$\pm$ 1.3	&14.2	$\pm$ 3.0	&0.73	&5.3\\
G040.430$+$00.697     &40.430&	$+$0.698	&3.23	$\pm$ 0.58	&5.7	$\pm$ 1.1	&12.9	$\pm$ 2.7	&0.72	&5.5\\
G040.797$-$00.132     &40.798&	$-$0.131	&6.95	$\pm$ 1.08	&68.2	$\pm$ 3.9	&20.3	$\pm$ 7.7	&1.53	&7.0\\
G040.855$-$00.224     &40.855&	$-$0.224	&4.61	$\pm$ 0.45	&61.3	$\pm$ 1.3	&26.1	$\pm$ 2.9	&0.80	&10.1\\
G040.965$-$00.621     &40.966&	$-$0.620	&13.72	$\pm$ 0.54	&65.1	$\pm$ 0.8	&39.7	$\pm$ 1.8	&1.17	&25.2\\
G041.074$-$00.162     &41.075&	$-$0.161	&28.71	$\pm$ 0.59	&61.7	$\pm$ 0.2	&23.3	$\pm$ 0.6	&0.99	&47.7\\
G041.129$+$00.112     &41.129&	$+$0.113	&3.19	$\pm$ 0.30	&65.3	$\pm$ 2.8	&60.7	$\pm$ 6.6	&0.80	&10.7\\
G041.235$+$00.367     &41.235&	$+$0.367	&12.90	$\pm$ 0.46	&74.1	$\pm$ 0.4	&23.0	$\pm$ 1.1	&0.75	&28.3\\
                      &41.235&	$+$0.367	&6.11	$\pm$ 0.52	&39.4	$\pm$ 1.0	&19.0	$\pm$ 2.8	&0.75	&12.2\\
                      &41.235&	$+$0.367	&4.59	$\pm$ 0.63	&18.1	$\pm$ 1.1	&12.3	$\pm$ 2.4	&0.75	&7.4\\
G041.246$-$00.168     &41.246&	$-$0.168	&14.23	$\pm$ 1.19	&59.1	$\pm$ 0.6	&14.8	$\pm$ 1.4	&1.59	&11.8\\
G041.375$+$00.032$^a$ &41.376&	$+$0.032	&25.62	$\pm$ 2.56	&62.3	$\pm$ 1.4	&27.4	$\pm$ 3.4	&4.57	&10.0\\
G041.382$+$00.037$^a$ &41.382&	$+$0.038	&25.62	$\pm$ 2.56	&62.3	$\pm$ 1.4	&27.4	$\pm$ 3.4	&4.57	&10.0\\
G041.512$+$00.021     &41.512&	$+$0.022	&27.55	$\pm$ 0.99	&15.9	$\pm$ 0.4	&20.1	$\pm$ 0.8	&1.53	&27.5\\
G041.516$-$00.142     &41.517&	$-$0.141	&13.98	$\pm$ 0.94	&60.6	$\pm$ 1.0	&26.1	$\pm$ 2.5	&1.65	&14.8\\
G041.659$-$00.020     &41.659&	$-$0.019	&11.33	$\pm$ 0.98	&49.9	$\pm$ 0.8	&18.0	$\pm$ 2.0	&1.42	&11.6\\
                      &41.659&	$-$0.019	&5.67	$\pm$ 1.00	&21.5	$\pm$ 1.6	&17.2	$\pm$ 3.9	&1.42	&5.7\\
G041.725$-$00.004     &41.725&	$-$0.004	&5.08	$\pm$ 0.60	&49.9	$\pm$ 1.0	&16.6	$\pm$ 2.3	&0.84	&8.4\\
G041.750$+$00.034     &41.750&	$+$0.035	&17.31	$\pm$ 1.22	&49.6	$\pm$ 0.7	&17.5	$\pm$ 1.7	&1.72	&14.4\\
                      &41.750&	$+$0.035	&9.94	$\pm$ 1.30	&25.0	$\pm$ 1.1	&14.9	$\pm$ 2.7	&1.72	&7.6\\
G041.762$+$01.479$^a$ &41.763&	$+$1.480	&5.97	$\pm$ 0.97	&118.8	$\pm$ 3.8	&47.3	$\pm$ 8.9	&1.89	&7.4\\
G041.880$+$00.492     &41.881&	$+$0.493	&12.27	$\pm$ 1.26	&23.9	$\pm$ 0.8	&16.0	$\pm$ 1.9	&1.74	&9.6\\
G041.927$-$00.125     &41.928&	$-$0.125	&7.63	$\pm$ 0.73	&18.8	$\pm$ 1.5	&39.9	$\pm$ 4.5	&1.04	&15.8\\
G041.929$+$00.030     &41.929&	$+$0.030	&9.63	$\pm$ 1.21	&33.1	$\pm$ 2.2	&35.6	$\pm$ 5.2	&2.50	&7.9\\
G042.065$+$00.244     &42.065&	$+$0.244	&18.94	$\pm$ 2.31	&21.8	$\pm$ 0.9	&14.8	$\pm$ 2.2	&3.07	&8.1\\
                      &42.065&	$+$0.244	&18.52	$\pm$ 1.82	&57.8	$\pm$ 1.2	&24.3	$\pm$ 2.9	&3.07	&10.2\\
G042.103$-$00.623     &42.104&	$-$0.623	&31.72	$\pm$ 0.88	&68.1	$\pm$ 0.3	&23.9	$\pm$ 0.8	&1.49	&35.7\\
G042.111$-$00.444     &42.111&	$-$0.444	&8.96	$\pm$ 1.39	&66.5	$\pm$ 1.2	&15.3	$\pm$ 2.7	&1.88	&6.4\\
G042.136$-$00.079     &42.137&	$-$0.079	&14.83	$\pm$ 1.57	&58.1	$\pm$ 0.9	&17.1	$\pm$ 2.1	&2.25	&9.3\\
G042.204$+$00.038     &42.204&	$+$0.038	&48.94	$\pm$ 3.67	&61.3	$\pm$ 0.7	&18.1	$\pm$ 1.6	&5.41	&13.2\\
G042.217$-$00.580     &42.217&	$-$0.579	&4.20	$\pm$ 0.42	&67.2	$\pm$ 1.5	&29.8	$\pm$ 3.5	&0.80	&9.8\\
G042.390$-$00.381     &42.390&	$-$0.381	&29.61	$\pm$ 1.55	&68.4	$\pm$ 1.1	&19.2	$\pm$ 2.0	&1.83	&24.3\\
                      &42.390&	$-$0.381	&11.88	$\pm$ 1.91	&48.7	$\pm$ 2.6	&17.5	$\pm$ 4.4	&1.83	&9.3\\
G042.434$-$00.275     &42.434&	$-$0.275	&14.52	$\pm$ 1.08	&67.7	$\pm$ 0.9	&24.4	$\pm$ 2.1	&1.85	&13.2\\
                      &42.434&	$-$0.275	&10.84	$\pm$ 1.50	&22.8	$\pm$ 0.9	&12.8	$\pm$ 2.0	&1.85	&7.2\\
G042.562$-$00.107     &42.563&	$-$0.107	&21.82	$\pm$ 0.63	&71.5	$\pm$ 0.3	&22.8	$\pm$ 0.8	&1.04	&34.2\\
G043.100$-$00.503     &43.100&	$-$0.502	&7.93	$\pm$ 0.83	&64.1	$\pm$ 1.7	&32.7	$\pm$ 4.0	&1.64	&9.5\\
G043.149$+$00.028     &43.149&	$+$0.029	&74.02	$\pm$ 15.23	&8.4	$\pm$ 1.3	&34.7	$\pm$ 3.9	&5.59	&26.6\\
                      &43.149&	$+$0.029	&32.11	$\pm$ 5.00	&46.2	$\pm$ 12.5	&65.0	$\pm$ 17.8	&5.59	&15.8\\
G043.154$-$00.039$^a$ &43.154&	$-$0.039	&425.47	$\pm$ 29.61	&10.1	$\pm$ 0.9	&32.8	$\pm$ 2.7	&30.44	&27.4\\
G043.165$-$00.031$^a$ &43.166&	$-$0.030	&425.47	$\pm$ 29.61	&10.1	$\pm$ 0.9	&32.8	$\pm$ 2.7	&30.44	&27.4\\
G043.170$-$00.004     &43.171&	$-$0.003	&244.92	$\pm$ 12.26	&10.0	$\pm$ 0.6	&31.6	$\pm$ 1.9	&14.39	&32.7\\
G043.237$-$00.044     &43.238&	$-$0.044	&134.52	$\pm$ 3.33	&9.7	$\pm$ 0.6	&31.3	$\pm$ 1.7	&6.20	&41.5\\
                      &43.238&	$-$0.044	&34.76	$\pm$ 3.29	&59.2	$\pm$ 1.6	&30.6	$\pm$ 4.0	&6.20	&10.6\\
                      &43.238&	$-$0.044	&23.32	$\pm$ 3.43	&-28.6	$\pm$ 3.5	&28.9	$\pm$ 7.6	&6.20	&6.9\\
G043.240$+$00.131     &43.240&	$+$0.131	&11.00	$\pm$ 1.34	&9.4	$\pm$ 1.0	&18.1	$\pm$ 2.8	&1.77	&9.1\\
G043.432$+$00.516     &43.432&	$+$0.517	&6.93	$\pm$ 0.41	&-9.6	$\pm$ 1.0	&35.3	$\pm$ 2.4	&0.84	&16.7\\
G043.730$+$00.114     &43.730&	$+$0.115	&9.40	$\pm$ 0.61	&75.9	$\pm$ 1.2	&36.6	$\pm$ 2.7	&1.21	&16.1\\
G043.774$+$00.057     &43.774&	$+$0.058	&8.50	$\pm$ 0.46	&75.9	$\pm$ 1.7	&64.9	$\pm$ 4.0	&1.03	&22.8\\
G043.792$-$00.122     &43.793&	$-$0.121	&5.07	$\pm$ 0.43	&75.4	$\pm$ 3.4	&80.5	$\pm$ 7.9	&1.21	&12.9\\
G043.794$-$00.129     &43.794&	$-$0.129	&3.31	$\pm$ 0.40	&62.1	$\pm$ 5.2	&87.5	$\pm$ 12.2	&1.26	&8.4\\
G043.818$+$00.395     &43.818&	$+$0.395	&6.86	$\pm$ 0.62	&-11.9	$\pm$ 1.3	&28.0	$\pm$ 3.1	&1.14	&10.9\\
                      &43.818&	$+$0.395	&3.42	$\pm$ 0.49	&53.4	$\pm$ 3.2	&45.1	$\pm$ 7.9	&1.14	&6.9\\
G043.894$+$00.197     &43.895&	$+$0.198	&5.74	$\pm$ 1.15	&66.1	$\pm$ 1.1	&12.1	$\pm$ 3.0	&1.09	&6.3\\
G043.968$+$00.993     &43.968&	$+$0.993	&4.54	$\pm$ 0.51	&94.9	$\pm$ 2.7	&50.1	$\pm$ 6.5	&1.05	&10.4\\
G043.999$+$00.978     &44.000&	$+$0.979	&5.49	$\pm$ 0.61	&103.0	$\pm$ 3.7	&37.9	$\pm$ 8.5	&0.90	&12.8\\
                      &44.000&	$+$0.979	&4.17	$\pm$ 1.38	&75.3	$\pm$ 2.0	&14.7	$\pm$ 5.4	&0.90	&6.0\\
G044.094$-$00.015     &44.095&	$-$0.014	&7.04	$\pm$ 0.47	&70.5	$\pm$ 0.8	&24.3	$\pm$ 1.9	&0.81	&14.7\\
G044.224$+$00.085     &44.225&	$+$0.085	&15.53	$\pm$ 0.74	&67.4	$\pm$ 0.4	&24.5	$\pm$ 1.3	&0.64	&40.8\\
                      &44.225&	$+$0.085	&1.97	$\pm$ 0.57	&115.8	$\pm$ 3.7	&33.3	$\pm$ 10.0	&0.64	&6.0\\
G044.331$-$00.837     &44.332&	$-$0.837	&4.81	$\pm$ 0.36	&74.4	$\pm$ 3.4	&91.9	$\pm$ 8.0	&0.93	&16.9\\
G044.379$-$00.327     &44.379&	$-$0.327	&6.59	$\pm$ 0.57	&65.6	$\pm$ 0.6	&18.2	$\pm$ 1.9	&0.61	&15.7\\
G044.501$+$00.332     &44.501&	$+$0.332	&13.70	$\pm$ 1.11	&-46.1	$\pm$ 0.9	&22.4	$\pm$ 2.1	&1.81	&12.2\\
G044.552$-$00.239     &44.553&	$-$0.239	&2.79	$\pm$ 0.25	&84.7	$\pm$ 2.6	&55.3	$\pm$ 6.7	&0.49	&14.5\\
G044.811$-$00.492     &44.812&	$-$0.492	&4.84	$\pm$ 1.46	&50.5	$\pm$ 3.2	&33.5	$\pm$ 7.2	&0.74	&13.0\\
                      &44.812&	$-$0.492	&3.66	$\pm$ 0.46	&93.0	$\pm$ 10.8	&61.1	$\pm$ 19.0	&0.74	&13.3\\
G044.904$-$00.733     &44.904&	$-$0.732	&4.19	$\pm$ 0.74	&67.0	$\pm$ 1.6	&20.2	$\pm$ 4.5	&0.89	&7.2\\
G045.002$-$00.611     &45.002&	$-$0.610	&6.51	$\pm$ 0.74	&66.8	$\pm$ 0.9	&19.1	$\pm$ 2.6	&0.56	&17.4\\
                      &45.002&	$-$0.610	&4.24	$\pm$ 0.48	&87.7	$\pm$ 3.9	&68.1	$\pm$ 6.0	&0.56	&21.4\\
G045.067$+$00.140$^a$ &45.067&	$+$0.141	&9.99	$\pm$ 1.27	&64.3	$\pm$ 1.1	&20.0	$\pm$ 3.2	&1.52	&10.0\\
G045.070$+$00.132$^a$ &45.070&	$+$0.132	&9.99	$\pm$ 1.27	&64.3	$\pm$ 1.1	&20.0	$\pm$ 3.2	&1.52	&10.0\\
G045.118$+$00.144$^a$ &45.119&	$+$0.144	&42.11	$\pm$ 2.17	&55.7	$\pm$ 0.5	&20.8	$\pm$ 1.2	&3.43	&19.1\\
G045.121$+$00.133     &45.122&	$+$0.133	&42.11	$\pm$ 2.17	&55.7	$\pm$ 0.5	&20.8	$\pm$ 1.2	&3.43	&19.1\\
G045.128$+$00.131$^a$ &45.129&	$+$0.131	&42.11	$\pm$ 2.17	&55.7	$\pm$ 0.5	&20.8	$\pm$ 1.2	&3.43	&19.1\\
G045.131$+$00.127$^a$ &45.132&	$+$0.128	&42.11	$\pm$ 2.17	&55.7	$\pm$ 0.5	&20.8	$\pm$ 1.2	&3.43	&19.1\\
G045.132$+$00.146$^a$ &45.133&	$+$0.146	&42.11	$\pm$ 2.17	&55.7	$\pm$ 0.5	&20.8	$\pm$ 1.2	&3.43	&19.1\\
G045.133$+$00.132$^a$ &45.133&	$+$0.133	&42.11	$\pm$ 2.17	&55.7	$\pm$ 0.5	&20.8	$\pm$ 1.2	&3.43	&19.1\\
G045.195$-$00.439     &45.195&	$-$0.439	&9.00	$\pm$ 0.55	&83.0	$\pm$ 1.9	&60.1	$\pm$ 4.8	&1.05	&22.7\\
G045.197$+$00.740     &45.197&	$+$0.740	&6.71	$\pm$ 0.98	&94.4	$\pm$ 2.9	&40.4	$\pm$ 6.8	&1.94	&7.5\\
G045.391$-$00.725     &45.391&	$-$0.724	&9.91	$\pm$ 0.52	&58.6	$\pm$ 0.7	&26.1	$\pm$ 1.6	&0.91	&19.0\\
G045.453$+$00.044     &45.453&	$+$0.045	&46.63	$\pm$ 1.87	&59.4	$\pm$ 0.4	&24.3	$\pm$ 1.2	&1.82	&43.2\\
G045.475$+$00.130     &45.475&	$+$0.130	&30.64	$\pm$ 2.15	&58.8	$\pm$ 0.7	&28.1	$\pm$ 1.9	&1.50	&37.0\\
G045.542$-$00.006     &45.542&	$-$0.006	&27.22	$\pm$ 1.21	&59.9	$\pm$ 0.8	&28.3	$\pm$ 1.8	&1.58	&31.2\\
                      &45.542&	$-$0.006	&7.85	$\pm$ 0.99	&108.4	$\pm$ 3.3	&43.1	$\pm$ 8.6	&1.58	&11.1\\
G045.634$-$00.016     &45.634&	$-$0.016	&13.01	$\pm$ 0.67	&58.9	$\pm$ 0.9	&33.5	$\pm$ 2.0	&1.34	&19.1\\
G045.825$-$00.291     &45.825&	$-$0.290	&9.92	$\pm$ 1.02	&56.8	$\pm$ 0.9	&17.4	$\pm$ 2.1	&1.48	&9.6\\
G045.838$-$00.296     &45.839&	$-$0.295	&7.13	$\pm$ 0.52	&59.7	$\pm$ 0.9	&24.3	$\pm$ 2.0	&0.88	&13.6\\
G045.933$-$00.403     &45.933&	$-$0.402	&3.12	$\pm$ 0.36	&71.1	$\pm$ 4.4	&77.3	$\pm$ 10.3	&1.09	&8.6\\
G046.088$+$00.254     &46.088&	$+$0.255	&3.21	$\pm$ 0.40	&16.0	$\pm$ 3.3	&54.3	$\pm$ 7.8	&1.02	&7.9\\
G046.203$+$00.532     &46.203&	$+$0.532	&11.48	$\pm$ 1.03	&102.5	$\pm$ 3.4	&75.2	$\pm$ 8.6	&2.11	&16.1\\
                      &46.203&	$+$0.532	&5.74	$\pm$ 1.41	&10.7	$\pm$ 4.9	&38.3	$\pm$ 11.7	&2.11	&5.8\\
G046.495$-$00.241     &46.495&	$-$0.240	&69.17	$\pm$ 1.18	&58.4	$\pm$ 0.2	&22.5	$\pm$ 0.4	&1.94	&57.8\\
G046.948$+$00.371$^a$ &46.948&	$+$0.371	&8.21	$\pm$ 0.77	&-42.5	$\pm$ 1.5	&32.8	$\pm$ 3.5	&1.52	&10.6\\
G047.028$+$00.216     &47.028&	$+$0.216	&4.41	$\pm$ 0.91	&61.3	$\pm$ 1.7	&17.8	$\pm$ 4.6	&1.07	&5.9\\
G048.456$+$00.123     &48.456&	$+$0.124	&14.65	$\pm$ 0.49	&17.7	$\pm$ 0.4	&23.9	$\pm$ 0.9	&0.83	&29.3\\
G048.547$-$00.005     &48.548&	$-$0.005	&51.99	$\pm$ 1.11	&19.8	$\pm$ 0.3	&28.8	$\pm$ 0.7	&2.07	&46.1\\
G048.599$+$00.044     &48.599&	$+$0.045	&53.93	$\pm$ 0.98	&19.7	$\pm$ 0.3	&28.4	$\pm$ 0.6	&1.80	&54.5\\
G048.604$+$00.022     &48.604&	$+$0.023	&114.93	$\pm$ 2.03	&20.2	$\pm$ 0.2	&25.4	$\pm$ 0.5	&3.55	&55.8\\
G048.630$+$00.230     &48.630&	$+$0.230	&27.22	$\pm$ 0.57	&10.7	$\pm$ 0.2	&21.7	$\pm$ 0.5	&0.92	&46.8\\
G048.922$-$00.285     &48.923&	$-$0.284	&285.12	$\pm$ 5.66	&67.0	$\pm$ 0.3	&30.8	$\pm$ 0.8	&7.38	&73.2\\
G049.002$-$00.303     &49.003&	$-$0.302	&207.99	$\pm$ 4.19	&66.4	$\pm$ 0.3	&28.4	$\pm$ 0.7	&7.76	&48.8\\
G049.051$-$00.255     &49.052&	$-$0.254	&257.09	$\pm$ 7.07	&65.9	$\pm$ 0.7	&27.3	$\pm$ 1.1	&6.58	&69.7\\
                      &49.052&	$-$0.254	&34.45	$\pm$ 7.31	&38.7	$\pm$ 5.4	&26.9	$\pm$ 8.2	&6.58	&9.3\\
G049.077$-$00.375     &49.078&	$-$0.375	&168.02	$\pm$ 9.95	&66.3	$\pm$ 0.9	&32.0	$\pm$ 2.1	&9.51	&34.1\\
                      &49.078&	$-$0.375	&39.98	$\pm$ 4.85	&118.6	$\pm$ 7.2	&64.5	$\pm$ 16.5	&9.51	&11.5\\
G049.163$-$00.066     &49.163&	$-$0.066	&25.66	$\pm$ 0.67	&63.9	$\pm$ 0.4	&28.1	$\pm$ 0.9	&1.23	&37.7\\
G049.201$-$00.365     &49.201&	$-$0.365	&98.37	$\pm$ 2.15	&65.3	$\pm$ 0.4	&35.9	$\pm$ 0.9	&4.45	&45.3\\
G049.384$-$00.298     &49.384&	$-$0.298	&287.30	$\pm$ 4.91	&54.4	$\pm$ 0.3	&32.0	$\pm$ 0.7	&9.43	&58.8\\
G049.399$-$00.490     &49.400&	$-$0.489	&83.48	$\pm$ 7.16	&62.7	$\pm$ 0.5	&23.6	$\pm$ 1.9	&3.71	&37.4\\
                      &49.400&	$-$0.489	&25.57	$\pm$ 3.41	&89.4	$\pm$ 6.7	&60.5	$\pm$ 8.9	&3.71	&18.3\\
G049.407$-$00.193     &49.407&	$-$0.193	&64.99	$\pm$ 4.24	&51.9	$\pm$ 0.2	&26.6	$\pm$ 1.1	&1.13	&101.7\\
G049.428$-$00.464     &49.428&	$-$0.464	&87.93	$\pm$ 6.80	&63.9	$\pm$ 0.6	&28.5	$\pm$ 2.0	&3.91	&41.1\\
                      &49.428&	$-$0.464	&19.49	$\pm$ 2.18	&107.8	$\pm$ 9.5	&67.9	$\pm$ 18.7	&3.91	&14.0\\
G049.484$-$00.391     &49.485&	$-$0.390	&470.70	$\pm$ 13.52	&62.8	$\pm$ 0.6	&45.5	$\pm$ 1.5	&29.50	&36.8\\
G049.489$-$00.378     &49.490&	$-$0.377	&161.89	$\pm$ 7.45	&62.6	$\pm$ 0.9	&37.7	$\pm$ 2.0	&15.68	&21.6\\
G049.501$-$00.524     &49.501&	$-$0.524	&43.04	$\pm$ 7.00	&65.6	$\pm$ 1.5	&27.1	$\pm$ 4.1	&1.87	&40.9\\
                      &49.501&	$-$0.524	&13.47	$\pm$ 2.35	&96.4	$\pm$ 14.2	&70.6	$\pm$ 20.2	&1.87	&20.6\\
                      &49.501&	$-$0.524	&10.84	$\pm$ 4.15	&40.1	$\pm$ 5.2	&21.3	$\pm$ 8.6	&1.87	&9.1\\
G049.592$-$00.456     &49.593&	$-$0.455	&43.74	$\pm$ 9.21	&62.3	$\pm$ 3.2	&52.3	$\pm$ 4.7	&2.93	&36.9\\
                      &49.593&	$-$0.455	&38.72	$\pm$ 9.52	&68.6	$\pm$ 1.2	&22.2	$\pm$ 4.1	&2.93	&21.2\\
                      &49.593&	$-$0.455	&10.11	$\pm$ 2.83	&112.7	$\pm$ 4.5	&24.1	$\pm$ 9.8	&2.93	&5.8\\
G049.690$-$00.166     &49.691&	$-$0.166	&15.86	$\pm$ 0.37	&60.5	$\pm$ 0.3	&29.8	$\pm$ 0.8	&0.70	&42.0\\
                      &49.691&	$-$0.166	&2.28	$\pm$ 0.33	&-21.3	$\pm$ 2.7	&39.1	$\pm$ 6.5	&0.70	&6.9\\
G049.828$+$00.366$^a$ &49.828&	$+$0.366	&7.73	$\pm$ 0.54	&6.5	$\pm$ 1.0	&29.7	$\pm$ 2.6	&1.01	&14.3\\
                      &49.828&	$+$0.366	&4.72	$\pm$ 0.51	&60.7	$\pm$ 1.8	&32.7	$\pm$ 4.4	&1.01	&9.2\\
G049.840$+$00.367$^a$ &49.840&	$+$0.368	&7.73	$\pm$ 0.54	&6.5	$\pm$ 1.0	&29.7	$\pm$ 2.6	&1.01	&14.3\\
                      &49.840&	$+$0.368	&4.72	$\pm$ 0.51	&60.7	$\pm$ 1.8	&32.7	$\pm$ 4.4	&1.01	&9.2\\
G049.997$-$00.087     &49.998&	$-$0.086	&16.55	$\pm$ 3.69	&73.9	$\pm$ 1.6	&21.1	$\pm$ 3.2	&0.81	&32.2\\
                      &49.998&	$-$0.086	&8.90	$\pm$ 1.58	&50.7	$\pm$ 5.4	&30.4	$\pm$ 8.3	&0.81	&20.8\\
                      &49.998&	$-$0.086	&4.54	$\pm$ 0.60	&105.2	$\pm$ 12.4	&67.1	$\pm$ 21.5	&0.81	&15.8\\
G049.997$-$00.130     &49.997&	$-$0.129	&28.96	$\pm$ 1.14	&74.3	$\pm$ 0.4	&17.7	$\pm$ 0.9	&1.65	&25.1\\
                      &49.997&	$-$0.129	&9.47	$\pm$ 1.02	&42.4	$\pm$ 1.3	&23.1	$\pm$ 3.3	&1.65	&9.4\\
G050.032$+$00.605     &50.033&	$+$0.606	&6.69	$\pm$ 1.26	&-0.1	$\pm$ 1.2	&14.0	$\pm$ 3.3	&1.24	&6.9\\
G050.038$-$00.274     &50.039&	$-$0.274	&5.30	$\pm$ 0.59	&97.7	$\pm$ 2.4	&44.6	$\pm$ 5.7	&1.13	&10.7\\
G050.079$+$00.571     &50.079&	$+$0.571	&5.55	$\pm$ 0.48	&75.7	$\pm$ 9.2	&88.0	$\pm$ 15.2	&0.71	&25.0\\
                      &50.079&	$+$0.571	&3.66	$\pm$ 0.83	&-2.9	$\pm$ 10.8	&70.8	$\pm$ 16.5	&0.71	&14.8\\
G050.137$-$00.660     &50.137&	$-$0.659	&18.93	$\pm$ 0.46	&75.1	$\pm$ 0.3	&23.1	$\pm$ 0.6	&0.76	&40.9\\
G050.262$-$00.409     &50.263&	$-$0.409	&5.50	$\pm$ 0.66	&7.8	$\pm$ 2.3	&39.4	$\pm$ 5.5	&1.43	&8.2\\
G050.287$-$00.392$^a$ &50.288&	$-$0.391	&4.03	$\pm$ 0.75	&11.7	$\pm$ 3.2	&34.9	$\pm$ 7.5	&1.53	&5.3\\
G050.298$+$00.674$^a$ &50.298&	$+$0.674	&3.86	$\pm$ 0.52	&115.9	$\pm$ 3.1	&46.1	$\pm$ 7.2	&0.86	&10.3\\
G050.489$+$00.992     &50.489&	$+$0.993	&4.02	$\pm$ 0.47	&106.9	$\pm$ 2.4	&41.7	$\pm$ 5.6	&0.84	&10.5\\
G050.785$+$00.167     &50.785&	$+$0.168	&28.68	$\pm$ 1.66	&47.8	$\pm$ 0.6	&19.6	$\pm$ 1.3	&2.54	&17.1\\
G051.010$+$00.060     &51.010&	$+$0.060	&15.51	$\pm$ 1.00	&47.1	$\pm$ 0.4	&24.3	$\pm$ 1.5	&0.47	&55.9\\
                      &51.010&	$+$0.060	&4.98	$\pm$ 1.01	&46.8	$\pm$ 2.7	&77.4	$\pm$ 11.0	&0.47	&32.1\\
G051.203$-$00.739     &51.204&	$-$0.738	&6.31	$\pm$ 0.32	&46.8	$\pm$ 0.7	&26.8	$\pm$ 1.6	&0.57	&19.7\\
G051.610$-$00.357     &51.610&	$-$0.357	&3.12	$\pm$ 0.42	&63.4	$\pm$ 1.3	&20.2	$\pm$ 3.1	&0.65	&7.4\\
G051.681$-$00.256     &51.681&	$-$0.256	&4.91	$\pm$ 0.35	&66.1	$\pm$ 1.0	&27.6	$\pm$ 2.2	&0.63	&14.0\\
G051.760$+$00.790     &51.760&	$+$0.791	&12.16	$\pm$ 0.66	&1.7	$\pm$ 0.6	&20.9	$\pm$ 1.3	&1.05	&18.1\\
G051.779$+$00.713     &51.779&	$+$0.714	&15.86	$\pm$ 0.73	&2.9	$\pm$ 0.6	&24.0	$\pm$ 1.3	&1.25	&21.3\\
G051.831$+$00.462     &51.831&	$+$0.463	&6.32	$\pm$ 1.60	&8.1	$\pm$ 1.7	&24.5	$\pm$ 3.7	&0.67	&16.1\\
                      &51.831&	$+$0.463	&5.11	$\pm$ 0.51	&-20.5	$\pm$ 5.9	&43.3	$\pm$ 9.3	&0.67	&17.2\\
G052.160$+$00.708     &52.160&	$+$0.708	&16.68	$\pm$ 0.92	&4.0	$\pm$ 0.7	&26.0	$\pm$ 1.7	&1.62	&17.9\\
G052.174$-$00.567     &52.175&	$-$0.567	&16.40	$\pm$ 0.48	&44.1	$\pm$ 0.8	&22.6	$\pm$ 1.4	&0.66	&40.4\\
                      &52.175&	$-$0.567	&11.52	$\pm$ 0.74	&65.8	$\pm$ 0.9	&18.2	$\pm$ 1.5	&0.66	&25.4\\
G052.201$+$00.752     &52.201&	$+$0.752	&22.22	$\pm$ 0.74	&4.4	$\pm$ 0.4	&25.1	$\pm$ 1.1	&1.25	&30.6\\
G052.232$+$00.735     &52.232&	$+$0.735	&22.55	$\pm$ 1.54	&0.9	$\pm$ 1.5	&30.2	$\pm$ 1.9	&1.45	&29.2\\
G052.256$+$00.702     &52.256&	$+$0.703	&17.08	$\pm$ 0.47	&4.5	$\pm$ 0.4	&25.3	$\pm$ 0.9	&0.81	&36.1\\
G052.398$-$00.591     &52.398&	$-$0.590	&13.12	$\pm$ 1.04	&60.5	$\pm$ 1.7	&40.3	$\pm$ 4.0	&1.88	&15.1\\
G052.537$-$00.946     &52.538&	$-$0.946	&8.43	$\pm$ 0.53	&62.3	$\pm$ 0.5	&15.5	$\pm$ 1.2	&0.57	&20.0\\
G052.799$-$00.534     &52.799&	$-$0.534	&12.44	$\pm$ 0.23	&51.6	$\pm$ 0.2	&24.1	$\pm$ 0.5	&0.39	&53.7\\
G052.980$+$00.133     &52.981&	$+$0.134	&5.59	$\pm$ 0.58	&4.0	$\pm$ 1.6	&31.7	$\pm$ 3.9	&0.99	&10.8\\
G053.095$+$00.212     &53.095&	$+$0.212	&19.69	$\pm$ 0.97	&4.6	$\pm$ 1.0	&38.6	$\pm$ 2.4	&1.25	&33.4\\
                      &53.095&	$+$0.212	&8.49	$\pm$ 0.63	&88.7	$\pm$ 3.8	&87.9	$\pm$ 9.9	&1.25	&21.7\\
G053.184$+$00.155     &53.184&	$+$0.155	&69.29	$\pm$ 2.66	&5.0	$\pm$ 0.7	&29.4	$\pm$ 1.7	&4.29	&29.9\\
                      &53.184&	$+$0.155	&18.16	$\pm$ 2.81	&44.1	$\pm$ 2.5	&25.6	$\pm$ 6.2	&4.29	&7.3\\
G053.541$-$00.011     &53.541&	$-$0.011	&9.79	$\pm$ 0.40	&32.6	$\pm$ 0.6	&29.2	$\pm$ 1.4	&0.75	&24.0\\
G053.644$+$00.014     &53.645&	$+$0.015	&6.86	$\pm$ 0.45	&31.5	$\pm$ 1.2	&37.3	$\pm$ 2.8	&0.95	&15.0\\
G053.822$-$00.057     &53.822&	$-$0.057	&10.28	$\pm$ 1.35	&48.4	$\pm$ 1.6	&26.6	$\pm$ 4.4	&1.44	&12.5\\
G053.935$+$00.228$^a$ &53.935&	$+$0.228	&7.32	$\pm$ 0.16	&38.5	$\pm$ 0.4	&34.1	$\pm$ 0.9	&0.33	&44.4\\
G054.099$-$00.068     &54.099&	$-$0.068	&28.66	$\pm$ 0.90	&41.9	$\pm$ 0.3	&18.9	$\pm$ 0.7	&1.36	&31.2\\
G055.158$-$00.297     &55.158&	$-$0.297	&6.78	$\pm$ 0.90	&105.0	$\pm$ 4.2	&64.0	$\pm$ 9.9	&2.06	&9.0\\
G056.083$-$00.176     &56.083&	$-$0.175	&5.13	$\pm$ 0.63	&46.8	$\pm$ 0.9	&15.6	$\pm$ 2.2	&0.86	&8.0\\
G057.541$-$00.279$^a$ &57.542&	$-$0.279	&6.66	$\pm$ 0.73	&-1.4	$\pm$ 1.0	&26.6	$\pm$ 3.4	&0.91	&12.9\\
G057.545$-$00.275$^a$ &57.546&	$-$0.274	&5.24	$\pm$ 0.75	&-17.2	$\pm$ 1.5	&16.1	$\pm$ 3.8	&0.96	&7.5\\
G059.321$-$00.223     &59.321&	$-$0.222	&12.41	$\pm$ 0.43	&26.2	$\pm$ 0.4	&23.7	$\pm$ 1.0	&0.72	&28.5\\
G059.600$+$00.316     &59.600&	$+$0.316	&2.34	$\pm$ 0.35	&35.0	$\pm$ 2.1	&28.6	$\pm$ 5.0	&0.65	&6.6\\
G059.796$+$00.241     &59.797&	$+$0.241	&12.45	$\pm$ 1.14	&-2.9	$\pm$ 1.5	&33.2	$\pm$ 3.5	&2.28	&10.7\\
G060.881$-$00.135     &60.881&	$-$0.135	&5.75	$\pm$ 0.81	&10.7	$\pm$ 4.2	&28.2	$\pm$ 6.0	&0.82	&12.7\\
G061.292$-$00.331     &61.292&	$-$0.331	&8.38	$\pm$ 1.72	&91.9	$\pm$ 4.2	&41.2	$\pm$ 9.8	&3.64	&5.0\\
G061.467$+$00.380     &61.467&	$+$0.380	&3.89	$\pm$ 0.30	&21.2	$\pm$ 0.8	&20.5	$\pm$ 1.8	&0.47	&12.9\\
G061.473$+$00.094     &61.473&	$+$0.095	&34.86	$\pm$ 2.69	&27.2	$\pm$ 1.1	&30.1	$\pm$ 2.7	&5.11	&12.8\\
G062.921$+$00.079     &62.921&	$+$0.079	&10.72	$\pm$ 0.40	&21.0	$\pm$ 0.4	&23.8	$\pm$ 1.0	&0.67	&26.7\\
G063.052$-$00.332     &63.052&	$-$0.332	&4.15	$\pm$ 0.36	&-2.9	$\pm$ 1.8	&43.5	$\pm$ 4.4	&0.81	&11.5\\
G063.164$+$00.449     &63.164&	$+$0.449	&69.45	$\pm$ 4.06	&19.5	$\pm$ 0.5	&16.5	$\pm$ 1.1	&5.71	&16.9\\
G064.130$-$00.475     &64.131&	$-$0.474	&4.54	$\pm$ 0.30	&26.8	$\pm$ 0.9	&27.9	$\pm$ 2.1	&0.55	&14.9\\
\end{longtable}
\begin{tablenotes}
	\item a. Detections arise from positions with multiple ``known'' or ``candidate'' \hii\ regions within the ALFA beam.
	\item b. Detections integrated over the entire \hii\ region also includes positions from overlapping ``known'' or ``candidate'' \hii\ regions.
	\item c. For the source G037.028$-$00.202, the V$_{LSR}$ component at $-40$ \kms\ is possible to be the helium conterpart of the H-RRL at 85.4 \kms.
\end{tablenotes}
\end{ThreePartTable}

\begin{ThreePartTable}
\begin{longtable}{lccccccc}
\caption{Catalog of 108 RRL detections towards 79 ``candidate'' HII regions from the WISE catalog\label{tab_HII_candidate}} \\
\hline
\hline
Name & $l$ & $b$ & Peak& V$_{LSR}$& FWHM& RMS& $S/N$\\
&$^\circ$&$^\circ$&mJy\,beam$^{-1}$&km\,s$^{-1}$&km\,s$^{-1}$&mJy\,beam$^{-1}$& \\
\hline
\endfirsthead
\caption{continued \ldots}\\
\hline
\hline
Name & $l$ & $b$ & Peak& V$_{LSR}$& FWHM& RMS& $S/N$\\
&$^\circ$&$^\circ$&mJy\,beam$^{-1}$&km\,s$^{-1}$&km\,s$^{-1}$&mJy\,beam$^{-1}$& \\
\hline
\endhead
\hline
\endfoot
\hline
\endlastfoot
G033.024$-$00.366     &33.025&$-$0.366&26.15	$\pm$2.02&50.8	$\pm$1.2	&28.2	$\pm$3.0	&2.61	&18.2\\
                      &33.025&$-$0.366&11.14	$\pm$1.76&104.6	$\pm$3.5	&43.1	$\pm$11.0	&2.61	&9.6\\
G033.306$-$00.150$^a$ &33.307&$-$0.150&8.23	$\pm$0.70&106.8	$\pm$1.2	&28.4	$\pm$2.8	&0.55	&27.5\\
G033.734$-$00.316$^a$ &33.735&$-$0.315&5.34	$\pm$0.68&52.7	$\pm$1.2	&20.3	$\pm$3.3	&0.71	&11.6\\
G034.130$-$00.174$^b$ &34.131&$-$0.173&12.77	$\pm$0.78&50.1	$\pm$0.6	&28.1	$\pm$2.0	&0.71	&32.4\\
                      &34.131&$-$0.173&3.12	$\pm$0.89&90.7	$\pm$1.8	&15.7	$\pm$5.3	&0.71	&5.9\\
G034.174$-$00.086$^b$ &34.175&$-$0.085&10.50	$\pm$0.99&48.2	$\pm$0.9	&29.1	$\pm$3.2	&0.80	&24.3\\
                      &34.175&$-$0.085&4.59	$\pm$1.06&92.2	$\pm$1.3	&13.1	$\pm$3.8	&0.80	&7.1\\
G034.190$-$00.063$^b$ &34.191&$-$0.063&10.26	$\pm$0.90&48.9	$\pm$0.8	&30.8	$\pm$2.8	&0.74	&26.4\\
                      &34.191&$-$0.063&4.32	$\pm$0.92&90.5	$\pm$1.2	&14.3	$\pm$3.6	&0.74	&7.6\\
G034.422$-$00.181     &34.423&$-$0.181&10.19	$\pm$0.49&88.8	$\pm$1.3	&27.9	$\pm$2.7	&0.43	&42.5\\
                      &34.423&$-$0.181&5.31	$\pm$0.45&56.8	$\pm$2.7	&29.7	$\pm$5.0	&0.43	&22.9\\
G034.443$+$00.103     &34.444&$+$0.104&23.64	$\pm$0.54&64.3	$\pm$0.3	&21.3	$\pm$0.7	&0.81	&46.0\\
                      &34.444&$+$0.104&6.87	$\pm$0.59&92.8	$\pm$0.8	&16.9	$\pm$2.0	&0.81	&11.9\\
G034.469$-$00.020     &34.470&$-$0.020&33.23	$\pm$1.97&92.2	$\pm$0.9	&17.7	$\pm$1.7	&2.18	&21.9\\
                      &34.470&$-$0.020&25.33	$\pm$4.60&54.6	$\pm$4.8	&31.8	$\pm$5.4	&2.18	&22.4\\
                      &34.470&$-$0.020&22.17	$\pm$9.55&65.5	$\pm$0.8	&12.7	$\pm$3.7	&2.18	&12.4\\
G034.687$+$00.012     &34.687&$+$0.012&16.44	$\pm$2.08&59.2	$\pm$1.3	&35.8	$\pm$4.7	&1.79	&18.8\\
G034.689$-$00.020$^b$ &34.690&$-$0.020&10.43	$\pm$1.08&59.4	$\pm$1.3	&32.2	$\pm$4.0	&1.18	&17.2\\
G035.349$+$00.004     &35.350&$+$0.005&19.49	$\pm$2.58&57.0	$\pm$0.9	&21.5	$\pm$2.4	&0.81	&38.3\\
                      &35.350&$+$0.005&3.50	$\pm$1.28&34.3	$\pm$4.9	&20.4	$\pm$8.3	&0.81	&6.7\\
                      &35.350&$+$0.005&3.40	$\pm$0.80&81.7	$\pm$16.6	&56.8	$\pm$24.0	&0.81	&10.8\\
G035.457$-$00.180     &35.457&$-$0.179&7.84	$\pm$0.89&48.6	$\pm$1.0	&17.0	$\pm$2.4	&1.26	&8.8\\
                      &35.457&$-$0.179&4.32	$\pm$0.82&81.1	$\pm$1.9	&20.1	$\pm$4.7	&1.26	&5.3\\
G035.489$-$00.079     &35.490&$-$0.079&19.84	$\pm$0.63&55.3	$\pm$0.6	&37.2	$\pm$1.6	&1.28	&32.2\\
G035.623$-$00.198     &35.624&$-$0.197&21.46	$\pm$2.74&52.5	$\pm$0.3	&20.1	$\pm$1.4	&0.65	&50.6\\
                      &35.624&$-$0.197&3.75	$\pm$1.19&34.0	$\pm$10.9	&38.6	$\pm$12.6	&0.65	&12.2\\
G035.629$+$00.074     &35.630&$+$0.075&16.24	$\pm$8.51&60.3	$\pm$2.3	&15.5	$\pm$3.2	&1.45	&15.0\\
                      &35.630&$+$0.075&16.00	$\pm$3.87&45.9	$\pm$5.1	&22.0	$\pm$6.4	&1.45	&17.6\\
G036.106$+$00.145     &36.106&$+$0.146&4.92	$\pm$2.02&89.2	$\pm$5.2	&53.7	$\pm$17.0	&0.71	&17.4\\
                      &36.106&$+$0.146&4.09	$\pm$1.27&99.8	$\pm$1.8	&14.7	$\pm$5.3	&0.71	&7.6\\
G036.262$-$00.721     &36.263&$-$0.720&3.34	$\pm$0.15&67.0	$\pm$1.7	&74.8	$\pm$3.9	&0.40	&24.4\\
G036.267$+$00.481$^b$ &36.268&$+$0.481&6.89	$\pm$0.40&78.0	$\pm$0.4	&12.1	$\pm$0.8	&0.48	&16.9\\
G036.495$+$00.061     &36.495&$+$0.062&5.52	$\pm$0.31&69.2	$\pm$1.1	&38.5	$\pm$2.5	&0.58	&20.1\\
G036.620$-$00.557     &36.620&$-$0.556&8.50	$\pm$4.59&57.3	$\pm$0.9	&20.3	$\pm$5.0	&0.55	&23.6\\
                      &36.620&$-$0.556&2.65	$\pm$2.31&41.6	$\pm$24.6	&38.0	$\pm$24.4	&0.55	&10.1\\
                      &36.620&$-$0.556&2.32	$\pm$0.33&106.1	$\pm$3.9	&39.2	$\pm$9.5	&0.55	&9.0\\
G037.029$-$00.025     &37.029&$-$0.024&6.33	$\pm$0.47&40.1	$\pm$1.0	&26.9	$\pm$2.4	&0.83	&13.5\\
                      &37.029&$-$0.024&6.06	$\pm$0.53&82.7	$\pm$0.9	&20.9	$\pm$2.2	&0.83	&11.4\\
G037.265$+$00.082$^b$ &37.266&$+$0.082&4.85	$\pm$0.85&72.7	$\pm$4.9	&56.6	$\pm$11.5	&2.23	&5.6\\
G037.419$+$01.513     &37.419&$+$1.514&19.55	$\pm$2.90&41.0	$\pm$1.3	&18.1	$\pm$3.1	&4.27	&6.6\\
G037.639$-$00.106$^a$ &37.640&$-$0.106&29.22	$\pm$0.96&52.4	$\pm$0.4	&25.4	$\pm$1.0	&1.65	&30.5\\
G037.838$-$00.510     &37.838&$-$0.509&10.59	$\pm$0.59&70.1	$\pm$0.4	&16.0	$\pm$1.0	&0.81	&17.8\\
G037.972$-$00.097     &37.972&$-$0.097&12.89	$\pm$1.05&65.2	$\pm$0.6	&15.4	$\pm$1.5	&1.43	&12.1\\
G038.017$-$00.385     &38.018&$-$0.385&10.53	$\pm$1.36&76.5	$\pm$3.0	&17.9	$\pm$4.6	&1.10	&13.8\\
G039.088$-$00.472     &39.088&$-$0.471&5.72	$\pm$0.79&61.5	$\pm$1.1	&16.2	$\pm$2.6	&1.11	&7.1\\
G039.294$-$00.311     &39.295&$-$0.311&53.52	$\pm$3.25&59.4	$\pm$1.1	&32.9	$\pm$2.3	&4.22	&24.9\\
                      &39.295&$-$0.311&14.59	$\pm$1.71&11.4	$\pm$6.6	&55.2	$\pm$14.3	&4.22	&8.8\\
G039.462$-$00.273$^b$ &39.462&$-$0.273&12.03	$\pm$0.47&66.5	$\pm$0.4	&22.2	$\pm$1.0	&0.77	&25.0\\
G039.506$-$00.280$^b$ &39.506&$-$0.280&8.81	$\pm$1.07&69.2	$\pm$0.9	&14.8	$\pm$2.1	&1.43	&8.1\\
G039.515$+$00.511$^c$ &39.516&$+$0.512&3.98	$\pm$0.21&-29.2	$\pm$0.9	&32.2	$\pm$2.2	&0.40	&19.4\\
                      &39.516&$+$0.512&1.70	$\pm$0.32&44.2	$\pm$1.3	&12.8	$\pm$3.2	&0.40	&5.2\\
G040.260$-$00.277     &40.261&$-$0.276&9.20	$\pm$1.65&46.7	$\pm$1.8	&19.8	$\pm$4.1	&2.55	&5.5\\
G040.418$-$00.055     &40.418&$-$0.054&4.82	$\pm$0.79&38.4	$\pm$3.5	&43.5	$\pm$8.3	&1.81	&6.0\\
G041.002$-$00.474$^a$ &41.002&$-$0.473&8.14	$\pm$0.49&63.8	$\pm$1.7	&21.4	$\pm$3.2	&0.67	&19.1\\
G041.042$+$00.306     &41.042&$+$0.307&9.24	$\pm$0.95&40.1	$\pm$1.9	&37.6	$\pm$4.5	&2.02	&9.6\\
G041.226$+$00.167     &41.227&$+$0.168&3.44	$\pm$0.53&63.3	$\pm$2.3	&30.4	$\pm$5.5	&1.02	&6.4\\
G041.373$+$00.086     &41.373&$+$0.087&5.58	$\pm$1.00&54.1	$\pm$5.5	&24.6	$\pm$17.7	&1.10	&8.6\\
                      &41.373&$+$0.087&4.54	$\pm$0.65&20.9	$\pm$5.4	&31.1	$\pm$10.2	&1.10	&7.8\\
G041.595$+$00.159     &41.595&$+$0.160&6.11	$\pm$0.82&23.0	$\pm$1.0	&15.3	$\pm$2.4	&1.10	&7.4\\
                      &41.595&$+$0.160&5.66	$\pm$0.78&53.1	$\pm$1.2	&17.0	$\pm$2.8	&1.10	&7.3\\
G041.800$+$00.094     &41.801&$+$0.094&7.78	$\pm$0.72&52.3	$\pm$2.2	&25.6	$\pm$4.5	&1.17	&11.5\\
                      &41.801&$+$0.094&5.27	$\pm$0.70&21.5	$\pm$3.3	&26.3	$\pm$6.8	&1.17	&7.9\\
G042.227$-$00.067$^a$ &42.228&$-$0.066&23.80	$\pm$1.22&55.6	$\pm$0.6	&22.1	$\pm$1.3	&1.99	&19.2\\
G042.239$+$00.343     &42.240&$+$0.343&24.14	$\pm$1.96&51.1	$\pm$1.8	&45.7	$\pm$4.3	&4.60	&12.1\\
G042.834$-$00.145$^b$ &42.834&$-$0.144&6.82	$\pm$0.86&70.5	$\pm$1.4	&22.0	$\pm$3.2	&1.39	&7.9\\
G042.834$-$00.157$^b$ &42.834&$-$0.157&8.27	$\pm$0.75&70.3	$\pm$0.8	&18.3	$\pm$1.9	&1.11	&10.9\\
G043.571$+$00.112$^b$ &43.572&$+$0.112&8.11	$\pm$0.63&84.3	$\pm$2.1	&55.3	$\pm$5.0	&1.34	&15.4\\
G043.617$+$00.059$^b$ &43.617&$+$0.059&5.96	$\pm$1.25&76.1	$\pm$1.6	&16.9	$\pm$4.4	&1.19	&7.0\\
G043.793$+$00.138     &43.794&$+$0.138&4.94	$\pm$0.76&76.2	$\pm$1.6	&20.9	$\pm$3.7	&1.20	&6.4\\
G043.969$-$00.277     &43.969&$-$0.277&2.09	$\pm$0.36&101.6	$\pm$2.9	&34.9	$\pm$6.9	&0.64	&6.6\\
G043.979$-$00.102     &43.979&$-$0.101&3.70	$\pm$0.51&93.0	$\pm$3.4	&50.0	$\pm$7.9	&1.08	&8.2\\
G044.243$-$00.129     &44.244&$-$0.129&8.26	$\pm$0.87&94.6	$\pm$3.3	&63.2	$\pm$7.7	&1.85	&12.1\\
G044.310$+$00.040$^a$ &44.310&$+$0.041&12.35	$\pm$0.79&75.1	$\pm$2.5	&81.5	$\pm$6.0	&1.91	&19.9\\
G044.375$-$00.076     &44.375&$-$0.076&1.87	$\pm$0.31&54.5	$\pm$2.6	&31.8	$\pm$6.1	&0.61	&6.0\\
G044.378$+$00.457$^b$ &44.379&$+$0.458&3.40	$\pm$0.69&97.0	$\pm$3.2	&31.8	$\pm$7.5	&1.24	&5.3\\
G045.191$-$00.486$^a$ &45.191&$-$0.485&9.30	$\pm$0.52&76.6	$\pm$2.2	&45.5	$\pm$5.1	&0.64	&33.7\\
                      &45.191&$-$0.485&4.14	$\pm$1.01&114.2	$\pm$2.7	&22.2	$\pm$6.2	&0.64	&10.5\\
G045.882$-$00.088     &45.883&$-$0.087&4.06	$\pm$0.56&68.3	$\pm$1.1	&17.5	$\pm$3.0	&0.69	&8.5\\
G046.033$-$00.097     &46.034&$-$0.096&4.72	$\pm$0.26&73.7	$\pm$2.4	&87.9	$\pm$5.6	&0.64	&23.6\\
G046.253$-$00.585     &46.253&$-$0.585&5.41	$\pm$0.40&89.9	$\pm$3.2	&87.1	$\pm$7.4	&0.89	&19.4\\
G046.392$+$00.861     &46.393&$+$0.862&2.19	$\pm$0.27&27.4	$\pm$4.3	&71.0	$\pm$10.2	&0.79	&8.0\\
G046.840$-$00.041     &46.841&$-$0.040&4.19	$\pm$0.43&112.4	$\pm$3.8	&74.6	$\pm$8.9	&0.89	&13.9\\
G047.580$-$00.075     &47.580&$-$0.075&4.24	$\pm$0.42&31.6	$\pm$2.4	&48.7	$\pm$5.5	&1.01	&10.0\\
G048.833$-$00.037     &48.834&$-$0.036&9.80	$\pm$0.42&54.5	$\pm$1.1	&39.3	$\pm$3.4	&0.66	&31.7\\
                      &48.834&$-$0.036&6.41	$\pm$0.67&16.3	$\pm$0.8	&15.5	$\pm$2.1	&0.66	&13.0\\
                      &48.834&$-$0.036&2.53	$\pm$0.37&112.3	$\pm$4.6	&45.2	$\pm$11.3	&0.66	&8.8\\
G049.048$-$00.886$^b$ &49.049&$-$0.886&18.83	$\pm$0.79&60.6	$\pm$1.5	&45.7	$\pm$3.3	&0.88	&49.7\\
G049.694$-$00.042$^b$ &49.695&$-$0.042&9.96	$\pm$0.37&54.1	$\pm$0.6	&34.2	$\pm$1.5	&0.74	&26.7\\
G049.775$-$00.951$^a$ &49.775&$-$0.951&2.66	$\pm$0.34&71.7	$\pm$1.6	&27.6	$\pm$3.9	&0.33	&14.4\\
G049.801$-$00.436     &49.801&$-$0.436&4.94	$\pm$0.81&68.5	$\pm$0.8	&21.8	$\pm$3.2	&0.48	&16.4\\
                      &49.801&$-$0.436&1.45	$\pm$0.35&94.3	$\pm$12.3	&55.3	$\pm$17.1	&0.48	&7.7\\
G051.457$-$00.286$^a$ &51.457&$-$0.285&2.52	$\pm$0.24&62.0	$\pm$0.8	&19.2	$\pm$2.3	&0.18	&20.7\\
G051.755$+$01.208     &51.755&$+$1.209&2.00	$\pm$0.21&28.2	$\pm$4.3	&83.1	$\pm$10.2	&0.68	&9.2\\
G052.369$-$01.050     &52.369&$-$1.050&9.16	$\pm$0.52&63.9	$\pm$0.7	&23.4	$\pm$1.5	&0.87	&17.5\\
G054.928$-$00.478     &54.928&$-$0.477&4.23	$\pm$0.50&35.7	$\pm$1.0	&17.5	$\pm$2.4	&0.73	&8.3\\
G055.567$+$00.701     &55.568&$+$0.701&7.53	$\pm$0.28&-1.3	$\pm$0.4	&19.1	$\pm$0.8	&0.42	&26.6\\
G056.725$-$00.179     &56.725&$-$0.178&7.81	$\pm$0.65&39.4	$\pm$3.1	&75.8	$\pm$7.3	&1.96	&11.8\\
G057.624$+$00.410$^b$ &57.625&$+$0.410&5.46	$\pm$0.58&94.1	$\pm$6.7	&74.4	$\pm$12.1	&0.74	&21.8\\
                      &57.625&$+$0.410&5.07	$\pm$1.13&38.3	$\pm$1.4	&14.5	$\pm$4.0	&0.74	&9.0\\
G057.715$+$00.173     &57.715&$+$0.174&1.95	$\pm$0.26&42.7	$\pm$2.6	&33.8	$\pm$7.2	&0.29	&13.3\\
                      &57.715&$+$0.174&1.80	$\pm$0.29&0.1	$\pm$2.0	&22.7	$\pm$4.9	&0.29	&10.0\\
                      &57.715&$+$0.174&1.71	$\pm$0.17&108.7	$\pm$4.4	&68.7	$\pm$11.3	&0.29	&16.6\\
G061.053$-$00.825$^b$ &61.054&$-$0.825&1.91	$\pm$0.32&5.1	$\pm$3.9	&47.9	$\pm$9.3	&0.77	&5.9\\
G061.144$-$00.891     &61.144&$-$0.891&2.51	$\pm$0.46&7.8	$\pm$1.7	&18.4	$\pm$3.9	&0.69	&5.4\\
G062.142$-$00.144     &62.142&$-$0.143&1.68	$\pm$0.26&14.1	$\pm$5.0	&63.9	$\pm$11.7	&0.73	&6.2\\
G063.050$-$00.124     &63.051&$-$0.124&6.53	$\pm$0.25&26.5	$\pm$0.5	&24.9	$\pm$1.1	&0.43	&26.0\\
G067.816$+$00.511     &67.816&$+$0.511&2.72	$\pm$0.49&2.0	$\pm$1.7	&19.1	$\pm$4.0	&0.75	&5.4\\
\end{longtable}
\begin{tablenotes}
	\item a. Detections arise from positions with multiple ``known'' or ``candidate'' \hii\ regions within the ALFA beam.
	\item b. Detections integrated over the entire \hii\ region also includes positions from overlapping ``known'' or ``candidate'' \hii\ regions.
\end{tablenotes}
\end{ThreePartTable}

\newpage

\bibliography{siggmaii}
\bibliographystyle{aasjournal}

\end{document}